\documentclass{article}

\PassOptionsToPackage{numbers,square,sort&compress}{natbib}
 \usepackage[preprint]{neurips_2026}

\usepackage[utf8]{inputenc} 
\usepackage[T1]{fontenc}    
\usepackage{hyperref}       
\usepackage{url}            
\usepackage{booktabs}       
\usepackage{amsfonts}       
\usepackage{nicefrac}       
\usepackage{microtype}      
\usepackage{xcolor}         

\usepackage{hyperref}

\title{\textsc{CityMPC}: A Large-Scale Physics-Informed Benchmark and Tool for Generative Complete Multipath Wireless Channel Modeling}

\author{%
Ashwin Natraj Arun$^{1*}$ \quad David R. Nickel$^{1}$ \quad Yaguang Zhang$^2$\\
\textbf{Yunchou Xing}$^3$ \quad  \textbf{Jie Chen} $^3$ \quad \textbf{Amitava Ghosh} $^3$\\
\textbf{Christopher G. Brinton}$^1$ \quad \textbf{David J. Love}$^1$ \quad \textbf{James V. Krogmeier}$^{1,2}$ \\
$^1$Purdue Univ. Elmore Family School of Electrical and Computer Engineering \\
$^2$Purdue Univ. School of Agricultural \& Biological Engineering\\
$^3$Nokia Standards, USA\\
\texttt{\{ashwin97,dnickel,ygzhang,cgb,djlove,jvk\}@purdue.edu}\\
\texttt{\{yunchou.xing,jie.chen,amitava.ghosh\}@nokia.com}\\
}

\usepackage[ruled,linesnumbered,noend]{algorithm2e}
\usepackage{algpseudocode}
\usepackage{amsmath,amssymb,amsfonts,amsthm}
\usepackage{balance,bm,booktabs,breqn}
\usepackage{cancel}
\usepackage{enumitem}
\usepackage{fancyhdr,float,footnote}
\usepackage{glossaries, graphicx}
\usepackage{ifpdf}
\usepackage[utf8]{inputenc}
\usepackage{mathtools,moresize,multirow}
\usepackage{soul,caption}
\usepackage{tabularx,textcomp}
\usepackage{ulem}
\usepackage{wrapfig}
\usepackage{xcolor,xfrac,xspace}

\makesavenoteenv{algorithm}

\newcommand{\C}{\mathbb{C}}

\newcommand{\R}{\mathbb{R}}

\newcommand{\bracks}[1]{\left[{#1}\right]}

\newcommand{\mcl}{\mathcal{L}}

\newcommand{\bma}{\bm{a}}

\newcommand{\bmc}{\bm{c}}

\newcommand{\bmm}{\bm{m}}

\newcommand{\bmx}{\bm{x}}

\newcommand{\bmz}{\bm{z}}

\newcommand{\bmmu}{\bm{\mu}}

\newcommand{\bmsigma}{\bm{\sigma}}

\newcommand{\whbmx}{\widehat{\bm{x}}}

\newcommand{\ttce}{\texttt{CE}\xspace}
\newcommand{\ttpe}{\texttt{PE}\xspace}
\newcommand{\ttpn}{\texttt{PN}\xspace}
\newcommand{\ttdec}{\texttt{DEC}\xspace}

\newcommand{\tauzero}{\tau_0}
\newcommand{\dtauell}{\widetilde{\tau}_\ell}
\newcommand{\daodell}{\hat{\mathbf{d}}_\ell^{\mathrm{d}}}
\newcommand{\daoaell}{\hat{\mathbf{d}}_\ell^{\mathrm{a}}}

\newcommand{\mell}{m_\ell}
\newcommand{\Ptot}{P_{\mathrm{tot}}}
\newcommand{\Prx}{P_{\mathrm{rx}}}

\newcommand{\Wdelay}{W}
\newcommand{\dtau}{\widetilde{\boldsymbol{\tau}}}

\allowdisplaybreaks

\newacronym{ltv}{LTV}{linear time-varying}
\newacronym{aod}{AoD}{angle of departure}
\newacronym{aoa}{AoA}{angle of arrival}
\newacronym{nlos}{nLoS}{non line-of-sight}
\newacronym{los}{LoS}{line-of-sight}
\newacronym{mimo}{MIMO}{multiple-input multiple-output}
\newacronym{gpu}{GPU}{graphics processing unit}
\newacronym{ris}{RIS}{reconfigurable intelligent surface}
\newacronym{isac}{ISAC}{integrated sensing and communication}
\newacronym{airan}{AI-RAN}{artificial intelligence-radio access network}
\newacronym{cvae}{cVAE}{conditional variational autoencoder}
\newacronym{vae}{VAE}{variational autoencoder}
\newacronym{gan}{GAN}{generative adversarial network}
\newacronym{ddpm}{DDPM}{denoising diffusion probabilistic model}
\newacronym{itur}{ITU-R}{International Telecommunication Union Radiocommunication Sector}
\newacronym{imt}{IMT}{International Mobile Telecommunications}
\newacronym{elaa}{ELAA}{extremely large aperture array}
\newacronym{ntn}{NTN}{non-terrestrial network}
\newacronym{mpc}{MPC}{multipath component}
\newacronym{gbsm}{GBSM}{geometry-based stochastic channel model}
\newacronym{rt}{RT}{ray tracing}
\newacronym{rssi}{RSSI}{received signal strength indicator}
\newacronym{pov}{PoV}{point-of-view}
\newacronym{umi}{UMi}{Urban Microcell}
\newacronym{uma}{UMa}{Urban Macrocell}
\newacronym{rma}{RMa}{Rural Macrocell}
\newacronym{aodt}{AODT}{NVIDIA Aerial Omniverse Digital Twin}
\newacronym{csi}{CSI}{channel state information}
\newacronym{mmwave}{mmWave}{millimeter wave}
\newacronym{ai}{AI}{artificial intelligence}
\newacronym{ran}{RAN}{radio access network}
\newacronym{6g}{6G}{sixth generation}
\newacronym{3gpp}{3GPP}{3rd Generation Partnership Project}
\newacronym{cir}{CIR}{channel impulse response}
\newacronym{bs}{BS}{base station}
\newacronym{ue}{UE}{user equipment}
\newacronym{cddpm}{cDDPM}{conditional denoising diffusion probabilistic model}
\newacronym{rsrp}{RSRP}{reference signal received power}
\newacronym{uav}{UAV}{unmanned aerial vehicle}
\newacronym{tx}{Tx}{transmitter}
\newacronym{rx}{Rx}{receiver}
\newacronym{mlp}{MLP}{multilayer perceptron}
\newacronym{cfr}{CFR}{crest factor reduction}
\newacronym{em}{EM}{electromagnetic}
\newacronym{rf}{RF}{radio frequency}
\newacronym{elbo}{ELBO}{evidence lower bound}
\newacronym{kl}{KL}{Kullback-Leibler}
\newacronym{mse}{MSE}{mean squared error}
\newacronym{bce}{BCE}{binary cross-entropy}
\newacronym{siso}{SISO}{single-input single-output}
\newacronym{tof}{ToF}{time-of-flight}
\newacronym{mae}{MAE}{mean absolute error}
\newacronym{f1}{F1}{F1 score}
\newacronym{cdf}{CDF}{cumulative distribution function}
\newacronym{llm}{LLM}{large language model}
\newacronym{hdf}{HDF5}{Hierarchical Data Format version 5}
\newacronym{uw}{UW}{uncertainty weighting}

\begin{document}

\maketitle


\begin{abstract}
    Multipath wireless channels are fully characterized by \glspl{mpc}, including complex channel gain, propagation delay, \gls{aod} and \gls{aoa} in azimuth and elevation.
    Generating these parameters with the fidelity of \gls{rt} remains an open problem.
    Existing methods either incur the computational cost of \gls{rt} or require explicit 3D scene geometry at inference.
    We present \textsc{CityMPC}, a \gls{cvae} that predicts the complete per-path \gls{mpc} parameter set from point-of-view imagery and terrain height maps alone, achieving environment-aware channel generation without access to any three-dimensional scene geometry at inference.
    Trained and evaluated across five urban environments spanning 427,397 links, \textsc{CityMPC} matches \gls{rt} ground truth to within 1.29\,dB received power \gls{mae} and 7.25\,ns $\tau_0$ \gls{mae}.
    \textsc{CityMPC} is a generative channel modeling framework\footnote{Code will be made available upon publication.} and reproducible benchmark, released together with a large-scale multi-city ray-traced dataset\footnote{Dataset will be made available upon publication.} to accelerate future scene-conditioned channel modeling research.
    We further analyze cross-city distribution shift to characterize the per-city diversity of the benchmark.
\end{abstract}

\section{Introduction}\label{sec:introduction}
The \gls{6g} of wireless communication systems is expected to fundamentally expand the role of the network beyond connectivity.
\gls{6g} is envisioned as a platform serving use cases such as immersive communication, massive communication, ubiquitous connectivity, hyper-reliable and low-latency communication, artificial intelligence and communication, and \gls{isac} \cite{itu_m2160,nga2023vertical,brinton2025_6g,11498538}.
Underpinning this is a requirement unmet by prior generations: the network must possess rich site-specific knowledge of the propagation environment at each link.
 
The physical layer functions demanded by \gls{6g} are directly limited by channel model fidelity.
At higher frequencies, \gls{6g} systems will utilize hundreds of antenna elements to overcome severe pathloss. 
Such systems require explicit knowledge of each individual \gls{mpc}, including the \gls{aod} and \gls{aoa} in azimuth and elevation, the excess propagation delay, and the complex path gain \cite{rappaport2022radio, samimi20163}. 
\gls{6g} is also seen as the first generation to embed AI natively into the radio access network, innately requiring a rich channel model to generate training data, validate algorithms offline, and evaluate system performance under various conditions. 
A channel model that cannot reproduce the environment's site-specific \glspl{mpc} will yield failure-prone AI components. 
Towards that end, wireless channel modeling aims to characterize how \gls{em} waves propagate (i.e., reflect, diffract, and scatter off of buildings, terrain, etc.) between a \gls{tx} and a \gls{rx} in a given locale. 
These interactions create multipath propagation, where multiple copies of the transmitted signal reach the \gls{rx} with different delays, angles, and gains, noting here that a wireless channel can be fully described by its multipath structure~\cite{tse2005fundamentals, samimi20163}. 
An explicit formulation of a multipath channel is given in eq. \eqref{eq:multipath_def}, but for now, we note that channel modeling can essentially be reduced to predicting each path's \glspl{mpc}.

Two paradigms have dominated channel modeling, each with structural trade-offs.
The first is the \gls{gbsm}, standardized in 3GPP TR~38.901 \cite{38901}. 
In the \gls{gbsm}, \glspl{mpc} are organized into clusters whose parameters are drawn from statistical distributions calibrated against measurement campaigns \cite{samimi20163}. 
The model is computationally efficient but environment-agnostic, and it neglects the fundamentally different characteristics of \gls{los} and \gls{nlos} links. 
Alternatively, deterministic \gls{rt}, implemented in tools such as Sionna \gls{rt} \cite{sionna}, is used to generate synthetic datasets for learning-based wireless models \cite{10465179} using scene-specific 3D geometries. 
\gls{rt} is computationally expensive, produces a single deterministic channel realization per link configuration, and presupposes the availability of a georeferenced 3D scene mesh. 
It produces no compact representation of the channel distribution that could be queried in real time or reused across different propagation environments. 
This inference-time dependency on scene geometry is the fundamental limitation that motivates a learned surrogate.
\begin{wrapfigure}{R}{0.28\textwidth}
  \vspace{-10mm}
  \begin{center}
    \includegraphics[width=0.27\textwidth]{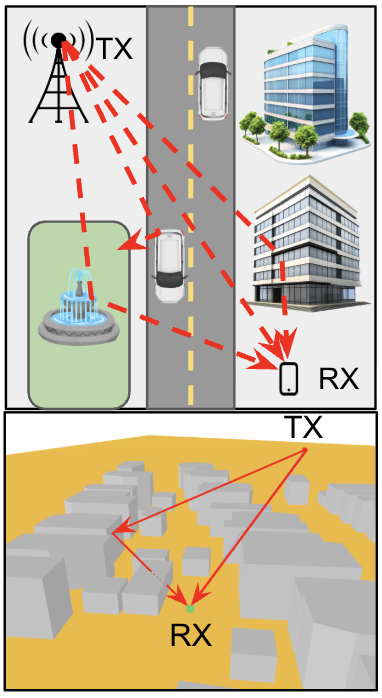}
  \end{center}
  \caption{Top: pictorial representation of multipath propagation. Bottom: sample Sionna \gls{rt} datapoint.}
  \label{fig:multipath}
  \vspace{-13mm}
\end{wrapfigure}

\paragraph{Related work.}
Stochastic channel models such as 3GPP TR 38.901~\cite{38901} and NYUSIM~\cite{sun2017nyusim} provide closed-form parametric distributions but lack site-specific conditioning.
Ray tracing tools such as Sionna \gls{rt}~\cite{sionna} and Wireless InSite~\cite{remcom2021wireless} produce physically accurate channels at high computational cost, requiring detailed 3D geometry and per-link execution.
Recent learning-based approaches fall into three categories.
Generative models~\cite{xiao2022channelgan, sengupta2023diffusion, kim2024diffusionchannel, lee2025diffusionchannel} learn channel distributions from data but do not condition on site-specific scene information.
Neural \gls{rt} surrogates such as WiNeRT~\cite{orekondy2023winert} and GeNeRT~\cite{bian2025genert} predict per-path \gls{mpc} parameters but require a 3D scene mesh at inference.
Coordinate-conditioned and radio map methods~\cite{wang2025radiodiff3d, jiang2025learnable, 11443807, levie2021radiounet} predict aggregate signal quantities or partial \gls{mpc} information without producing the complete per-path parameter set.
A detailed comparison and full discussion of prior work is provided in Appendix~\ref{app:model_comparison}.

\subsection{Summary of contributions}

\paragraph{Complete \gls{mpc} characterization.}
We formulate the geometry-free \gls{mpc} prediction task of predicting the full per-path parameter set from \gls{pov} multi-channel imagery and a terrain heightmap alone, with no access to ray geometry or 3D mesh at inference.
The predicted parameters include \gls{aoa} and \gls{aod} in azimuth and elevation, excess delay, complex gain, \gls{tof} time, and received power.
To our knowledge, this is the first work to predict per-path \gls{aoa} and \gls{aod} elevation angles conditioned only on scene imagery, making \textsc{CityMPC} a benchmark for future research into scene-conditioned channel models.

\paragraph{Generative matching of \gls{rt} performance.}
We propose a transformer-based \gls{cvae} architecture that generates the full structured \gls{mpc} parameter set via learned slot queries~\cite{carion2020detr}, with multi-task loss balancing via Kendall uncertainty weighting~\cite{kendall2018multi}.
The model is trained and evaluated on roughly half a million \gls{tx}-\gls{rx} links spanning five US cities generated via DeepMIMO~\cite{alkhateeb2019deepmimo} and Sionna \gls{rt}~\cite{sionna}.
\textsc{CityMPC} learns site-specific channel distributions whose generated \gls{mpc} realizations match ray-traced ground truth across all predicted parameters.

\paragraph{Dataset generator and dataset release.}
Once trained on ray-traced data from a city, \textsc{CityMPC} generates arbitrarily many physically valid \gls{mpc} realizations at inference time without ray tracing or a 3D scene mesh, at lower computational cost.
This makes \textsc{CityMPC} a practical dataset generation tool for researchers who need large-scale site-specific channel data without repeated \gls{rt} runs.
Along with \textsc{CityMPC}, we also release a large-scale multi-city ray traced dataset to accelerate future scene-conditioned channel modeling research.

\section{System model and dataset}
\subsection{Multipath wireless channels}\label{sec:channel}
\Gls{rf} propagation between the \gls{tx} and \gls{rx} undergoes reflection, refraction, diffraction, and scattering.
This produces multiple discrete propagation paths arriving at the receiver with distinct delays and directions.
A basic illustration of this principle is seen in Fig. \ref{fig:multipath}.
We model the wireless channel as a \gls{ltv} system \cite{tse2005fundamentals,rappaport2013mmwave} expressed as
\begin{equation}\label{eq:multipath_def}
    h(t,\Theta,\Phi)=\sum_{\ell=1}^{L}m_{\ell} (t)\,\alpha_{\ell}(t)\,
    \delta(t-\tau_{\ell}(t))\,
    \delta \big(\Theta - \Theta_\ell(t)\big)\,
    \delta \big(\Phi - \Phi_\ell(t)\big),
\end{equation}
where $\bmm=\bracks{m_1,\dots,m_{L}}$ is the path presence mask, with $m_\ell=1$ for active paths and $m_\ell=0$ otherwise.
Each active path $\ell$ carries a complex baseband gain $\alpha_\ell\in\C$, an absolute propagation delay $\tau_\ell\in\R_+$, an azimuth and elevation \gls{aod} $\Theta_\ell=(\theta_\ell^{\mathrm{az}}, \theta_\ell^{\mathrm{el}})\in[-\pi,\pi)\times[0,\pi]$, and an azimuth and elevation \gls{aoa} $\Phi_\ell=(\phi_\ell^{\mathrm{az}}, \phi_\ell^{\mathrm{el}})$.
The function $\delta(\cdot)$ denotes the Dirac delta.
Under the quasi-static assumption \cite{tse2005fundamentals}, the channel parameters are approximately constant over a coherence interval, 
so we drop the time dependence and model each link as a single static snapshot $\{m_\ell,\alpha_\ell,\tau_\ell,\Theta_\ell,\Phi_\ell\}_{\ell=1}^{L}$.
Channel modeling then reduces to predicting, for each link, these per-path parameters alongside the \gls{tof} time $\tauzero=\min\{\tau_\ell : m_\ell=1\}$ and the aggregate received power $P_{\mathrm{rx}}=10\log_{10}\!\bigl(\sum_\ell m_\ell|\alpha_\ell|^2\bigr)$~dB (see Appendix~\ref{app:normalization} for normalization details).

\subsection{Construction of training/testing dataset}\label{sec:dataset}

\paragraph{Channel extraction.}
We employ DeepMIMO~\cite{alkhateeb2019deepmimo} to extract per-link \gls{mpc} parameters across 20 urban US city scenarios, selecting only those that use Sionna \gls{rt}~\cite{10465179} as their ray tracing backend at $3.5$\,GHz.
Each scenario covers a $512 \times 512$\,m urban area, yielding a consistent geographic footprint across all cities.
We configure DeepMIMO to generate up to $L=25$ paths per link, each carrying a complex gain $\alpha_\ell$, absolute delay $\tau_\ell$, \gls{aod} $\Theta_\ell$, and \gls{aoa} $\Phi_\ell$.
Channels are power-ordered such that $|\alpha_1| \geq |\alpha_2| \geq \cdots \geq |\alpha_L|$, which does not imply a corresponding ordering on arrival times $\tau_\ell$.
Link- and path-level filtering criteria are detailed in Appendix~\ref{app:dataset}.

\paragraph{Scene rendering.}
For each city, we convert the DeepMIMO scenario geometry to a Sionna-compatible Mitsuba XML scene \cite{Mitsuba3} and render two 12-channel \gls{pov} image stacks, one from the \gls{tx} and one from the \gls{rx}, at $128 \times 128$ resolution.
Each \gls{pov} stack captures the local geometric perspective of what the \gls{tx} or \gls{rx} sees, encoding the 3D objects from which rays reflect, diffract, and scatter on their propagation path.
Alongside RGB appearance, depth, and surface normals, each stack encodes the radio material properties of visible surfaces, namely relative permittivity $\varepsilon_r$, conductivity $\sigma$, scattering coefficient, cross-polarization coefficient, and material thickness, which govern ray-surface interactions through Maxwell's equations \cite{10465179}.
Along with the \gls{pov} stacks, we also generate a global height map at $128 \times 128$ resolution by capturing the macro-scale building layout of the full scenario, together providing a complete geometric and electromagnetic description of the environment without requiring the 3D scene mesh at inference.
The global height map has a 4\,m per pixel spatial resolution as our 3D mesh covers an area of $512 \times 512$\,m.
Complete details of the conversion and rendering pipeline are given in Appendix~\ref{app:scene_conversion}.

\begin{figure}[t]
\centering
\scriptsize
\setlength{\tabcolsep}{1pt}
\begin{tabular}{ccccccccc}
& RGB & Normal & Depth & $\varepsilon_r$ & $\sigma$ & Scatter & XPD & Thickness \\[2pt]
&
\includegraphics[height=1.3cm]{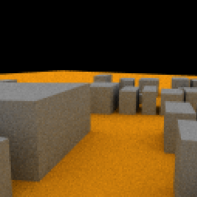} &
\includegraphics[height=1.3cm]{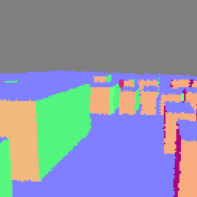} &
\includegraphics[height=1.3cm]{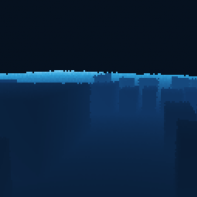} &
\includegraphics[height=1.3cm]{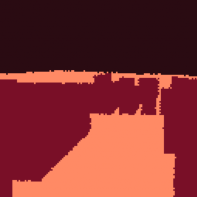} &
\includegraphics[height=1.3cm]{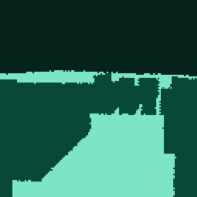} &
\includegraphics[height=1.3cm]{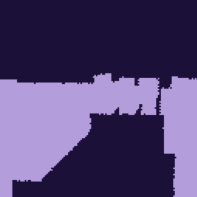} &
\includegraphics[height=1.3cm]{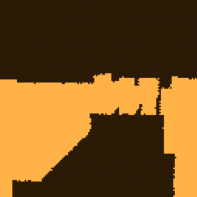} &
\includegraphics[height=1.3cm]{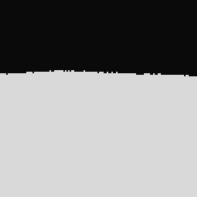} \\
& \multicolumn{8}{c}{\textbf{\Gls{tx} \gls{pov}}} \\[1pt]
\multirow{-10}{*}{
    \begin{minipage}{0.17\linewidth}
    \centering
    \includegraphics[width=\linewidth]{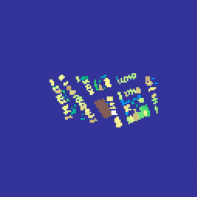}\\[1pt]
    {\tiny Global map}
    \end{minipage}
} &
\includegraphics[height=1.3cm]{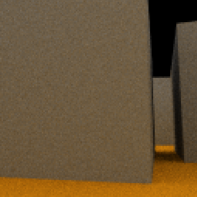} &
\includegraphics[height=1.3cm]{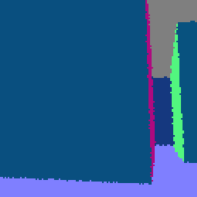} &
\includegraphics[height=1.3cm]{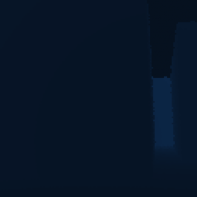} &
\includegraphics[height=1.3cm]{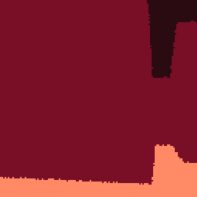} &
\includegraphics[height=1.3cm]{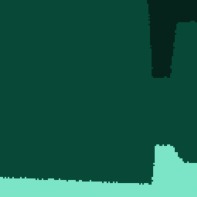} &
\includegraphics[height=1.3cm]{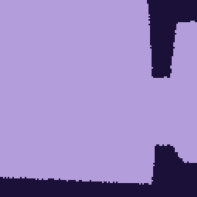} &
\includegraphics[height=1.3cm]{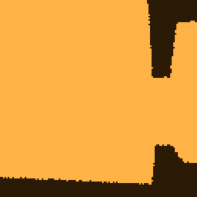} &
\includegraphics[height=1.3cm]{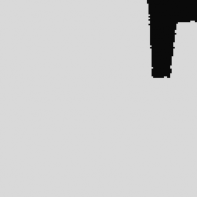} \\
& \multicolumn{8}{c}{\textbf{\Gls{rx} \gls{pov}}} \\
\end{tabular}
\caption{
Conditioning inputs for the \gls{tx}/\gls{rx} link shown in Fig.~\ref{fig:multipath}.
The global height map (left) encodes macro-scale building layout across the full $512{\times}512$\,m scenario.
Each \gls{pov} stack provides 12 channels capturing the local geometric and electromagnetic environment: RGB appearance, depth, surface normals $(N_x, N_y, N_z)$, and five radio material properties ($\varepsilon_r$, $\sigma$, scattering coefficient, cross-polarization coefficient, and thickness) that govern ray-surface interactions~\cite{10465179}.
No 3D scene mesh is required at inference.
}
\label{fig:pov_inputs}
\end{figure}

\paragraph{Dataset statistics and splits.}
The DeepMIMO dataset used in this work spans 5 US cities -- Dallas, Fort Worth, New York, Denver, and Austin -- and a total of $427{,}397$ \gls{tx}-\gls{rx} links generated via Sionna \gls{rt} at 3.5\,GHz, as detailed in Table~\ref{tab:per_city_results}.
These cities span diverse urban morphologies, from dense high-rise environments to sprawling low-rise grids, providing varied propagation conditions for training and evaluation.
Each city is partitioned 70/15/15 into train, validation, and test splits using spatially disjoint \gls{tx} and \gls{rx} grid regions to prevent leakage.
Normalization statistics ($\mu_{\log}$, $\sigma_{\log}$, $\mu_{\mathrm{rx}}$, $\sigma_{\mathrm{rx}}$) are computed exclusively from each city's training split and applied independently per city.

\paragraph{Inputs and outputs.}
Let $\bmc = \{\mathbf{I}_{\mathrm{tx}}\in\R^{12\times128\times128},\, \mathbf{I}_{\mathrm{rx}}\in\R^{12\times128\times128},\, \mathbf{I}_{g}\in\R^{1\times128\times128},\, \mathbf{s}\in\R^{6}\}$ denote the conditioning information, comprising the \gls{tx} and \gls{rx} \gls{pov} image stacks, the global heightmap, and the raw \gls{tx} and \gls{rx} Cartesian coordinates in metres.
The \gls{pov} stacks $\mathbf{I}_{\mathrm{tx}}$ and $\mathbf{I}_{\mathrm{rx}}$ encode the local geometric and electromagnetic  perspective of what the \gls{tx} and \gls{rx} see, capturing the 3D surfaces from which rays interact on each propagation path (see Section~\ref{sec:dataset}).
The global heightmap $\mathbf{I}_{g}$ provides the macro-scale building layout of the full scenario.
The coordinates $\mathbf{s}$ are embedded via a sinusoidal Fourier feature encoding \cite{tancik2020fourier} as described in Appendix~\ref{app:conditioning}.
Full details of the normalization and preprocessing applied to each conditioning input are given in Appendix~\ref{app:conditioning}.
Let $\bmx=\bracks{x_1,\dots,x_{L}}\in\R^{L\times10}$ denote the per-path channel data, where each slot vector is
\begin{equation}\label{eq:y_def}
    x_{\ell} = \Bigl[\,
    \tilde{\alpha}_\ell^{\Re},\,
    \tilde{\alpha}_\ell^{\Im},\,
    \dtauell,\,
    \daodell,\,
    \daoaell,\,
    \mell
    \,\Bigr] \in \R^{10}.
\end{equation}
Here $\dtauell$ is the normalized excess delay of path $\ell$ relative to the \gls{tof} time $\tauzero$.
The terms $\daodell$ and $\daoaell$ are unit-vector encodings of the \gls{aod} $\Theta_\ell$ and the \gls{aoa} $\Phi_\ell$.
The term $[\tilde{\alpha}_\ell^{\Re}, \tilde{\alpha}_\ell^{\Im}]$ is the normalized complex baseband gain of path $\ell$.
The term $\mell \in \{0,1\}$ is the path presence indicator for path $\ell$.
Full normalization details for all quantities are given in Appendix~\ref{app:normalization}.
The goal of the trained network is to output a valid channel realization
\begin{equation}
    \whbmx = \bigl\{
    \hat{\tau}_0,\,
    \hat{P}_{\mathrm{rx}},\,
    \hat{\bm{m}},\,
    \widehat{\dtau},\,
    \Re(\widehat{\tilde{\bm{\alpha}}}),\,
    \Im(\widehat{\tilde{\bm{\alpha}}}),\,
    \widehat{\mathbf{D}}^{d},\,
    \widehat{\mathbf{D}}^{a}
    \bigr\} \in \R^{2+10\cdot L}.
\end{equation}
The per-path components mirror $\bmx$ in eq.~\eqref{eq:y_def}, with the additional scalar link-level outputs $\hat{\tau}_0$ for the predicted \gls{tof} time and $\hat{P}_{\mathrm{rx}}$ for the predicted received power in dB.
All outputs are normalized per Appendix~\ref{app:normalization}, from which absolute-unit values are recovered via per-city inverse transforms.

\section{Model architecture}
\begin{figure}
    \centering
    \includegraphics[width=.95\linewidth]{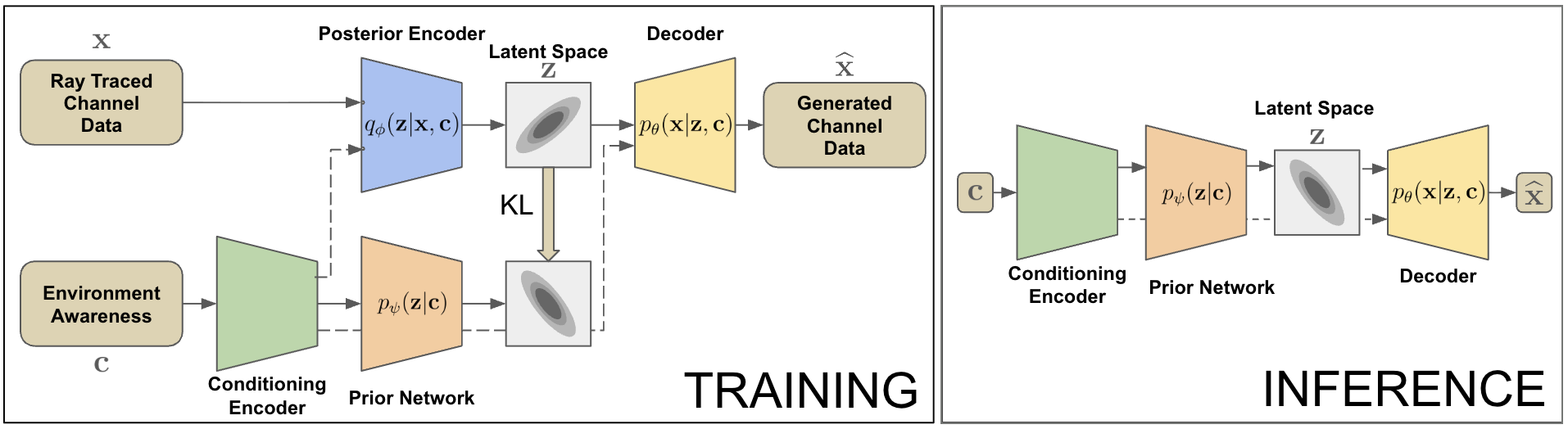}
    \caption{System diagram for \textsc{CityMPC} multipath channel generation. During training, samples are drawn from the latent space generated by the posterior encoder. For inference, the prior network, conditioned only on scene information $\bmc$, is used for generating a sample from the latent space.}
    \label{fig:module_diag}
\end{figure}
The goal of the architecture is to learn a mapping $\whbmx = f(\bmc)$ that generates a physically valid channel realization from scene conditioning alone, without access to ray geometry at inference.
We learn this mapping via a \gls{cvae}~\cite{kingma2013auto, sohn2015cvae}, which factorizes the generation of $\whbmx$ through a latent variable $\bmz \in \R^{d_z}$ and is trained by optimizing the \gls{elbo} of $p(\bmx|\bmc)$.
The overall architecture, illustrated in Fig.~\ref{fig:module_diag}, comprises four modules: the conditioning encoder (\ttce), the prior network (\ttpn), the posterior encoder (\ttpe), and the decoder (\ttdec).
Model dimensions are given in Appendix~\ref{app:model_dims}.

\paragraph{Conditioning encoder.}
\ttce maps the raw environmental inputs $\bmc$ into scene tokens and a compact scene embedding $\bmc' \in \R^{d_{\mathrm{scene}}}$ used by the remaining modules.\footnote{For clarity of presentation, we present the distributions of subsequent modules as being conditioned on $\bmc$ rather than the output of \ttce.}
Three independent ChannelViT~\cite{bao2024channelvit} towers process the global heightmap, \gls{tx} \gls{pov} stack, and \gls{rx} \gls{pov} stack respectively, preserving per-channel identity across modalities of heterogeneous physical meaning.
The \gls{tx} and \gls{rx} coordinates are Fourier-encoded~\cite{tancik2020fourier} and projected to a scalar token that is appended to the scene tokens when constructing the cross-attention memory.
The global, \gls{tx}, and \gls{rx} tokens are mean-pooled and fused via an \gls{mlp} with the scalar feature to produce $\bmc'$.

\paragraph{Prior network.}
\ttpn learns the conditional prior $p_\psi(\bmz \mid \bmc)$ from $\bmc'$ via an \gls{mlp} that outputs $(\bmmu_p, \log \bmsigma_p^2) \in \R^{d_z}$.
Unlike a standard \gls{cvae}~\cite{sohn2015cvae} with a fixed $\mathcal{N}(\mathbf{0}, \mathbf{I})$ prior, the scene-dependent prior here better captures the structured variability of the multipath channel across environments~\cite{tomczak2018vampprior}.
At inference, $\bmz \sim \mathcal{N}(\bmmu_p, \mathrm{diag}(\bmsigma_p^2))$ is sampled without channel observations.

\paragraph{Posterior encoder.}
\ttpe operates only during training and learns the approximate posterior $q_\phi(\bmz|\bmx, \bmc)$.
Per-path parameter vectors are projected into a sequence of $L$ tokens with sinusoidal positional encoding, which cross-attend to the scene token memory via stacked transformer layers.
The output path tokens are pooled by path presence, concatenated with $\Prx$, $\tauzero$, and $\bmc'$, and projected to $(\bmmu_q, \log \bmsigma_q^2) \in \R^{d_z}$.
For training, $\bmz$ is sampled via the reparameterization trick~\cite{kingma2013auto}, and the \gls{kl} divergence to $p_\psi$ is regularized with free bits~\cite{kingma2016improved} to prevent posterior collapse.

\paragraph{Decoder.}
\ttdec generates $\whbmx$ from $L$ learned slot queries~\cite{carion2020detr}, each initialized by adding a projection of $\bmz$ to a learned embedding so the latent influences every slot from the first layer.
Since \glspl{mpc} are sorted by received power, slot queries internalize this ordinal structure, assigning each slot a role that reflects path rank rather than sequence position.
The slot queries cross-attend to the scene token memory via stacked transformer layers, producing $L$ slot tokens.
Per-path heads predict path presence, excess delay, complex gain, and \gls{aod}/\gls{aoa} unit vectors, while link-level scalar heads predict $\hat{\tau}_0$ and $\hat{P}_{\mathrm{rx}}$ from mean-pooled slot tokens.
The \gls{aod} and \gls{aoa} unit vectors are produced by separate cross-attention layers that attend only to \gls{tx} and \gls{rx} scene tokens respectively, providing a physics-informed routing of geometric information to each angle prediction.

\section{Training}\label{sec:training}
The model is trained by maximizing the \gls{elbo} of $p(\bmx \mid \bmc)$,
\begin{equation}\label{eq:elbo}
    \mathcal{L} = \mathbb{E}_{q_\phi(\bmz \mid \bmx, \bmc)}\!\left[\log p_\theta(\bmx \mid \bmz, \bmc)\right]
    - \beta \cdot D_{\mathrm{KL}}\!\left(q_\phi(\bmz \mid \bmx, \bmc) \;\|\; p_\psi(\bmz \mid \bmc)\right),
\end{equation}
where $\beta$ is linearly annealed from zero over the first training steps to prevent \gls{kl} collapse~\cite{bowman2016generating}.
The reconstruction term decomposes into $7$ per-task losses, one per output head, yielding the total loss
\begin{equation}\label{eq:loss_tot}
    \mcl(\bmx, \whbmx) = \beta \cdot D_{\mathrm{KL}}\!\left(q_\phi(\bmz \mid \bmx, \bmc) \;\|\; p_\psi(\bmz \mid \bmc)\right) + \sum_{k=1}^{7} w_k\,\mcl_k(\bmx, \whbmx),
\end{equation}
where each task uses an output-appropriate loss (\gls{bce} for path presence, \gls{mse} for scalar regression, masked \gls{mse} for per-path regression, and masked cosine loss for direction unit vectors), and the weights $w_k$ are learned via Kendall uncertainty weighting~\cite{kendall2018multi}.
Full per-task loss definitions are given in Appendix~\ref{app:kendall_loss}.
The model is optimized with AdamW~\cite{loshchilov2017decoupled} using a cosine decay learning rate schedule with linear warmup, and free bits~\cite{kingma2016improved} per \gls{kl} dimension prevent posterior collapse early in training.
Full training hyperparameters are given in Appendix~\ref{app:training_hparams}.

\section{Experimental results}
\label{sec:results}
We evaluate \textsc{CityMPC} on five urban environments drawn from the DeepMIMO dataset: Austin, Dallas, Fort Worth, Denver, and New York.
Each city uses a spatially disjoint 70/15/15 train/validation/test split over \gls{tx} and \gls{rx} grid locations, ensuring no link seen during training appears at test time.
Per-city training and hardware details are reported in Appendix~\ref{app:hardware}.

\paragraph{Baseline.}
We compare \textsc{CityMPC} against an \gls{mlp} baseline that shares the same \gls{cvae} framework and Kendall uncertainty weighting~\citep{kendall2018multi} but replaces the ChannelViT conditioning encoder and transformer decoder with a ResNet-18 scene encoder, a fusion \gls{mlp} posterior encoder of hidden width 512, and a flat \gls{mlp} decoder trunk of hidden width 512 with per-path output heads.
All other training hyperparameters are identical across both models and are reported in Appendix~\ref{app:training_hparams}.

\paragraph{Metrics.}
We evaluate channel generation quality using eight metrics computed on the held-out test set of each city.
Path presence \gls{f1} measures the accuracy of the binary active-path mask over all $L = 25$ slots.
\Gls{tof} \gls{mae} measures the absolute error in the predicted \gls{tof} time $\tau_0$ in nanoseconds.
Average delay \gls{mae} measures the absolute error in the power-weighted mean absolute delay $\bar{\tau} = \sum_\ell |a_\ell|^2 \tau_\ell / \sum_\ell |a_\ell|^2$ in nanoseconds, consistent with the definition used in \citet{orekondy2023winert} and \citet{bian2025genert}.
Received power \gls{mae} measures the absolute error in total received power in decibels.
Average \gls{aod} azimuth \gls{mae}, average \gls{aod} elevation \gls{mae}, average \gls{aoa} azimuth \gls{mae}, and average \gls{aoa} elevation \gls{mae} each measure the absolute error in the power-weighted mean angle over all active paths at the link level, computed analogously to the average delay metric above~\cite{orekondy2023winert, bian2025genert}.
The detailed evaluation metrics are reported in Appendix~\ref{app:metrics}. 
Since \textsc{CityMPC} is generative, non-zero \gls{mae} values are expected even for a well-trained model, reflecting the stochastic nature of the learned channel distribution rather than a deterministic regression error.

\subsection{Channel generation}
\label{subsec:channel_gen}
We evaluate the ability of \textsc{CityMPC} to generate physically consistent \gls{mpc} parameter sets across five urban environments.
Each model is trained and evaluated on the same city, with test links spatially disjoint from training.
Table~\ref{tab:per_city_results} reports quantitative results for \textsc{CityMPC} and the \gls{mlp} baseline.

\begin{table*}[t]
  \caption{Per-city channel generation results for \textsc{CityMPC} and the \gls{mlp} baseline.
  \textsc{CityMPC} values are mean $\pm$ standard deviation over seed sweeps after filtering Kendall \gls{uw} divergent runs (see App.~\ref{app:kendall_loss}).
  \gls{mlp} values are mean $\pm$ standard deviation over runs.
  All angular \glspl{mae} are power-weighted mean angles over active paths.
  Lower is better for all metrics except \gls{f1}, where higher is better.
  The best result per city per metric is in \textbf{bold}.
  \textsc{CityMPC} is abbreviated as CMPC.}
  \label{tab:per_city_results}
  \centering
  \small
  \setlength{\tabcolsep}{4pt}

  \begin{tabular}{@{}llccccc@{}}
    \toprule
    Metric & Model & Austin & Dallas & Denver & Fort Worth & New York \\
    \midrule
    \multirow{2}{*}{\gls{rx} Pwr. (dB)}
      & CMPC & $\mathbf{1.29\!\pm\!0.30}$ & $\mathbf{1.49\!\pm\!0.08}$ & $2.72\!\pm\!1.05$            & $\mathbf{1.95\!\pm\!0.14}$ & $\mathbf{2.56\!\pm\!0.08}$ \\
      & \gls{mlp} & $1.80\!\pm\!0.07$         & $2.42\!\pm\!0.36$         & $\mathbf{2.70\!\pm\!0.07}$  & $2.60\!\pm\!0.08$         & $3.47\!\pm\!0.15$ \\
    \midrule
    \multirow{2}{*}{\gls{tof} (ns)}
      & CMPC & $\mathbf{7.25\!\pm\!1.96}$ & $\mathbf{3.89\!\pm\!1.46}$ & $\mathbf{4.73\!\pm\!0.79}$ & $\mathbf{4.98\!\pm\!1.35}$ & $\mathbf{7.56\!\pm\!0.33}$ \\
      & \gls{mlp} & $14.51\!\pm\!1.71$         & $10.32\!\pm\!1.56$         & $12.34\!\pm\!1.06$         & $15.93\!\pm\!0.59$         & $11.11\!\pm\!0.96$ \\
    \midrule
    \multirow{2}{*}{Av.\ Del. (ns)}
      & CMPC & $\mathbf{18.74\!\pm\!3.47}$ & $\mathbf{8.45\!\pm\!0.92}$ & $\mathbf{11.88\!\pm\!2.24}$ & $\mathbf{11.99\!\pm\!1.02}$ & $\mathbf{19.23\!\pm\!0.70}$ \\
      & \gls{mlp} & $29.21\!\pm\!1.38$           & $15.39\!\pm\!0.90$         & $18.51\!\pm\!1.04$           & $22.76\!\pm\!0.66$           & $25.33\!\pm\!0.42$ \\
    \midrule
    \multirow{2}{*}{\gls{aod} Az.\ ($^\circ$)}
      & CMPC & $\mathbf{3.56\!\pm\!0.97}$ & $\mathbf{1.50\!\pm\!0.05}$ & $\mathbf{2.10\!\pm\!0.40}$ & $\mathbf{1.65\!\pm\!0.03}$ & $\mathbf{3.05\!\pm\!0.07}$ \\
      & \gls{mlp} & $5.83\!\pm\!0.01$           & $3.13\!\pm\!0.56$           & $3.23\!\pm\!0.03$           & $2.79\!\pm\!0.04$           & $4.85\!\pm\!0.12$ \\
    \midrule
    \multirow{2}{*}{\gls{aod} El.\ ($^\circ$)}
      & CMPC & $\mathbf{0.67\!\pm\!0.14}$ & $\mathbf{0.42\!\pm\!0.01}$ & $\mathbf{0.54\!\pm\!0.06}$ & $\mathbf{0.46\!\pm\!0.01}$ & $\mathbf{0.77\!\pm\!0.01}$ \\
      & \gls{mlp} & $1.52\!\pm\!0.06$           & $1.11\!\pm\!0.06$           & $1.13\!\pm\!0.13$           & $1.02\!\pm\!0.07$           & $1.25\!\pm\!0.04$ \\
    \midrule
    \multirow{2}{*}{\gls{aoa} Az.\ ($^\circ$)}
      & CMPC & $\mathbf{10.06\!\pm\!2.09}$ & $\mathbf{8.54\!\pm\!0.35}$ & $\mathbf{16.06\!\pm\!4.46}$ & $\mathbf{12.53\!\pm\!0.17}$ & $\mathbf{18.19\!\pm\!0.79}$ \\
      & \gls{mlp} & $12.92\!\pm\!0.21$           & $11.50\!\pm\!1.31$           & $16.60\!\pm\!0.33$           & $15.33\!\pm\!0.12$           & $21.81\!\pm\!0.80$ \\
    \midrule
    \multirow{2}{*}{\gls{aoa} El.\ ($^\circ$)}
      & CMPC & $\mathbf{2.08\!\pm\!0.41}$ & $\mathbf{1.87\!\pm\!0.05}$ & $\mathbf{3.15\!\pm\!0.63}$ & $\mathbf{2.53\!\pm\!0.05}$ & $\mathbf{3.76\!\pm\!0.09}$ \\
      & \gls{mlp} & $3.06\!\pm\!0.04$           & $2.97\!\pm\!0.37$           & $4.44\!\pm\!0.19$           & $3.90\!\pm\!0.11$           & $5.91\!\pm\!0.48$ \\
    \midrule
    \multirow{2}{*}{\gls{f1}}
      & CMPC & $0.888\!\pm\!0.017$        & $0.874\!\pm\!0.003$        & $0.846\!\pm\!0.029$        & $0.877\!\pm\!0.003$        & $0.870\!\pm\!0.003$ \\
      & \gls{mlp} & $\mathbf{0.893\!\pm\!0.002}$ & $\mathbf{0.880\!\pm\!0.010}$ & $\mathbf{0.869\!\pm\!0.004}$ & $\mathbf{0.888\!\pm\!0.002}$ & $\mathbf{0.883\!\pm\!0.003}$ \\
    \bottomrule
  \end{tabular}
  \label{tab:quant_results}
\end{table*}

\paragraph{Quantitative results.}
For each city we run a seed sweep and report mean $\pm$ standard deviation over the surviving runs in Table~\ref{tab:per_city_results}.
A subset of \textsc{CityMPC} seeds exhibits the well-documented \gls{uw} failure mode in which one task's noise parameter $\sigma_k$ diverges and effectively removes that task from the joint loss~\cite{kirchdorfer2024analytical, liebel2018auxiliary}, and we filter these seeds out using a training-time criterion described in Appendix~\ref{app:kendall_loss}.
On the surviving runs, \textsc{CityMPC} outperforms the \gls{mlp} baseline on received power \gls{mae}, \gls{tof} \gls{mae}, average delay \gls{mae}, and all four angular \glspl{mae} on every city, with the only exception being Denver received power where the two models are within $0.02$\,dB.
Path presence \gls{f1} is comparable between both models, with differences below $0.025$ in all cities and a higher seed variance for \textsc{CityMPC}, reflecting the residual sensitivity of the \gls{uw} formulation even after divergent runs are excluded.
The \gls{tof} \gls{mae} gap is the most pronounced, with \textsc{CityMPC} achieving between $1.5\times$ and $3.2\times$ lower error than the \gls{mlp} baseline, demonstrating the advantage of the transformer decoder for capturing the geometric structure of first-arrival propagation.
Received power \gls{mae} is consistently lower for \textsc{CityMPC} across four of five cities, with New York yielding the highest error for both models due to the greater received power variance of its dense urban geometry.
\Gls{aod} azimuth and elevation \glspl{mae} are consistently lower than their \gls{aoa} counterparts across all cities and both models.
In the DeepMIMO scenario, transmitters are mounted atop buildings while receivers are placed $1.5$\,m above ground level, so \gls{aod} elevation angles exceed $90^\circ$ as signals propagate downward while \gls{aoa} elevation angles remain below $90^\circ$ as signals arrive from above.
The \gls{rx}-side \gls{pov} imagery is more heavily occluded by surrounding buildings than the elevated \gls{tx} view, making \gls{aoa} prediction inherently harder than \gls{aod} prediction.
Despite this asymmetry, \textsc{CityMPC} achieves \gls{aod} azimuth \glspl{mae} of at most $3.56^\circ$ and \gls{aoa} azimuth \glspl{mae} of at most $18.19^\circ$ across all cities, demonstrating that the physics-informed angle heads effectively leverage the separate \gls{tx}/\gls{rx} scene token memories.
\textsc{CityMPC} achieves these results with $15.8$\,M parameters compared to $35.5$\,M for the \gls{mlp} baseline, with the parameter efficiency stemming from the ChannelViT architecture being more compact than the ResNet-18 towers used in the baseline.
The variation in performance across cities reflects genuine differences in per-city channel distributions, which we analyze in detail in Section~\ref{subsec:cross_city}.

\begin{figure*}[t]
\centering
\scriptsize
\setlength{\tabcolsep}{2pt}
\begin{tabular}{ccccc}
  \Gls{cir} & \gls{aod} Azimuth & \gls{aod} Elevation & \gls{aoa} Azimuth & \gls{aoa} Elevation \\[2pt]
  \includegraphics[width=0.24\textwidth]{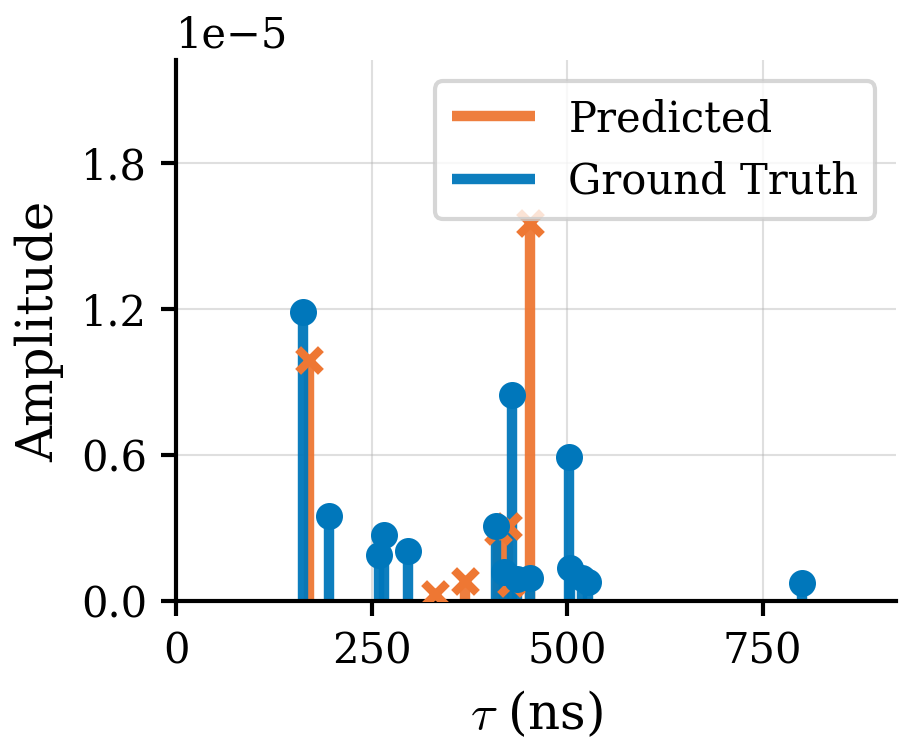} &
  \includegraphics[width=0.19\textwidth]{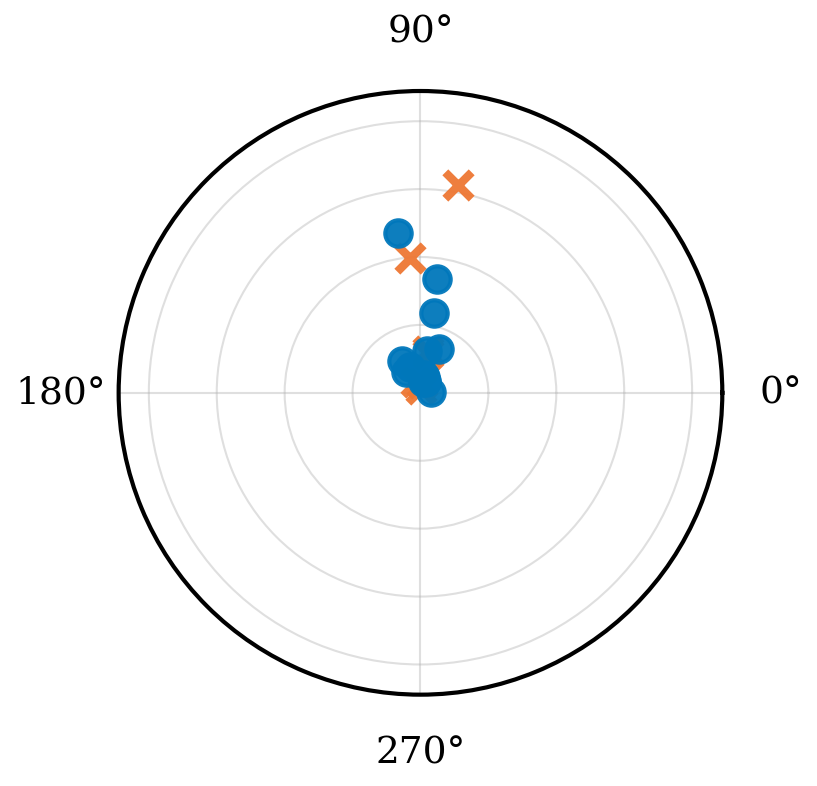} &
  \includegraphics[width=0.15\textwidth]{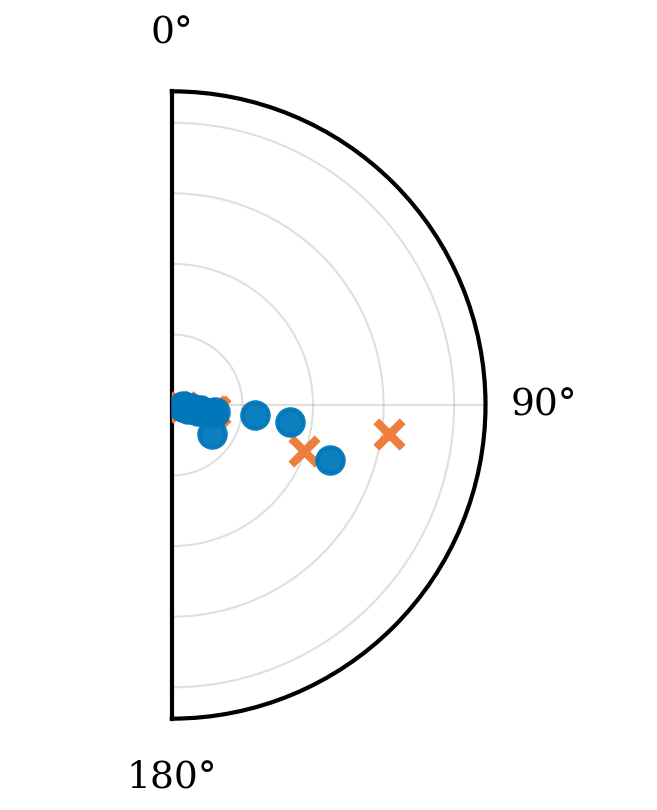} &
  \includegraphics[width=0.19\textwidth]{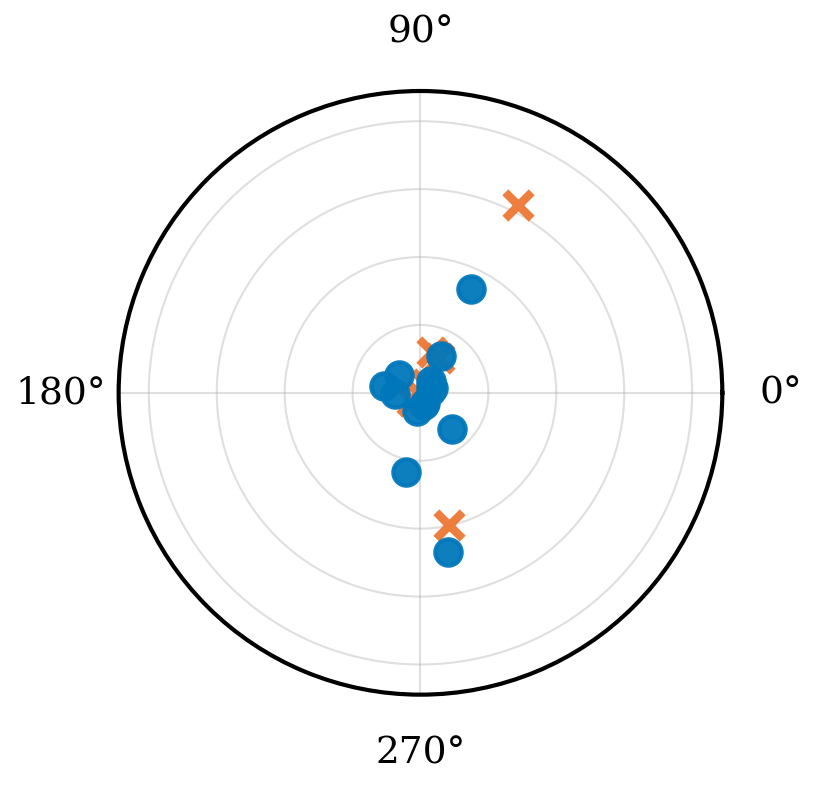} &
  \includegraphics[width=0.15\textwidth]{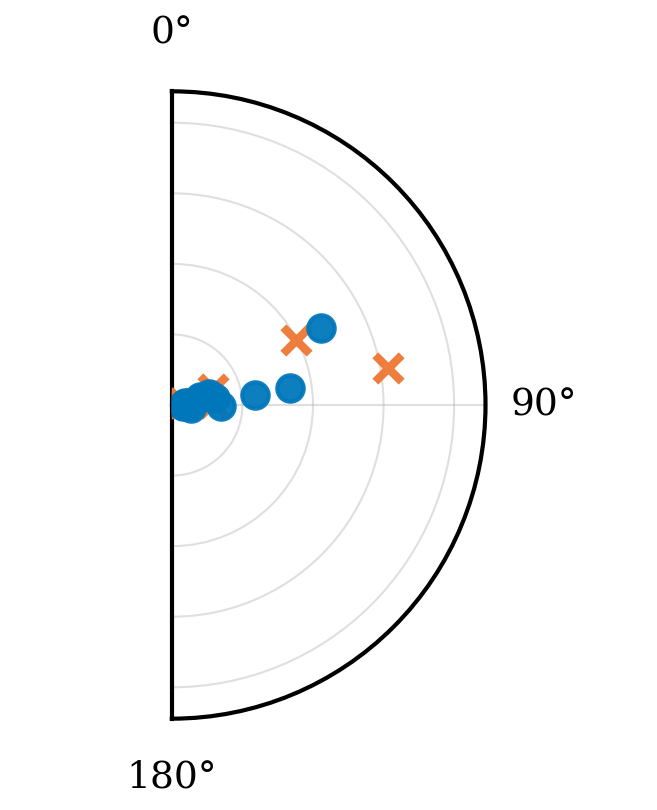} \\
\end{tabular}
\caption{A channel realization generated by \textsc{CityMPC} for a single Austin \gls{tx}-\gls{rx} link.
Columns show from left to right: \gls{cir} amplitude $|h(\tau)|$ in dB versus absolute delay $\tau$ (ns), \gls{aod} azimuth polar plot, \gls{aod} elevation polar plot, \gls{aoa} azimuth polar plot, and \gls{aoa} elevation polar plot.
Ground truth is shown in blue and predicted paths in orange in all panels.}
\label{fig:cir}
\end{figure*}

\paragraph{Qualitative results.}
Figure~\ref{fig:cir} shows one realization generated by \textsc{CityMPC} for a single Austin \gls{tx}-\gls{rx} link, with additional independent realizations for the same link shown in Appendix~\ref{app:realizations}.
Each realization differs in its specific path configuration, yet all share the same aggregate channel statistics, including received power, average delay, and mean angles, consistent with the ground truth.
This stochastic diversity is a direct consequence of sampling from the learned prior $p_\psi(\bmz \mid \bmc)$ and confirms that \textsc{CityMPC} captures the channel distribution rather than overfitting to a single output.

\begin{figure*}[h]
\centering
\includegraphics[width=0.85\textwidth]{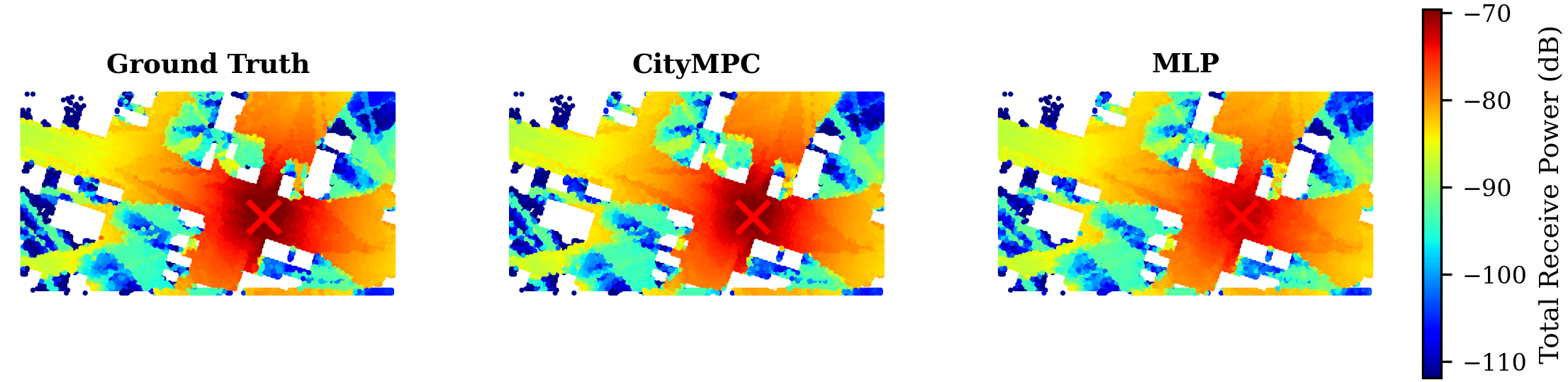}
\caption{Spatial maps of total received power (dB) across all \gls{rx} locations for a fixed \gls{tx} position (red $\times$) in Austin.
Left: DeepMIMO ground truth.
Center: \textsc{CityMPC} prediction.
Right: \gls{mlp} baseline prediction.
\textsc{CityMPC} accurately reproduces the spatial power distribution including street canyon effects, building shadows, and the near-far power gradient, closely matching the ground truth across the full coverage area.}
\label{fig:spatial_power}
\end{figure*}

Figure~\ref{fig:spatial_power} shows the predicted total received power across all \gls{rx} locations for a fixed \gls{tx} in Austin.
\textsc{CityMPC} reproduces the spatial structure of the ground-truth power map closely, capturing the street canyon propagation pattern, building shadow regions, and the near-far power gradient.
The \gls{mlp} baseline produces a broadly similar spatial structure but exhibits smoother transitions and less accurate reproduction of fine-grained shadowing near building boundaries.

\begin{figure*}[t]
\centering
\scriptsize
\setlength{\tabcolsep}{2pt}
\begin{tabular}{cccc}
\includegraphics[width=0.24\textwidth]{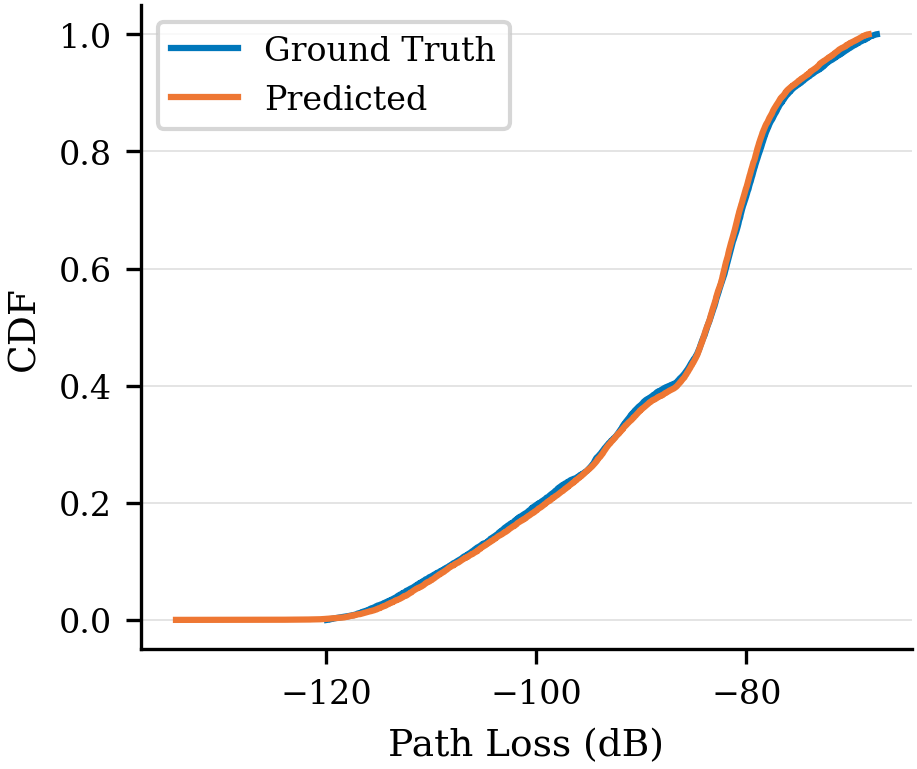} &
\includegraphics[width=0.24\textwidth]{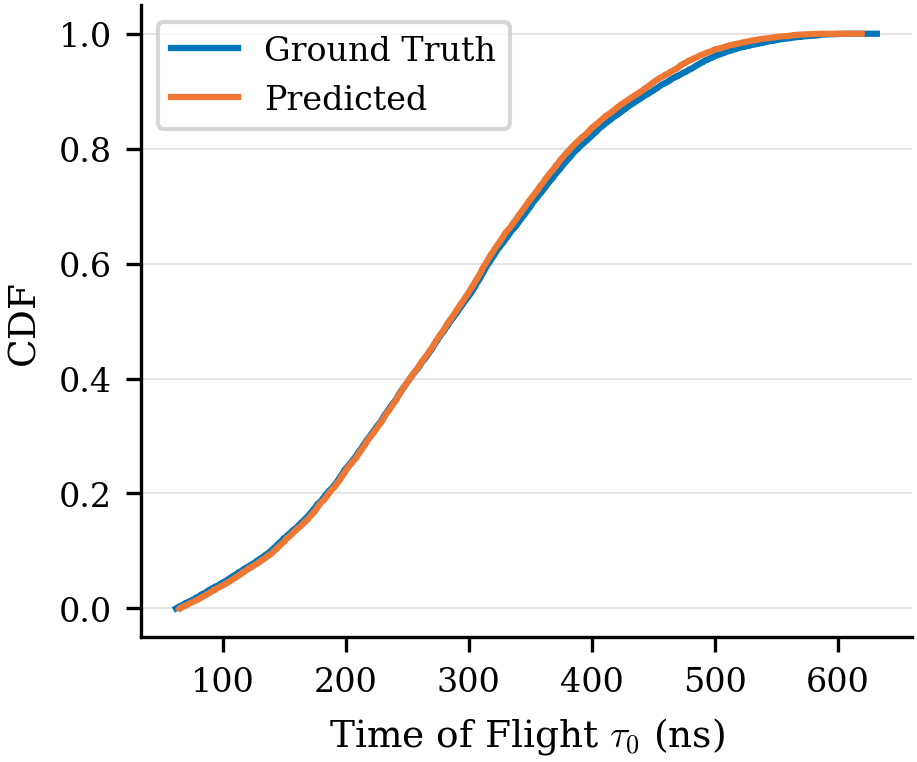} &
\includegraphics[width=0.24\textwidth]{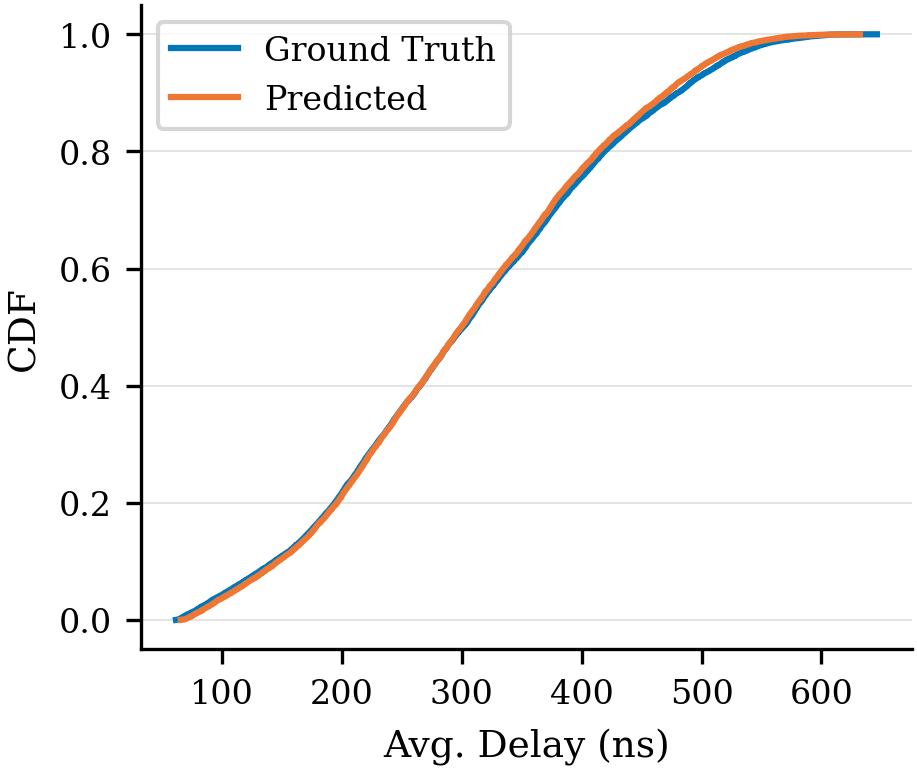} &
\includegraphics[width=0.24\textwidth]{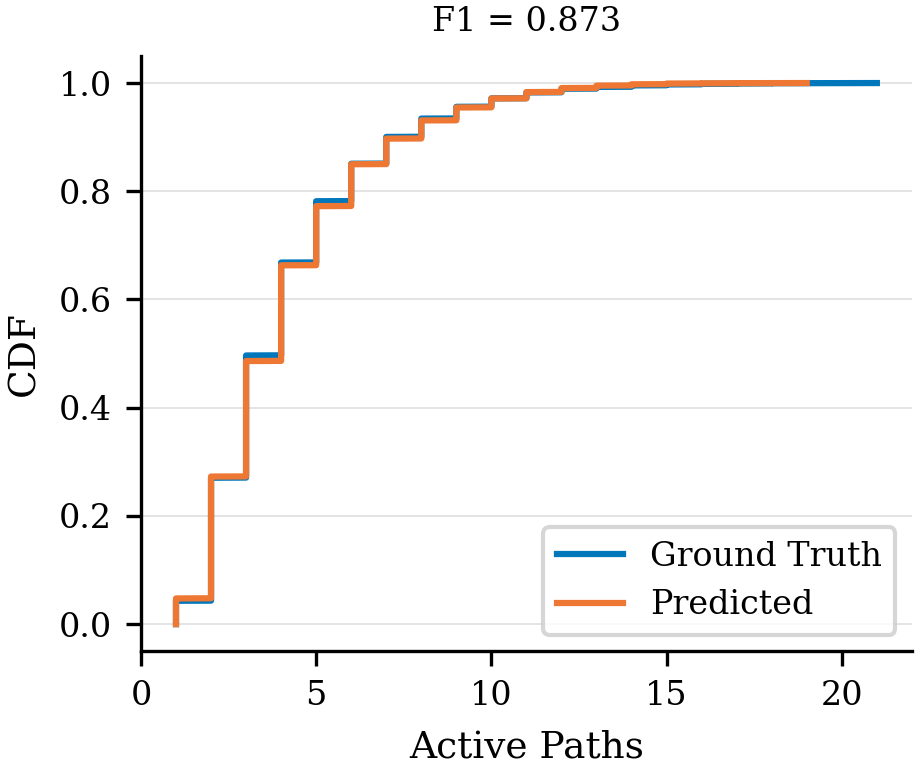} \\
Received Power (dB) & Time of Flight $\tau_0$ (ns) & Avg.\ Delay (ns) & Active Paths \\[4pt]
\includegraphics[width=0.24\textwidth]{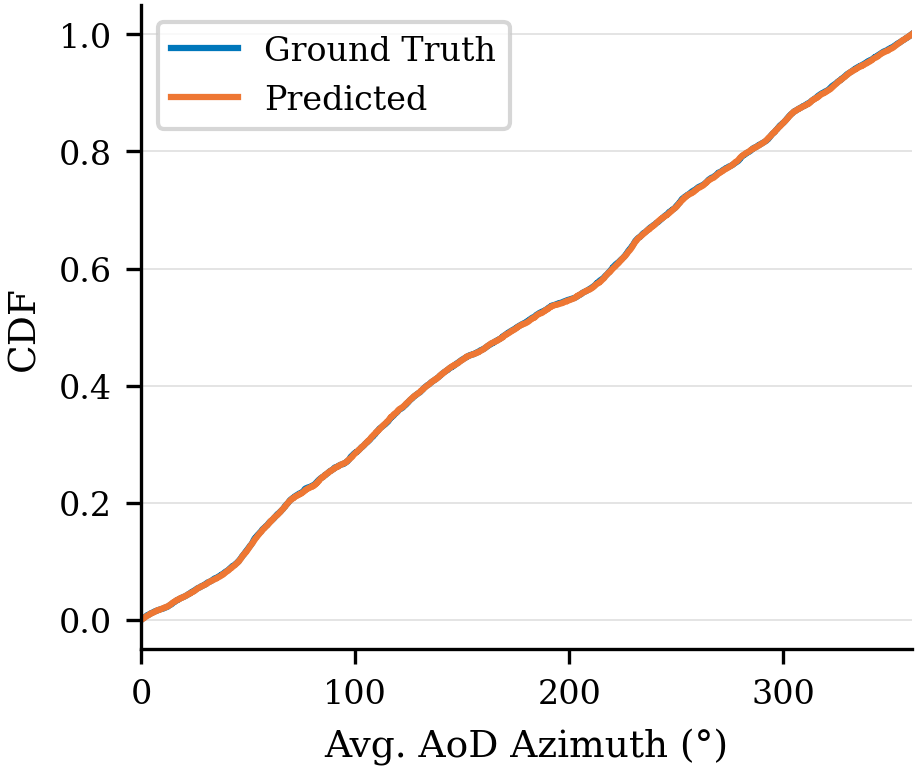} &
\includegraphics[width=0.24\textwidth]{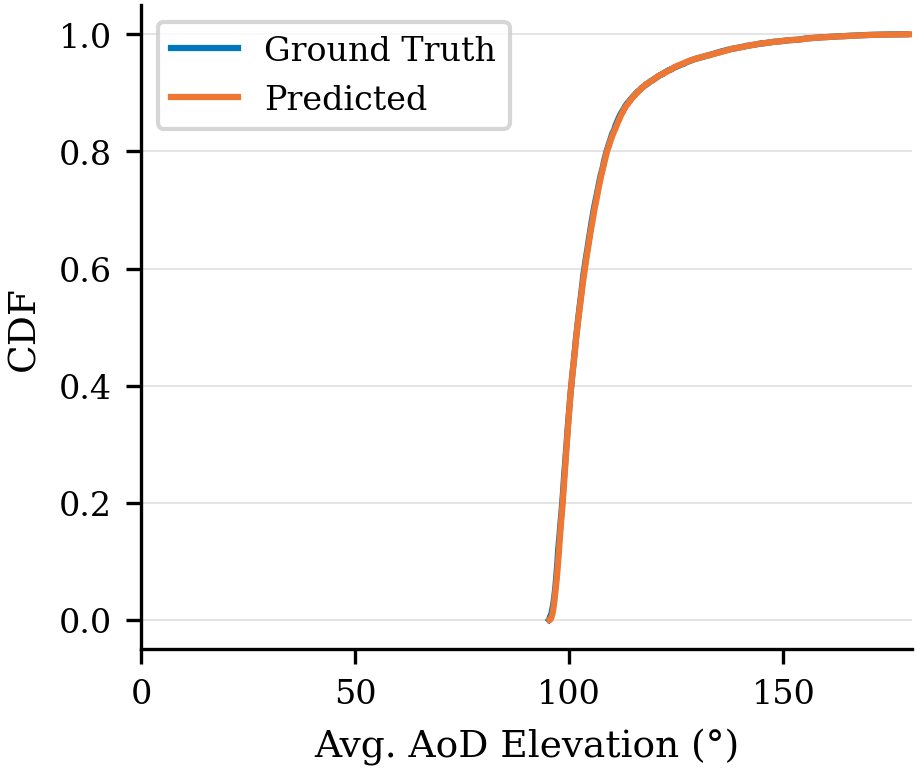} &
\includegraphics[width=0.24\textwidth]{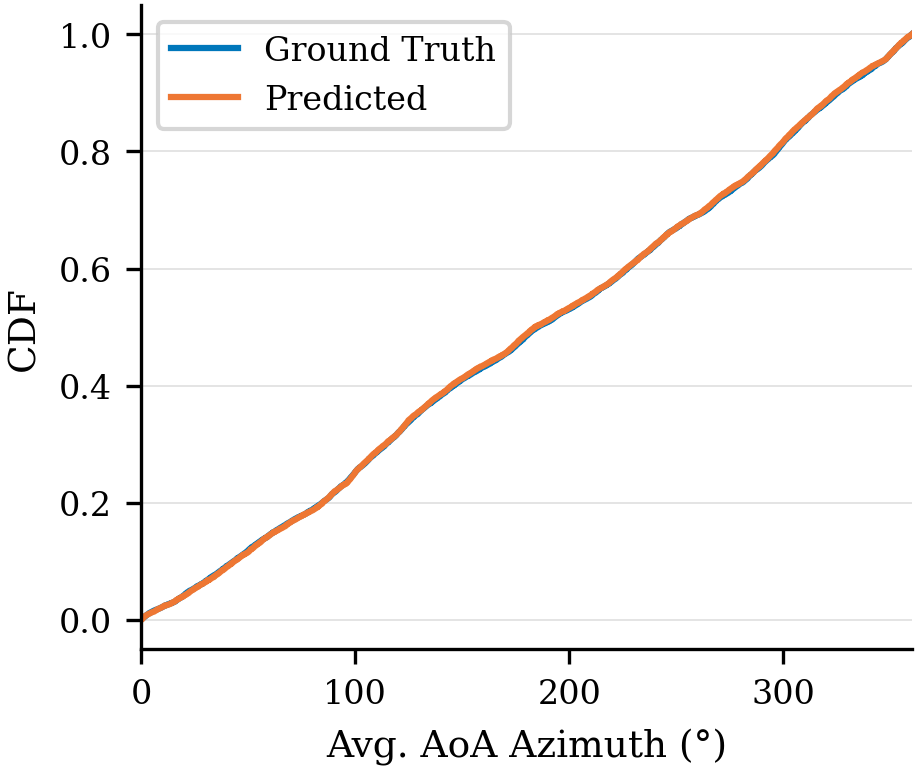} &
\includegraphics[width=0.24\textwidth]{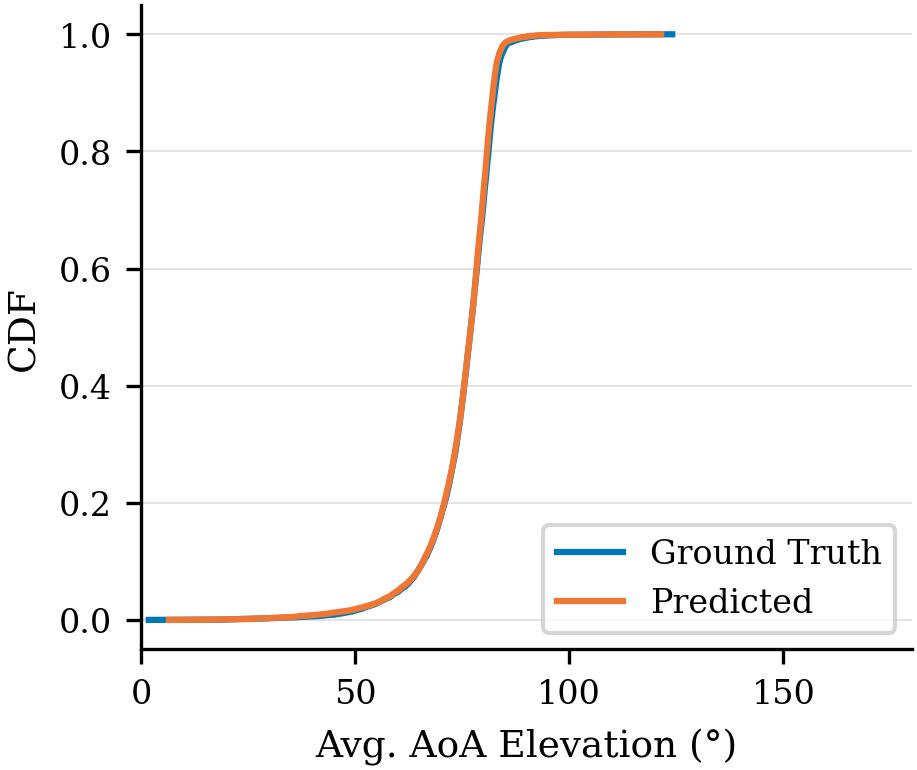} \\
Avg.\ \gls{aod} Azimuth ($^\circ$) & Avg.\ \gls{aod} Elevation ($^\circ$) & Avg.\ \gls{aoa} Azimuth ($^\circ$) & Avg.\ \gls{aoa} Elevation ($^\circ$) \\
\end{tabular}
\caption{Empirical \glspl{cdf} of ground-truth (blue) and generated (orange) channel parameters for Dallas.
The top row shows scalar and delay parameters.
The bottom row shows power-weighted mean angular parameters.
The near-perfect overlap across all parameters confirms that \textsc{CityMPC} reproduces the complete marginal channel statistics from scene imagery alone, without 3D geometry at inference.
The \gls{aod} elevation \gls{cdf} is sharply concentrated above $90^\circ$, consistent with downward-facing rooftop-mounted transmitters, while \gls{aoa} elevation is concentrated below $90^\circ$, consistent with upward-facing ground-level receivers.
Per-city \glspl{cdf} for all cities are given in Appendix~\ref{app:distributions}.}
\label{fig:dallas_cdfs}
\end{figure*}

Figure~\ref{fig:dallas_cdfs} shows the empirical \glspl{cdf} of ground-truth and generated channel parameters for Dallas across all eight channel dimensions.
The predicted distributions closely overlap with the ray-traced ground truth, confirming that \textsc{CityMPC} reproduces the full marginal channel statistics from scene imagery alone.
The active path count \gls{cdf} matches precisely, consistent with the \gls{f1} score of $0.873$ reported in Table~\ref{tab:per_city_results}.
The \gls{aod} elevation \gls{cdf} is sharply concentrated above $90^\circ$, reflecting the downward-looking geometry of rooftop-mounted transmitters, and the \gls{aoa} elevation \gls{cdf} is concentrated below $90^\circ$, consistent with upward-looking ground-level receivers.
The model recovers both physical constraints from scene imagery alone, without access to 3D geometry at inference.
Per-city \glspl{cdf} for all five cities are provided in Appendix~\ref{app:distributions}.

\subsection{Cross-city generalization}
\label{subsec:cross_city}
\begin{figure*}[h]
\centering
\scriptsize
\setlength{\tabcolsep}{2pt}
\begin{tabular}{ccc}
\includegraphics[width=0.33\textwidth]{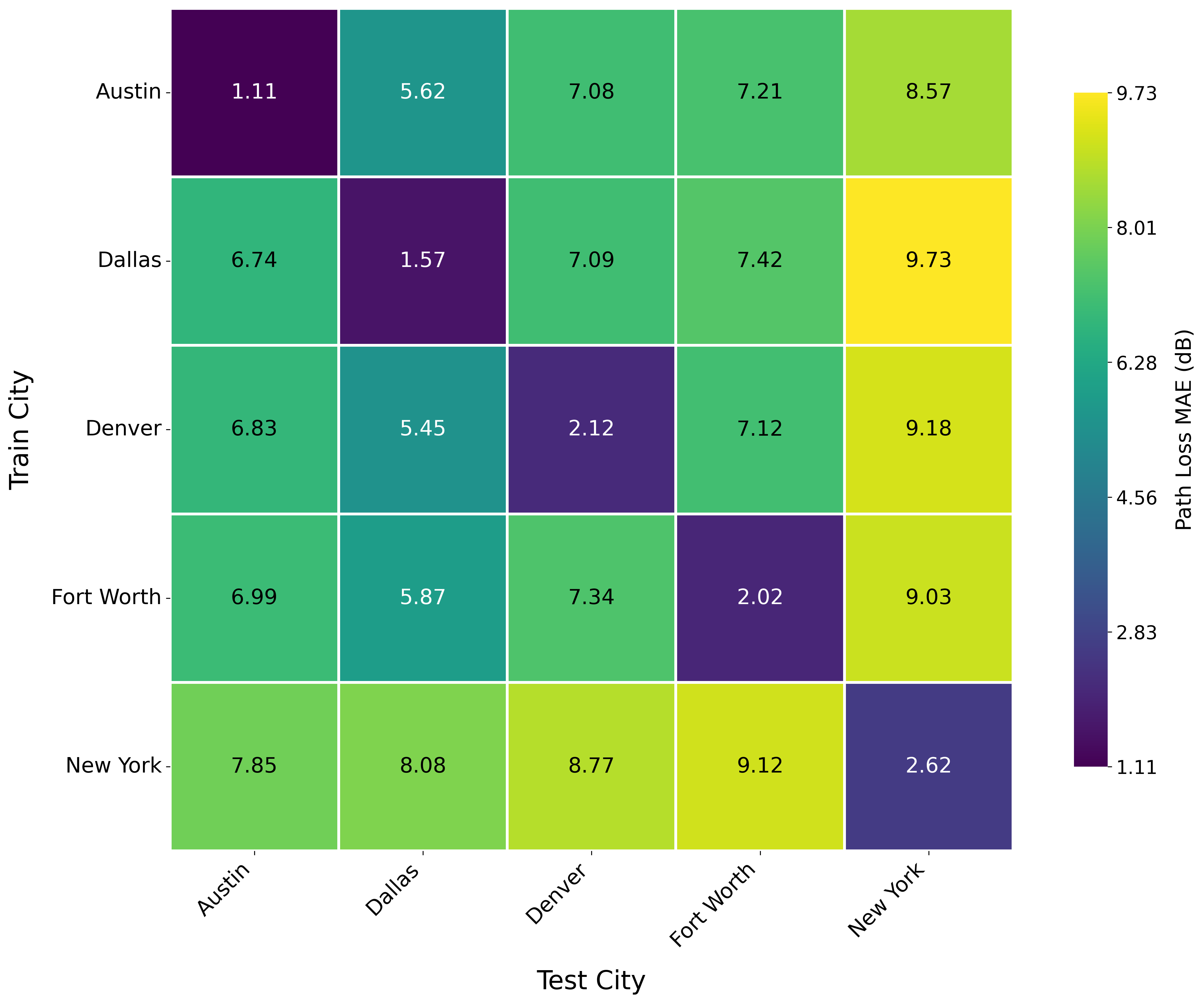} &
\includegraphics[width=0.33\textwidth]{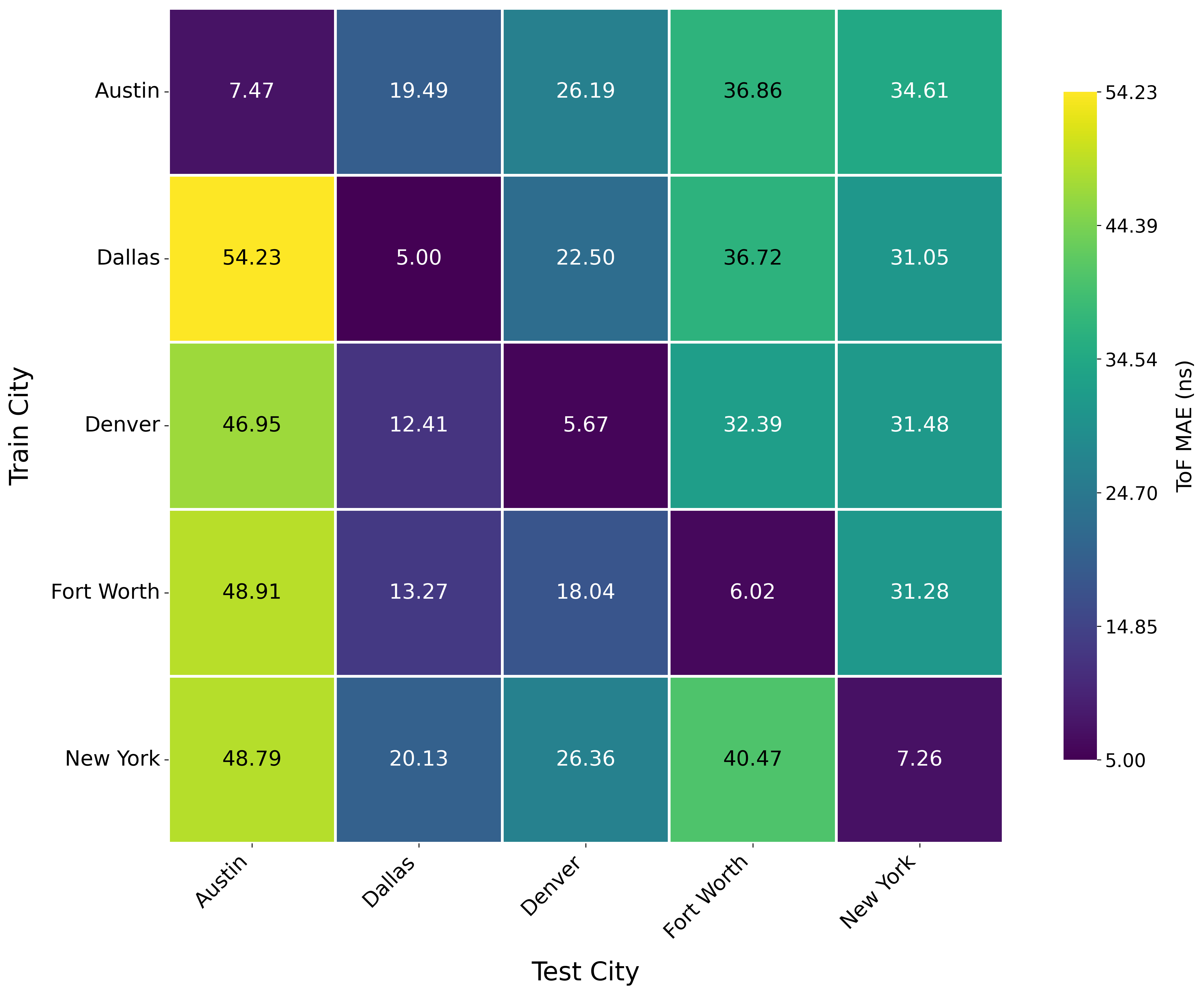} &
\includegraphics[width=0.33\textwidth]{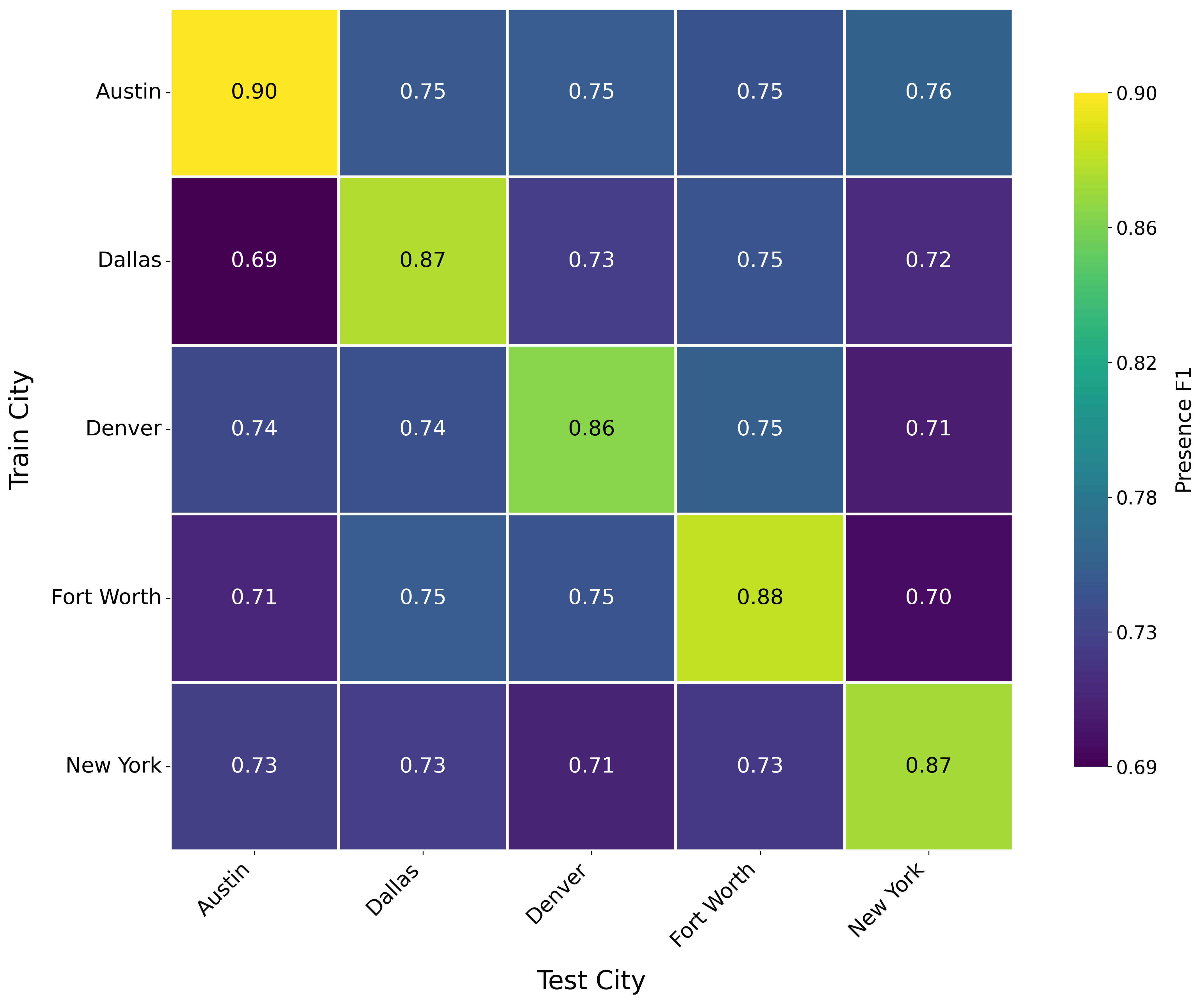} \\
Received Power \gls{mae} (dB) & Time of Flight \gls{mae} (ns) & Presence \gls{f1} \\
\end{tabular}
\caption{Cross-city transfer matrices for received power \gls{mae}, \gls{tof} \gls{mae}, and presence \gls{f1}.
Rows indicate the city used for training and columns give the test city.
Diagonal entries map to in-distribution evaluation.
Off-diagonal entries reflect the per-city distribution shift shown in Table~\ref{tab:city_stats}.}
\label{fig:heatmap_pl_tof_f1}
\end{figure*}

Table~\ref{tab:city_stats} summarizes the per-city test-set channel statistics, showing substantial differences in received power, propagation delay, active path count, and link distance across the five cities, motivating independent per-city training.
\begin{table}[h]
  \centering
  \caption{Per-city channel statistics computed on the test split.}
  \label{tab:city_stats}
  \small      
  \begin{tabular}{lcccc}  
    \toprule
    City & Mean $\tau_0$ (ns) & Mean Active Paths & Mean Dist.\ (m) & Mean Received Power (dB) \\
    \midrule
    Austin     & $329.8 \pm 156.1$ & $5.87 \pm 3.44$ & $95.4  \pm 43.8$ & $-88.7 \pm 12.2$ \\
    Dallas     & $289.0 \pm 115.1$ & $4.17 \pm 2.48$ & $85.8  \pm 34.0$ & $-88.1 \pm 12.1$ \\
    Denver     & $314.0 \pm 130.6$ & $4.82 \pm 2.85$ & $93.3  \pm 38.8$ & $-90.6 \pm 12.5$ \\
    Fort Worth & $342.2 \pm 156.9$ & $4.67 \pm 2.74$ & $101.7 \pm 46.8$ & $-91.6 \pm 13.0$ \\
    New York   & $303.3 \pm 146.8$ & $5.77 \pm 3.24$ & $88.1  \pm 42.9$ & $-93.1 \pm 13.8$ \\
    \bottomrule
  \end{tabular}
\end{table}
To assess how a city-trained model transfers to other cities without any domain adaptation, we evaluate each model on the test sets of all other cities.
Figure~\ref{fig:heatmap_pl_tof_f1} shows the resulting transfer matrices for received power \gls{mae}, \gls{tof} \gls{mae}, and presence \gls{f1}, where diagonal entries correspond to the in-distribution performance reported in Table~\ref{tab:per_city_results}.
Off-diagonal entries reflect the effect of per-city distribution shift, with received power \gls{mae} increasing from below $2.1$\,dB to above $7$\,dB and \gls{tof} \gls{mae} increasing from below $7.5$\,ns to above $20$\,ns.
Despite this shift, transferred models remain meaningful predictors of the channel structure, with presence \gls{f1} remaining above $0.7$ across all city pairs.
Transfer matrices for the remaining five metrics are provided in Appendix~\ref{app:crosscity_extra}.

\subsection{Model efficiency}
\label{subsec:efficiency}
\begin{wraptable}{r}{0.5\textwidth}
  \vspace{-1.5em}
  \caption{Per-link inference latency.}
  \label{tab:model_times}
  \centering
  \small
  \setlength{\tabcolsep}{4pt}
  \begin{tabular}{lcc}
    \toprule
    Method & ms/link & Speedup \\
    \midrule
    Sionna \gls{rt}        & $1.41 \pm 0.22$   & $-$ \\
    \textsc{CityMPC} (e2e)          & $0.74 \pm 0.15$   & $1.90\times$ \\
    \textsc{CityMPC} (model)        & $0.143 \pm 0.009$ & $9.88\times$ \\
    \bottomrule
  \end{tabular}
  \vspace{-1em}
\end{wraptable}
After a one-time \gls{rt} simulation per city to generate training data, \textsc{CityMPC} produces channel realizations at inference time without any further ray tracing or 3D scene mesh.
Table~\ref{tab:model_times} reports per-link wall-clock latency averaged over the test sets of all five cities, with three paired \gls{rt} and \textsc{CityMPC} seeds per city.
Test sets range from $9.4$k to $15.4$k links per city.
End-to-end inference is $1.90\times$ faster than \gls{rt} and includes data loading and host-to-device transfer, while the model forward pass alone is $9.88\times$ faster.
Hardware details and per-city training times are reported in Appendix~\ref{app:hardware}.

\section{Conclusions and Limitations}
\label{sec:conclusions_limitaitons}
We presented \textsc{CityMPC}, the first generative model to predict the complete per-path \gls{mpc} parameter set from scene imagery alone, jointly predicting path presence, complex gain, \gls{aod} and \gls{aoa} in azimuth and elevation, excess delay, \gls{tof} time, and received power without access to 3D scene geometry at inference.
\textsc{CityMPC} uses a transformer-based \gls{cvae} conditioned on multi-channel \gls{pov} imagery and a terrain heightmap, trained with Kendall uncertainty weighting across seven prediction tasks.
Evaluated across five US cities, \textsc{CityMPC} achieves strong channel generation fidelity, with generated \gls{mpc} distributions closely matching ray-traced ground truth across all predicted parameters.
The accompanying dataset of $427{,}397$ ray-traced links across five cities, together with the released model code and generation pipeline, provides a reproducible benchmark for future generative channel modeling work.
Our cross-city transfer analysis reveals that each city presents a distinct channel distribution.
Per-city trained models establish a strong in-distribution baseline for future multi-city generalization research.

\textsc{CityMPC} has not been validated against real-world channel measurements.
The sim-to-real gap remains an open challenge.
Per-city training is required because cross-city generalization degrades with distributional shift, as shown in Section~\ref{subsec:cross_city}.
Fine-tuning from a multi-city pretrained model is a natural next step toward reducing per-city data requirements.
The current model targets outdoor urban macro-cell environments at 3.5\,GHz with omnidirectional antennas.
Extension to indoor scenarios, mmWave frequencies, and directional antenna configurations remains an open problem.
Because of Kendall uncertainty weighting, the model may diverge during training.
We acknowledge these limitations and plan to address each of them in our future work.
Improving cross-city transfer through domain adaptation, stabilizing multi-task training through alternative loss weighting strategies, and validating against real-world channel measurements are the key directions we plan to pursue.

\clearpage
\bibliographystyle{unsrtnat} 
\bibliography{references}

@book{tse2005fundamentals,
  title     = {{Fundamentals of Wireless Communication}},
  author    = {Tse, David and Viswanath, Pramod},
  year      = {2005},
  publisher = {Cambridge University Press}
}

@book{rappaport2022radio,
  title     = {{Radio Propagation Measurements and Channel Modeling}},
  author    = {Rappaport, Theodore S and Remley, Kate A and Gentile, Camillo
               and Molisch, Andreas F and Zaji{\'c}, Alenka},
  year      = {2022},
  publisher = {Cambridge University Press}
}

@online{38901,
  author = {{3rd Generation Partnership Project (3GPP)}},
  title  = {{Study on Channel Model for Frequencies from 0.5 to 100~GHz}},
  year   = {2026},
  url    = {https://www.3gpp.org/ftp/Specs/archive/38_series/38.901/38901-j20.zip},
  note   = {TR 38.901 v19.1.0, accessed: 2026-01-24}
}

@article{samimi20163,
    author={Samimi, Mathew K. and Rappaport, Theodore S.},
  journal={IEEE Transactions on Microwave Theory and Techniques}, 
  title={{3-D Millimeter-Wave Statistical Channel Model for 5G Wireless System Design}}, 
  year={2016},
  volume={64},
  number={7},
  pages={2207-2225},
  doi={10.1109/TMTT.2016.2574851}
}

@software{sionna,
  title   = {{Sionna}},
  author  = {Hoydis, Jakob and Cammerer, Sebastian and {Ait Aoudia}, Fay{\c{c}}al
             and Nimier-David, Merlin and Maggi, Lorenzo and Marcus, Guillermo
             and Vem, Avinash and Keller, Alexander},
  year    = {2022},
  version = {1.2.1},
  note    = {\url{https://nvlabs.github.io/sionna/}}
}

@misc{remcom2021wireless,
  title  = {{Wireless InSite: 3D Wireless Prediction Software}},
  author = {Remcom},
  year   = {2021}
}

@INPROCEEDINGS{10465179,
  author={Hoydis, Jakob and Aoudia, Faycal Ait and Cammerer, Sebastian and Nimier-David, Merlin and Binder, Nikolaus and Marcus, Guillermo and Keller, Alexander},
  booktitle={2023 IEEE Globecom Workshops (GC Wkshps)}, 
  title={{Sionna RT: Differentiable Ray Tracing for Radio Propagation Modeling}}, 
  year={2023},
  volume={},
  number={},
  pages={317-321},
  doi={10.1109/GCWkshps58843.2023.10465179}
}

@inproceedings{orekondy2023winert,
    title={{WiNe{RT}: Towards Neural Ray Tracing for Wireless Channel Modelling and Differentiable Simulations}},
    author={Tribhuvanesh Orekondy and Pratik Kumar and Shreya Kadambi and Hao Ye and Joseph Soriaga and Arash Behboodi},
    booktitle={The Eleventh International Conference on Learning Representations },
    year={2023},
    url={https://openreview.net/forum?id=tPKKXeW33YU}
}

@misc{bian2025genert,
  title={{GeNeRT: A Physics-Informed Approach to Intelligent Wireless Channel Modeling via Generalizable Neural Ray Tracing}}, 
      author={Kejia Bian and Meixia Tao and Shu Sun and Jun Yu},
      year={2025},
      eprint={2506.18295},
      archivePrefix={arXiv},
      primaryClass={cs.LG},
      url={https://arxiv.org/abs/2506.18295},
}

@inproceedings{zhao2023nerf2,
author = {Zhao, Xiaopeng and An, Zhenlin and Pan, Qingrui and Yang, Lei},
title = {{NeRF2: Neural Radio-Frequency Radiance Fields}},
year = {2023},
isbn = {9781450399906},
publisher = {Association for Computing Machinery},
address = {New York, NY, USA},
url = {https://doi.org/10.1145/3570361.3592527},
doi = {10.1145/3570361.3592527},
booktitle = {Proceedings of the 29th Annual International Conference on Mobile Computing and Networking},
articleno = {27},
numpages = {15},
location = {Madrid, Spain},
series = {ACM MobiCom '23}
}

@misc{lu2024newrf,
author = {Lu, Haofan and Vattheuer, Christopher and Mirzasoleiman, Baharan and Abari, Omid},
title = {NeWRF: a deep learning framework for wireless radiation field reconstruction and channel prediction},
year = {2024},
publisher = {JMLR.org},
booktitle = {Proceedings of the 41st International Conference on Machine Learning},
articleno = {1345},
numpages = {13},
location = {Vienna, Austria},
series = {ICML'24}
}

@misc{zhang2025rf3dgs,
  author={Zhang, Lihao and Sun, Haijian and Berweger, Samuel and Gentile, Camillo and Qingyang Hu, Rose},
  journal={IEEE Transactions on Wireless Communications}, 
  title={RF-3DGS: Wireless Channel Modeling With Radio Radiance Field and 3D Gaussian Splatting}, 
  year={2026},
  volume={25},
  number={},
  pages={10419-10433},
  doi={10.1109/TWC.2026.3652154}
}

@article{jiang2025learnable,
  author={Jiang, Shuaifeng and Qu, Qi and Pan, Xiaqing and Agrawal, Abhishek K. and Newcombe, Richard and Alkhateeb, Ahmed},
  journal={IEEE Open Journal of the Communications Society}, 
  title={Learnable Wireless Digital Twins: Reconstructing Electromagnetic Field With Neural Representations}, 
  year={2025},
  volume={6},
  number={},
  pages={1568-1590},
  doi={10.1109/OJCOMS.2025.3535959}
}

@inproceedings{li2024geo2sigmap,
  author={Li, Yiming and Li, Zeyu and Gao, Zhihui and Chen, Tingjun},
  booktitle={2024 IEEE International Symposium on Dynamic Spectrum Access Networks (DySPAN)}, 
  title={{Geo2SigMap: High-Fidelity RF Signal Mapping Using Geographic Databases}}, 
  year={2024},
  volume={},
  number={},
  pages={277-285},
  doi={10.1109/DySPAN60163.2024.10632773}
}

@article{wang2025radiodiff3d,
  author={Wang, Xiucheng and Zhang, Qiming and Cheng, Nan and Chen, Junting and Zhang, Zezhong and Li, Zan and Cui, Shuguang and Shen, Xuemin},
  journal={IEEE Transactions on Network Science and Engineering}, 
  title={{RadioDiff-3D: A 3D× 3D Radio Map Dataset and Generative Diffusion Based Benchmark for 6G Environment-Aware Communication}}, 
  year={2026},
  volume={13},
  number={},
  pages={3773-3789},
  doi={10.1109/TNSE.2025.3590545}
}

@inproceedings{goodfellow2014generative,
author = {Goodfellow, Ian J. and Pouget-Abadie, Jean and Mirza, Mehdi and Xu, Bing and Warde-Farley, David and Ozair, Sherjil and Courville, Aaron and Bengio, Yoshua},
title = {Generative adversarial nets},
year = {2014},
publisher = {MIT Press},
address = {Cambridge, MA, USA},
booktitle = {Proceedings of the 28th International Conference on Neural Information Processing Systems - Volume 2},
pages = {2672–2680},
numpages = {9},
location = {Montreal, Canada},
series = {NIPS'14}
}

@misc{kingma2013auto,
      title={Auto-Encoding Variational Bayes}, 
      author={Diederik P Kingma and Max Welling},
      year={2022},
      eprint={1312.6114},
      archivePrefix={arXiv},
      primaryClass={stat.ML},
      url={https://arxiv.org/abs/1312.6114}, 
}

@inproceedings{sohn2015cvae,
 author = {Sohn, Kihyuk and Lee, Honglak and Yan, Xinchen},
 booktitle = {Advances in Neural Information Processing Systems},
 editor = {C. Cortes and N. Lawrence and D. Lee and M. Sugiyama and R. Garnett},
 pages = {},
 publisher = {Curran Associates, Inc.},
 title = {Learning Structured Output Representation using Deep Conditional Generative Models},
 url = {https://proceedings.neurips.cc/paper_files/paper/2015/file/8d55a249e6baa5c06772297520da2051-Paper.pdf},
 volume = {28},
 year = {2015}
}

@inprceedings{ho2020denoising,
author = {Ho, Jonathan and Jain, Ajay and Abbeel, Pieter},
title = {Denoising diffusion probabilistic models},
year = {2020},
isbn = {9781713829546},
publisher = {Curran Associates Inc.},
address = {Red Hook, NY, USA},
booktitle = {Proceedings of the 34th International Conference on Neural Information Processing Systems},
articleno = {574},
numpages = {12},
location = {Vancouver, BC, Canada},
series = {NIPS '20}
}

@article{xiao2022channelgan,
  author={Xiao, Han and Tian, Wenqiang and Liu, Wendong and Shen, Jia},
  journal={IEEE Wireless Communications Letters}, 
  title={{ChannelGAN: Deep Learning-Based Channel Modeling and Generating}}, 
  year={2022},
  volume={11},
  number={3},
  pages={650-654},
  doi={10.1109/LWC.2021.3140102}
}

@misc{orekondy2022mimogan,
  author={Orekondy, Tribhuvanesh and Behboodi, Arash and Soriaga, Joseph B.},
  booktitle={ICC 2022 - IEEE International Conference on Communications}, 
  title={{MIMO-GAN: Generative MIMO Channel Modeling}}, 
  year={2022},
  volume={},
  number={},
  pages={5322-5328},
  doi={10.1109/ICC45855.2022.9839123}
}

@inproceedings{sengupta2023diffusion,
  author={Sengupta, Ushnish and Jao, Chinkuo and Bernacchia, Alberto and Vakili, Sattar and Shiu, Da-shan},
  booktitle={GLOBECOM 2023 - 2023 IEEE Global Communications Conference}, 
  title={{Generative Diffusion Models for Radio Wireless Channel Modelling and Sampling}}, 
  year={2023},
  volume={},
  number={},
  pages={4779-4784},
  doi={10.1109/GLOBECOM54140.2023.10437154}
}

@misc{kim2024diffusionchannel,
      title={Diffusion Models for Accurate Channel Distribution Generation}, 
      author={Muah Kim and Rick Fritschek and Rafael F. Schaefer},
      year={2024},
      eprint={2309.10505},
      archivePrefix={arXiv},
      primaryClass={cs.IT},
      url={https://arxiv.org/abs/2309.10505}, 
}

@article{lee2025diffusionchannel,
  author={Lee, Taekyun and Park, Juseong and Kim, Hyeji and Andrews, Jeffrey G.},
  journal={IEEE Transactions on Wireless Communications}, 
  title={Generating High Dimensional User-Specific Wireless Channels Using Diffusion Models}, 
  year={2026},
  volume={25},
  number={},
  pages={2907-2921},
  doi={10.1109/TWC.2025.3600286}
}

@inproceedings{deshpande2025hwcond,
title={Hardware-Conditioned Generative Channel Modeling: A Diffusion-Based Approach for Location and Hardware-Aware Wireless Dataset Synthesis},
author={Nitish Deshpande and Sanjay Ganapathy and Viraj Shah},
booktitle={Open Conference of AI Agents for Science 2025},
year={2025},
url={https://openreview.net/forum?id=ExGHHgTM2p}
}

@misc{bock2025physicsinformed,
    title={Physics-Informed Generative Modeling of Wireless Channels},
    author={Benedikt B{\"o}ck and Andreas Oeldemann and Timo Mayer and Francesco Rossetto and Wolfgang Utschick},
    booktitle={Forty-second International Conference on Machine Learning},
    year={2025},
    url={https://openreview.net/forum?id=FFJFT93oa7}
}

@misc{wagle2025physicsinformed,
      title={Physics-Informed Generative Approaches for Wireless Channel Modeling}, 
      author={Satyavrat Wagle and Akshay Malhotra and Shahab Hamidi-Rad and Aditya Sant and David J. Love and Christopher G. Brinton},
      year={2025},
      eprint={2503.05988},
      archivePrefix={arXiv},
      primaryClass={eess.SP},
      url={https://arxiv.org/abs/2503.05988}, 
}

@misc{kang2024gbsm,
      title={A Geometry-based Stochastic Wireless Channel Model using Channel Images}, 
      author={Seongjoon Kang},
      year={2024},
      eprint={2312.06637},
      archivePrefix={arXiv},
      primaryClass={eess.SP},
      url={https://arxiv.org/abs/2312.06637}, 
}

@inproceedings{bao2024channelvit,
title={Channel Vision Transformers: An Image Is Worth 1 x 16 x 16 Words},
author={Yujia Bao and Srinivasan Sivanandan and Theofanis Karaletsos},
booktitle={The Twelfth International Conference on Learning Representations},
year={2024},
url={https://openreview.net/forum?id=CK5Hfb5hBG}
}

@inproceedings{carion2020detr,
author="Carion, Nicolas
and Massa, Francisco
and Synnaeve, Gabriel
and Usunier, Nicolas
and Kirillov, Alexander
and Zagoruyko, Sergey",
editor="Vedaldi, Andrea
and Bischof, Horst
and Brox, Thomas
and Frahm, Jan-Michael",
title="End-to-End Object Detection with Transformers",
booktitle="Computer Vision -- ECCV 2020",
year="2020",
publisher="Springer International Publishing",
address="Cham",
pages="213--229",
isbn="978-3-030-58452-8"
}

@misc{kendall2018multi,
      title={Multi-Task Learning Using Uncertainty to Weigh Losses for Scene Geometry and Semantics}, 
      author={Alex Kendall and Yarin Gal and Roberto Cipolla},
      year={2018},
      eprint={1705.07115},
      archivePrefix={arXiv},
      primaryClass={cs.CV},
      url={https://arxiv.org/abs/1705.07115}, 
}

@misc{chen_gradnorm_2018,
title={GradNorm: Gradient Normalization for Adaptive Loss Balancing in Deep Multitask Networks},
author={Zhao Chen and Vijay Badrinarayanan and Chen-Yu Lee and Andrew Rabinovich},
year={2018},
url={https://openreview.net/forum?id=H1bM1fZCW},
}

@misc{tomczak2018vampprior,
      title={VAE with a VampPrior}, 
      author={Jakub M. Tomczak and Max Welling},
      year={2018},
      eprint={1705.07120},
      archivePrefix={arXiv},
      primaryClass={cs.LG},
      url={https://arxiv.org/abs/1705.07120}, 
}

@inproceedings{alkhateeb2019deepmimo,
author = {Alkhateeb, A.},
title = {{DeepMIMO}: A Generic Deep Learning Dataset for Millimeter Wave and Massive {MIMO} Applications},
booktitle = {Proc. of Information Theory and Applications Workshop (ITA)},
year = {2019},
pages = {1-8},
month = {Feb},
Address = {San Diego, CA},
}

@article{rappaport2013mmwave,
  author={Rappaport, Theodore S. and Sun, Shu and Mayzus, Rimma and Zhao, Hang and Azar, Yaniv and Wang, Kevin and Wong, George N. and Schulz, Jocelyn K. and Samimi, Mathew and Gutierrez, Felix},
  journal={IEEE Access}, 
  title={Millimeter Wave Mobile Communications for 5G Cellular: It Will Work!}, 
  year={2013},
  volume={1},
  number={},
  pages={335-349},
  doi={10.1109/ACCESS.2013.2260813}
}

@online{itu_m2160,
  author  = {{ITU-R}},
  title   = {{Recommendation M.2160: Framework and Overall Objectives of
             the Future Development of IMT for 2030 and Beyond}},
  year    = {2023},
  url     = {https://www.itu.int/rec/R-REC-M.2160/en},
  note    = {Accessed: 2026-01-24}
}

@ARTICLE{brinton2025_6g,
  author={Brinton, Christopher G. and Chiang, Mung and Kim, Kwang Taik and Love, David J. and Beesley, Michael and Repeta, Morris and Roese, John and Beming, Per and Ekudden, Erik and Li, Clara and Wu, Geng and Batra, Nishant and Ghosh, Amitava and Ziegler, Volker and Ji, Tingfang and Prakash, Rajat and Smee, John},
  journal={IEEE Communications Magazine}, 
  title={{Key Focus Areas and Enabling Technologies for 6G}}, 
  year={2025},
  volume={63},
  number={3},
  pages={84-91},
  doi={10.1109/MCOM.002.2400150}}

@online{nga2023vertical,
  author  = {{Next G Alliance}},
  title   = {{6G Roadmap for Vertical Industries}},
  year    = {2023},
  url     = {https://nextgalliance.org/white_papers/6g-roadmap-vertical-industries/},
  note    = {ATIS, Accessed: 2026-01-24}
}

@inproceedings{11443807,
  author={Park, Juseong and Lee, Taekyun and Xing, Yunchou and Chen, Jie and Ghosh, Amitava and Andrews, Jeffrey G.},
  booktitle={2025 59th Asilomar Conference on Signals, Systems, and Computers}, 
  title={Learning Ray-Tracing for Radio Propagation via Cross-Attention-Based Diffusion Models}, 
  year={2025},
  volume={},
  number={},
  pages={1210-1214},
  doi={10.1109/IEEECONF67917.2025.11443807}
}

@inproceedings{sun2017nyusim,
  author={Sun, Shu and MacCartney, George R. and Rappaport, Theodore S.},
  booktitle={2017 IEEE International Conference on Communications (ICC)}, 
  title={A novel millimeter-wave channel simulator and applications for 5G wireless communications}, 
  year={2017},
  volume={},
  number={},
  pages={1-7},
  doi={10.1109/ICC.2017.7996792}
}

@article{levie2021radiounet,
  author={Levie, Ron and Yapar, Çağkan and Kutyniok, Gitta and Caire, Giuseppe},
  journal={IEEE Transactions on Wireless Communications}, 
  title={{RadioUNet: Fast Radio Map Estimation With Convolutional Neural Networks}}, 
  year={2021},
  volume={20},
  number={6},
  pages={4001-4015},
  doi={10.1109/TWC.2021.3054977}
}

@software{Mitsuba3,
    title = {Mitsuba 3 renderer},
    author = {Wenzel Jakob and Sébastien Speierer and Nicolas Roussel and Merlin Nimier-David and Delio Vicini and Tizian Zeltner and Baptiste Nicolet and Miguel Crespo and Vincent Leroy and Ziyi Zhang},
    note = {https://mitsuba-renderer.org},
    version = {3.8.0},
    year = 2022
}

@inproceedings{tancik2020fourier,
author = {Tancik, Matthew and Srinivasan, Pratul P. and Mildenhall, Ben and Fridovich-Keil, Sara and Raghavan, Nithin and Singhal, Utkarsh and Ramamoorthi, Ravi and Barron, Jonathan T. and Ng, Ren},
title = {Fourier features let networks learn high frequency functions in low dimensional domains},
year = {2020},
isbn = {9781713829546},
publisher = {Curran Associates Inc.},
address = {Red Hook, NY, USA},
booktitle = {Proceedings of the 34th International Conference on Neural Information Processing Systems},
articleno = {632},
numpages = {11},
location = {Vancouver, BC, Canada},
series = {NIPS '20}
}

@inproceedings{kingma2016improved,
author = {Kingma, Diederik P. and Salimans, Tim and Jozefowicz, Rafal and Chen, Xi and Sutskever, Ilya and Welling, Max},
title = {Improved variational inference with inverse autoregressive flow},
year = {2016},
isbn = {9781510838819},
publisher = {Curran Associates Inc.},
address = {Red Hook, NY, USA},
booktitle = {Proceedings of the 30th International Conference on Neural Information Processing Systems},
pages = {4743–4751},
numpages = {9},
location = {Barcelona, Spain},
series = {NIPS'16}
}

@inproceedings{bowman2016generating,
  title={Generating sentences from a continuous space},
  author={Bowman, Samuel R and Vilnis, Luke and Vinyals, Oriol and Dai, Andrew M and Jozefowicz, Rafal and Bengio, Samy},
  booktitle={20th SIGNLL Conference on Computational Natural Language Learning, CoNLL 2016},
  pages={10--21},
  year={2016},
  organization={Association for Computational Linguistics (ACL)}
}

@misc{loshchilov2017decoupled,
      title={Decoupled Weight Decay Regularization}, 
      author={Ilya Loshchilov and Frank Hutter},
      year={2019},
      eprint={1711.05101},
      archivePrefix={arXiv},
      primaryClass={cs.LG},
      url={https://arxiv.org/abs/1711.05101}, 
}

@techreport{ITURP20403,
  author      = {{ITU-R}},
  title       = {Recommendation {ITU-R P.2040-3}: Effects of building materials
                 and structures on radiowave propagation above about 100\,{MHz}},
  institution = {International Telecommunication Union},
  year        = {2023},
  month       = aug,
  url         = {https://www.itu.int/rec/R-REC-P.2040-3-202308-I/en}
}

@ARTICLE{11498538,
  author={Kim, Hwanjin and Choi, Junil and Love, David J.},
  journal={IEEE Wireless Communications}, 
  title={{Machine-Learning Techniques for Wireless Channel Prediction: Insights and Practical Guidance}}, 
  year={2026},
  volume={},
  number={},
  pages={1-8},
  doi={10.1109/MWC.2026.3678210}}

@misc{liebel2018auxiliary,
  title={Auxiliary Tasks in Multi-task Learning},
  author={Lukas Liebel and Marco K\"orner},
  year={2018},
  eprint={1805.06334},
  archivePrefix={arXiv},
  primaryClass={cs.CV},
  url={https://arxiv.org/abs/1805.06334}
}

@article{kirchdorfer2024analytical,
  title={Analytical Uncertainty-Based Loss Weighting in Multi-Task Learning},
  author={Kirchdorfer, Lukas and Elich, Cathrin and Kutsche, Simon and Stuckenschmidt, Heiner and Schott, Lukas and K\"ohler, Jan M.},
  journal={International Journal of Computer Vision},
  year={2025},
  doi={10.1007/s11263-025-02625-x},
  note={Preprint: arXiv:2408.07985}
}

\clearpage

\appendix
\section{Extended related work and comparison}
\label{app:model_comparison}
Deep learning approaches to channel modeling fall into three categories. We explore each in turn to give a more complete characterization of where \textsc{CityMPC} falls within this landscape.
\paragraph{Generative channel models.}
Early works applied \glspl{gan}~\cite{goodfellow2014generative}, \glspl{vae}~\cite{kingma2013auto}, and \glspl{ddpm}~\cite{ho2020denoising} to learn channel distributions from data.
\cite{xiao2022channelgan} and \cite{orekondy2022mimogan} learn distributions over \gls{mimo} channel matrices.
Diffusion-based models achieve strong results for channel distribution learning~\cite{sengupta2023diffusion, kim2024diffusionchannel, lee2025diffusionchannel, deshpande2025hwcond}.
Physics-informed generative models incorporate physical or geometric structure into generation~\cite{bock2025physicsinformed, wagle2025physicsinformed, kang2024gbsm}.
These methods do not condition on site-specific scene information, so samples from the same location are drawn from a single global distribution regardless of the local propagation environment.

\paragraph{Radio map and channel prediction.}
Other works predict aggregate signal quantities or channel responses from geographic or scene inputs.
\citet{wang2025radiodiff3d} generates 3D radio maps containing received power, direction of arrival, and time of arrival via a diffusion model.
Learnable digital twins~\cite{jiang2025learnable} reconstruct the \gls{em} field by combining 3D geometry with learned per-object \gls{em} properties, predicting the \gls{mimo} channel matrix.
These methods predict aggregate or location-level channel quantities and do not produce discrete per-path gains, delays, \gls{aod}, and \gls{aoa} at both \gls{tx} and \gls{rx}.

\paragraph{Neural ray tracing surrogates.}
More recent works use neural networks as surrogates for \gls{rt}, keeping the 3D scene mesh as input.
WiNeRT~\cite{orekondy2023winert}, validated on indoor environments, trains a \gls{mlp} to predict per-path gain, delay, \gls{aod} azimuth, and \gls{aoa} azimuth via a time-angle \gls{cir} representation, but does not predict elevation angles.
GeNeRT~\cite{bian2025genert} improves upon WiNeRT using a polarization-driven dual-branch network to predict per-path gain, delay, and \gls{aod} and \gls{aoa} in azimuth and elevation, demonstrating zero-shot transfer to unseen outdoor environments.
\citet{11443807} train a \gls{cddpm} conditioned on \gls{tx} and \gls{rx} coordinates to synthesize channels in the angular-delay domain without 3D scene geometry at inference.
The synthesis uses a 2D image representation where one axis encodes propagation delay and the other encodes \gls{tx}-side \gls{aod} azimuth.
\Gls{rx}-side \gls{aoa} is not predicted, and evaluation is limited to a single environment.

Our proposed \gls{cvae}-based model, \textsc{CityMPC}, addresses the remaining gaps in this landscape.
It conditions on \gls{pov} imagery from both the \gls{tx} and \gls{rx} sides alongside a terrain heightmap, encoding local scattering geometry at each endpoint without requiring a 3D mesh.
\textsc{CityMPC} predicts the full per-path parameter set: \gls{aod} and \gls{aoa} in azimuth and elevation, per-path excess delay, \gls{tof} time, received power, and complex baseband gain.
Table~\ref{tab:comparison} compares \textsc{CityMPC} against representative prior methods 
across five properties relevant to per-path \gls{mpc} generation.
``No 3D'' indicates whether the method operates without a 3D scene mesh at inference.
``Per-path Delay'' indicates prediction of discrete per-path delays rather than 
delay spread or aggregate maps.
``Gen'' indicates whether the model is generative.
``Cross-scene'' indicates whether a single trained model generalizes to held-out scenes without per-scene retraining or fine-tuning.
\begin{table}[h]
\centering
\caption{Comparison of channel modeling approaches.
\checkmark: supported. \texttimes: not supported.
Az: azimuth only. Az+El: azimuth and elevation. Sp: continuous angular spectrum, not discrete per-path.}
\label{tab:comparison}
\setlength{\tabcolsep}{4pt}
\renewcommand{\arraystretch}{1.2}
\footnotesize
\begin{tabular}{lcccccc}
\toprule
\textbf{Method} 
  & \textbf{No 3D} 
  & \textbf{\shortstack{Per-path\\AoD}} 
  & \textbf{\shortstack{Per-path\\AoA}} 
  & \textbf{\shortstack{Per-path\\Delay}} 
  & \textbf{Gen} 
  & \textbf{Cross-scene} \\
\midrule

\multicolumn{7}{l}{\textit{Stochastic / parametric}} \\
3GPP TR 38.901~\cite{38901}
  & \checkmark & Az+El & Az+El & \checkmark & \checkmark & \checkmark \\
Sionna \gls{rt}~\cite{sionna}
  & $\times$ & Az+El & Az+El & \checkmark & $\times$ & \checkmark \\

\midrule
\multicolumn{7}{l}{\textit{Neural RT surrogates (require 3D scene)}} \\
WiNeRT~\cite{orekondy2023winert}
  & $\times$ & Az & Az & \checkmark & $\times$ & $\times$ \\
GeNeRT~\cite{bian2025genert}
  & $\times$ & Az+El & Az+El & \checkmark & $\times$ & \checkmark \\

\midrule
\multicolumn{7}{l}{\textit{Scene-implicit neural fields (per-site, per-Tx)}} \\
NeRF$^{2}$~\cite{zhao2023nerf2}
  & $\times$ & $\times$ & Sp & $\times$ & $\times$ & $\times$ \\
RF-3DGS~\cite{zhang2025rf3dgs}
  & $\times$ & Sp & Sp & Sp & $\times$ & $\checkmark$ \\
NeWRF~\cite{lu2024newrf}
  & $\times$ & $\times$ & Sp & $\times$ & $\times$ & $\times$ \\

\midrule
\multicolumn{7}{l}{\textit{Radio map prediction (received power maps)}} \\
RadioUNet~\cite{levie2021radiounet}
  & \checkmark & $\times$ & $\times$ & $\times$ & $\times$ & \checkmark \\
Geo2SigMap~\cite{li2024geo2sigmap}
  & \checkmark & $\times$ & $\times$ & $\times$ & $\times$ & $\checkmark$ \\
RadioDiff-3D~\cite{wang2025radiodiff3d}
  & \checkmark & $\times$ & $\times$ & $\times$ & \checkmark & $\times$ \\

\midrule
\multicolumn{7}{l}{\textit{Unconditional generative}} \\
ChannelGAN / Diffusion~\cite{xiao2022channelgan,sengupta2023diffusion}
  & \checkmark & $\times$ & $\times$ & $\times$ & \checkmark & $\times$ \\

\midrule
\multicolumn{7}{l}{\textit{Coordinate-conditioned (no 3D)}} \\
\citet{11443807}
  & \checkmark & Az & $\times$ & \checkmark & \checkmark & $\times$ \\

\midrule
\textbf{\textsc{CityMPC} (ours)}
  & \checkmark & \textbf{Az+El} & \textbf{Az+El} & \checkmark & \checkmark & \checkmark \\

\bottomrule
\end{tabular}
\end{table}

\section{Channel parameter normalization}\label{app:normalization}
All channel quantities are normalized per-link prior to training.
Table~\ref{tab:norm_summary} summarizes the transforms applied to each output.

\paragraph{Complex gain.}
Let $\Ptot = \sum_{\ell:\,m_\ell=1}|\alpha_\ell|^2$ denote the total received power.
The normalized path coefficient is
\begin{equation}
    \tilde{\alpha}_\ell = \frac{\alpha_\ell}{\sqrt{\Ptot}},
    \label{eq:gain_normalized}
\end{equation}
so that $\sum_\ell m_\ell |\tilde{\alpha}_\ell|^2 = 1$ by construction.
The original gain is recovered as $\alpha_\ell = \tilde{\alpha}_\ell \sqrt{\Ptot}$.

\paragraph{Excess delay.}
Delays are expressed relative to the \gls{tof} time $\tauzero = \min\{\tau_\ell : m_\ell=1\}$ and clipped to a $\Wdelay=1\,\mu\mathrm{s}$ window consistent with the 3GPP TR~38.901 delay-spread definition \cite{38901}
\begin{equation}\label{eq:delay_normalized}
    \dtauell = \mathrm{clip}\!\left(\frac{\tau_\ell - \tauzero}{\Wdelay},\,0,\,1\right).
\end{equation}
The absolute delay is recovered as $\tau_\ell = \dtauell\cdot\Wdelay + \tauzero$.

\paragraph{Angle encoding.}
Each \gls{aod} direction $\Theta_\ell=(\theta_\ell^{\mathrm{az}},\theta_\ell^{\mathrm{el}})$ and \gls{aoa} direction $\Phi_\ell=(\phi_\ell^{\mathrm{az}},\phi_\ell^{\mathrm{el}})$ is encoded as a unit vector on $S^2$ via the standard spherical-to-Cartesian map:
\begin{equation}\label{eq:angle_def}
    \hat{\mathbf{d}}(\theta_{\mathrm{az}},\theta_{\mathrm{el}})=
    \begin{bmatrix}
        \sin(\theta_{\mathrm{el}})\cos(\theta_{\mathrm{az}})\\
        \sin(\theta_{\mathrm{el}})\sin(\theta_{\mathrm{az}})\\
        \cos(\theta_{\mathrm{el}})
    \end{bmatrix}\in S^2\subset\R^3.
\end{equation}
This avoids the $\pm180^\circ$ wrap-around discontinuity of azimuth-only parameterizations and admits a smooth cosine loss on the unit sphere.
The original angles are recovered via $\theta_{\mathrm{az}}=\mathrm{atan2}(\hat{d}_2,\hat{d}_1)$ and $\theta_{\mathrm{el}}=\arccos(\hat{d}_3)$.

\paragraph{\gls{tof} time.}
The distribution of $\tauzero$ across urban links is well approximated by a log-normal \cite{rappaport2013mmwave}, so we apply a log z-score transform using per-city training statistics $(\mu_{\log},\sigma_{\log})$
\begin{equation}
    \widetilde{\tau}_0 = \frac{\log(\tauzero^{\mathrm{ns}} + \epsilon) - \mu_{\log}}{\sigma_{\log}},
\end{equation}
where $\tauzero^{\mathrm{ns}} = \tauzero \times 10^9$ is in nanoseconds and $\epsilon=10^{-12}$ prevents numerical issues.

\paragraph{Received power.}
The received power $\Prx = 10\log_{10}(\Ptot)$~dB is approximately Gaussian-distributed conditioned on city, so we apply a z-score using per-city statistics $(\mu_{\mathrm{rx}},\sigma_{\mathrm{rx}})$:
\begin{equation}
    \widetilde{P}_{\mathrm{rx}} = \frac{P_{\text{rx}} - \mu_{\mathrm{rx}}}{\sigma_{\mathrm{rx}}}.
\end{equation}
\begin{table}[h]
\centering
\caption{Summary of channel normalization transforms.}
\label{tab:norm_summary}
\begin{tabular}{llcc}
\toprule
Quantity & Transform & Range & City-dependent \\
\midrule
Path presence $m_\ell$              & None                        & $\{0,1\}$      & No  \\
Complex gain $\tilde{\alpha}_\ell$  & Eq.~\eqref{eq:gain_normalized}    & $[0,1]$   & No  \\
Excess delay $\dtauell$             & Eq.~\eqref{eq:delay_normalized} & $[0,1]$    & No  \\
Angle directions $\daodell,\daoaell$& Eq.~\eqref{eq:angle_def}    & $[-1,1]^3$     & No  \\
\gls{tof} $\tauzero$            & Log z-score                 & $\mathbb{R}$   & Yes \\
Received power $\Prx$               & Z-score                     & $\mathbb{R}$   & Yes \\
\bottomrule
\end{tabular}
\end{table}

\section{DeepMIMO to Sionna \gls{rt} scene conversion}\label{app:scene_conversion}

DeepMIMO distributes scenario geometry as four per-scenario files.
\texttt{objects.json} contains object metadata and semantic labels.
\texttt{params.json} contains the scenario coordinate frame and grid parameters.
\texttt{vertices.npy} contains the 3D mesh vertex arrays per object.
\texttt{materials.mat} contains the per-object electromagnetic material properties.
Sionna \gls{rt} requires scene geometry in Mitsuba XML format \cite{Mitsuba3}, with each surface assigned a radio material BSDF encoding the properties required for physically-based ray-surface interaction.
We implement a conversion pipeline (\texttt{scenegen}) that bridges these two representations without manual scene authoring.
Critically, DeepMIMO provides the electromagnetic material properties ($\varepsilon_r$, $\sigma$, scattering coefficient, cross-polarization coefficient, and material thickness) for each 3D object in the scene.
We assign these same properties to the corresponding Sionna shapes, ensuring that the 3D geometry used for ray tracing is electromagnetically consistent with the DeepMIMO channel simulation.

\paragraph{Geometry assembly.}
Each entry in \texttt{objects.json} references a named mesh object with an associated material label.
We load the corresponding vertex arrays from \texttt{vertices.npy} and triangulate each object's faces using a standard fan triangulation from the per-face vertex lists.
Objects are written as individual Mitsuba \texttt{shape} nodes with vertex coordinates expressed in the scenario's local Cartesian frame defined by \texttt{params.json}.
The coordinate origin and axis alignment are preserved from DeepMIMO so that \gls{tx} and \gls{rx} positions are directly compatible with the rendered scene without any additional coordinate transform.

\paragraph{Radio material assignment.}
Each object carries a material label that maps to the \gls{em} properties stored in \texttt{materials.mat}.
We register a custom Mitsuba BSDF plugin (\texttt{radio\_material}) via \texttt{register\_radio\_material\_bsdf()} that stores the five scalar electromagnetic fields on each shape.
The BSDF is instantiated once per unique material label and assigned to all shapes sharing that label so that the Sionna \gls{rt} intersection engine can query per-surface electromagnetic properties during ray tracing.
Background geometry such as sky and ground plane is assigned default properties of $\varepsilon_r=1$ and $\sigma=0$ with zero scattering and cross-polarization coefficients.

\paragraph{POV rendering.}
Once the Mitsuba XML scene is loaded into Sionna \gls{rt}, we render two 12-channel \gls{pov} stacks per link.
A camera is placed at the \gls{tx} position looking toward the \gls{rx} (and vice versa) with a fixed field of view of $45^\circ$ and $128$ samples per pixel for the RGB pass.
Mitsuba directly exposes per-pixel color, metric depth, geometric surface normals, and a shape index image from each rendered viewpoint.
The shape index identifies which 3D object is visible at each pixel of the rendered image.
We use this shape index to look up the radio material BSDF assigned to each visible shape, thereby transferring the electromagnetic properties of the 3D objects directly into the pixel space of the \gls{pov} image.
The 12 channels are assembled as follows.
Channels 0 to 2 are RGB rendered via Sionna's visual scene pipeline.
Channel 3 is metric depth.
Channels 4 to 6 are geometric surface normals.
Channels 7 to 11 are the per-pixel radio material properties ($\varepsilon_r$, $\sigma$, scattering coefficient, cross-polarization coefficient, and material thickness) of the intersected shape at each pixel.
This design means the \gls{pov} stack encodes not only the geometry visible from the \gls{tx} or \gls{rx} viewpoint but also the full electromagnetic character of every visible surface, without requiring the 3D mesh to be available at inference time.
The global heightmap is rendered as a top-down orthographic projection of the scene, capturing building heights at 4\,m per pixel resolution over the full $512\times512$\,m scenario footprint.
All rendered data is packed alongside the channel tensors into per-city NPZ archives by \texttt{export\_channels\_and\_env.py}.

\section{Dataset details}\label{app:dataset}
\subsection{Channel generation parameters}
Channels are generated using \gls{siso} links with H-polarization and omnidirectional antennas at both \gls{tx} and \gls{rx}.
Links whose total received power falls below $-120$\,dBm are discarded entirely.
Paths more than $25$\,dB below the strongest active path per link are pruned, following 3GPP TR~38.901~\cite{38901}.

\subsection{Radio materials}
\label{app:materials}
All city scenes use the same set of three radio materials, applied uniformly across all scenarios.
The materials follow the electromagnetic model implemented in Sionna \gls{rt}~\cite{10465179}, in which each surface is parameterized by a relative permittivity $\varepsilon_r$, a conductivity $\sigma$\,(S/m), a scattering coefficient $S \in [0,1]$, and a cross-polarization discrimination coefficient $K_x \in [0,1]$.
The permittivity and conductivity values are taken from ITU-R P.2040-3~\cite{ITURP20403}, which tabulates measured electrical properties of common building materials at radio frequencies.
Table~\ref{tab:materials} lists the three materials and their parameters at $3.5$\,GHz.

\begin{table}[h]
\centering
\caption{Radio material parameters used in all 20 city scenes at $3.5$\,GHz.
Permittivity and conductivity follow ITU-R P.2040-3~\cite{ITURP20403}.
Scattering uses the directive pattern of Sionna \gls{rt}~\cite{10465179}.}
\label{tab:materials}
\begin{tabular}{lcccc}
\toprule
Material & $\varepsilon_r$ & $\sigma$ (S/m) & $S$ & $K_x$ \\
\midrule
Concrete & 18.18 & 0.765 & 0.00 & 0.00 \\
Wood     &  5.72 & 0.001 & 0.40 & 0.40 \\
Glass    &  5.24 & 0.123 & 0.40 & 0.40 \\
\bottomrule
\end{tabular}
\end{table}

Concrete is assigned to building facades and ground surfaces.
Wood and glass are assigned to secondary structural elements.
Scattering for wood and glass follows the directive pattern with parameters $\alpha_r = \alpha_i = 4$ and $\lambda = 0.75$, consistent with the default Sionna \gls{rt} scattering configuration~\cite{10465179}.

\subsection{Per-city link counts}
Table~\ref{tab:per_city} reports the number of links in each city's train, validation, and test splits for every dataset which DeepMIMO is capable of creating. While only the cities found in Tab. \ref{tab:per_city_results} are used for training and evaluation in this paper, we provide the larger listing of available cities to elucidate other possibilities.
The five cities selected for training and evaluation in this work — Austin, Dallas, Fort Worth, Denver, and New York — were chosen to represent diverse urban morphologies spanning grid-based suburban layouts, mid-density urban cores, and dense high-rise environments.

\begin{table}[h]
\centering
\caption{Per-city link counts (70/15/15 train/val/test split).}
\label{tab:per_city}
\small
\begin{tabular}{lrrrr}
\toprule
City & Train & Val & Test & Total \\
\midrule
New York          & 43{,}738 & 9{,}372  & 9{,}373  & 62{,}483  \\
Los Angeles       & 41{,}841 & 8{,}965  & 8{,}967  & 59{,}773  \\
Chicago           & 17{,}138 & 3{,}672  & 3{,}673  & 24{,}483  \\
Houston           & 47{,}902 & 10{,}264 & 10{,}266 & 68{,}432  \\
Phoenix           & 39{,}595 & 8{,}484  & 8{,}486  & 56{,}565  \\
Philadelphia      & 20{,}203 & 4{,}329  & 4{,}330  & 28{,}862  \\
Miami             & 54{,}521 & 11{,}683 & 11{,}684 & 77{,}888  \\
San Diego         & 42{,}672 & 9{,}144  & 9{,}144  & 60{,}960  \\
Dallas            & 69{,}029 & 14{,}791 & 14{,}793 & 98{,}613  \\
San Francisco     & 40{,}557 & 8{,}690  & 8{,}692  & 57{,}939  \\
Austin            & 55{,}636 & 11{,}922 & 11{,}923 & 79{,}481  \\
Santa Clara       & 49{,}658 & 10{,}641 & 10{,}641 & 70{,}940  \\
Fort Worth        & 58{,}890 & 12{,}619 & 12{,}621 & 84{,}130  \\
Columbus          & 31{,}065 & 6{,}656  & 6{,}658  & 44{,}379  \\
Charlotte         & 68{,}152 & 14{,}604 & 14{,}605 & 97{,}361  \\
Indianapolis      & 55{,}202 & 11{,}829 & 11{,}829 & 78{,}860  \\
San Francisco (2) & 68{,}970 & 14{,}779 & 14{,}780 & 98{,}529  \\
Seattle           & 32{,}540 & 6{,}973  & 6{,}974  & 46{,}487  \\
Denver            & 71{,}883 & 15{,}403 & 15{,}404 & 102{,}690 \\
Oklahoma City     & 53{,}069 & 11{,}372 & 11{,}373 & 75{,}814  \\
\midrule
\textbf{Total} & \textbf{962{,}261} & \textbf{206{,}192} & \textbf{206{,}216} & \textbf{1{,}374{,}669} \\
\bottomrule
\end{tabular}
\end{table}

\begin{table}[h]
\centering
\caption{Link counts for the five cities used in evaluation.}
\label{tab:per_city_results}
\small
\begin{tabular}{lrrrr}
\toprule
City & Train & Val & Test & Total \\
\midrule
Dallas         & 69{,}029 & 14{,}791 & 14{,}793 & 98{,}613 \\
Fort Worth     & 58{,}890 & 12{,}619 & 12{,}621 & 84{,}130 \\
New York       & 43{,}738 & 9{,}372  & 9{,}373  & 62{,}483 \\
Denver         & 71{,}883 & 15{,}403 & 15{,}404 & 102{,}690 \\
Austin         & 55{,}636 & 11{,}922 & 11{,}923 & 79{,}481 \\
\midrule
\textbf{Total} & \textbf{299{,}176} & \textbf{64{,}107} & \textbf{64{,}114} & \textbf{427{,}397} \\
\bottomrule
\end{tabular}
\end{table}

\section{Conditioning input preprocessing}\label{app:conditioning}

\paragraph{Point-of-view image stacks.}
Each of the 12 \gls{pov} channels is independently normalized before being passed to the conditioning encoder.
RGB channels are divided by 255 to map pixel values to $[0,1]$.
Depth is transformed as $\log(1 + d) / \log(1 + d_{\max})$ with $d_{\max} = 500$\,m, mapping all values to $[0,1]$.
Surface normals are kept in $[-1,1]$ as they are already unit-vector components.
Relative permittivity $\varepsilon_r$ is transformed as $\log(1 + \varepsilon_r) / \log(1 + \varepsilon_{r,\max})$ with $\varepsilon_{r,\max} = 25$.
Conductivity $\sigma$ is transformed as $\log(1 + \sigma) / \log(1 + \sigma_{\max})$ with $\sigma_{\max} = 10$\,S/m.
Scattering coefficient, cross-polarization coefficient, and material thickness are already in $[0,1]$ and are passed through unchanged.

\paragraph{Global heightmap.}
The global heightmap is a single-channel $128 \times 128$ image capturing building height at each spatial location over the full $512 \times 512$\,m scenario footprint at 4\,m per pixel resolution.
Building heights are normalized as $\log(1 + h) / \log(1 + h_{\max})$ with $h_{\max} = 500$\,m, mapping all values to $[0,1]$.

\paragraph{\gls{tx} and \gls{rx} coordinates.}
The \gls{tx} and \gls{rx} Cartesian coordinates are provided as the raw scalar vector $\mathbf{s} \in \R^6$ in metres.
Each scalar coordinate is embedded independently using a sinusoidal Fourier feature encoding \cite{tancik2020fourier} with $L_f = 8$ log-spaced frequency bands and a coordinate scale of $1000$\,m.
This maps each scalar to a $2L_f = 16$-dimensional vector, so that all six coordinates together produce a $96$-dimensional embedding.
The embedding is then projected to $\R^{128}$ via a learned linear layer before being concatenated with the image features in the conditioning encoder.

\section{Model dimensions and training hyperparameters}\label{app:model_dims}
\paragraph{Token counts.}
Each ChannelViT tower divides its input image into non-overlapping spatial patches of size $32 \times 32$ pixels.
For an input resolution of $128 \times 128$, this yields $\lfloor 128/32 \rfloor^2 = 16$ spatial patches per image channel.
The global heightmap has $C=1$ channel, producing 16 global tokens.
The \gls{tx} and \gls{rx} \gls{pov} stacks each have $C=12$ channels, producing 192 tokens per modality.
The \gls{tx} and \gls{rx} coordinate scalar feature vector is projected to a single scalar token, which is appended to the image tokens when constructing the cross-attention memory in \ttpe and \ttdec.
The full cross-attention memory therefore comprises $16 + 192 + 192 + 1 = 401$ tokens of dimension $d_{\mathrm{model}}$.
For the physics-informed angle heads, \gls{aod} tokens attend to a \gls{tx}-only memory of $16 + 192 + 1 = 209$ tokens, and \gls{aoa} tokens attend to an \gls{rx}-only memory of $16 + 192 + 1 = 209$ tokens.

\paragraph{Training hyperparameters.}
\label{app:training_hparams}
Tables~\ref{tab:hparams_txvit} and~\ref{tab:hparams_mlp} report the full hyperparameters for \textsc{CityMPC} and the \gls{mlp} baseline respectively.
Both models use AdamW~\citep{loshchilov2017decoupled} with cosine learning rate decay, Kendall uncertainty weighting~\citep{kendall2018multi}, and identical KL scheduling and regularization.

\begin{table}[h]
  \centering
  \caption{Hyperparameters for \textsc{CityMPC}.}
  \label{tab:hparams_txvit}
  \begin{tabular}{lc}
    \toprule
    Hyperparameter & Value \\
    \midrule
    \multicolumn{2}{l}{\textit{Model}} \\
    \midrule
    $L$ & 25 \\
    Latent dimension $d_z$ & 64 \\
    Scene token dimension $d_{\text{scene}}$ & 128 \\
    Transformer hidden dimension $d_{\text{model}}$ & 256 \\
    Attention heads & 8 \\
    ChannelViT layers & 3 \\
    Posterior encoder layers & 2 \\
    Decoder layers & 3 \\
    Patch size & $32 \times 32$ \\
    FFN width & 1024 \\
    Dropout & 0.1 \\
    Free bits $\lambda$ & 0.1 \\
    \midrule
    \multicolumn{2}{l}{\textit{Training}} \\
    \midrule
    Optimizer & AdamW \\
    Weight decay & $10^{-4}$ \\
    Peak learning rate & $3.5 \times 10^{-4}$ \\
    LR schedule & Cosine decay \\
    LR warmup steps & 1{,}000 \\
    LR minimum ratio & 0.01 \\
    Gradient clip & 1.0 \\
    Batch size (effective) & 256 \\
    Epochs & 500 \\
    Precision & BF16 \\
    \midrule
    \multicolumn{2}{l}{\textit{KL Schedule}} \\
    \midrule
    $\beta_{\max}$ & 0.1 \\
    $\beta$ warmup steps & 11{,}000 \\
    \midrule
    Total parameters & 15.8\,M \\
    \bottomrule
  \end{tabular}
\end{table}

\begin{table}[h]
  \centering
  \caption{Hyperparameters for the \gls{mlp} baseline.}
  \label{tab:hparams_mlp}
  \begin{tabular}{lc}
    \toprule
    Hyperparameter & Value \\
    \midrule
    \multicolumn{2}{l}{\textit{Model}} \\
    \midrule
    $L$ & 25 \\
    Latent dimension $d_z$ & 128 \\
    Scene embedding dimension $d_{\text{scene}}$ & 256 \\
    Path embedding dimension $d_{\text{path}}$ & 64 \\
    Decoder trunk hidden width $d_{\text{dec}}$ & 512 \\
    Posterior encoder hidden width & 512 \\
    Prior hidden width & 256 \\
    Dropout & 0.1 \\
    \midrule
    \multicolumn{2}{l}{\textit{Training}} \\
    \midrule
    Optimizer & AdamW \\
    Weight decay & $10^{-4}$ \\
    Peak learning rate & $1.4 \times 10^{-4}$ \\
    LR schedule & Cosine decay \\
    LR warmup steps & 1{,}000 \\
    LR minimum ratio & 0.1 \\
    Gradient clip & 1.0 \\
    Batch size (effective) & 256 \\
    Epochs & 500 \\
    Precision & BF16 \\
    \midrule
    \multicolumn{2}{l}{\textit{KL Schedule}} \\
    \midrule
    $\beta_{\max}$ & 0.5 \\
    $\beta$ warmup steps & 11{,}000 \\
    \midrule
    Total parameters & 35.5\,M \\
    \bottomrule
  \end{tabular}
\end{table}

\section{Per-task losses, Kendall uncertainty weighting, and training hyperparameters}\label{app:kendall_loss}
\paragraph{Per-task loss functions.}
Let $\bmm = [m_1, \dots, m_L]$ be the ground-truth presence mask with $m_\ell \in \{0,1\}$.
The presence-masked \gls{mse} for a per-path scalar target $\bma \in \R^L$ and prediction $\widehat{\bma} \in \R^L$ is
\begin{equation}\label{eq:loss_mse}
    \mcl_{m}(\bma, \widehat{\bma};\, \bmm)
    = \frac{1}{\|\bmm\|_1} \sum_{\ell=1}^{L} m_\ell \left(a_\ell - \hat{a}_\ell\right)^2.
\end{equation}
The presence-masked cosine loss for unit-vector matrices $\mathbf{D} \in \R^{L \times 3}$ and $\widehat{\mathbf{D}} \in \R^{L \times 3}$ is
\begin{equation}\label{eq:loss_cos}
    \mcl_{c}(\mathbf{D}, \widehat{\mathbf{D}};\, \bmm)
    = \frac{1}{\|\bmm\|_1} \sum_{\ell=1}^{L} m_\ell \left(1 - \hat{\mathbf{d}}_\ell^\intercal \mathbf{d}_\ell \right).
\end{equation}
Table~\ref{tab:loss} lists the loss function assigned to each of the $K=7$ tasks.
\begin{table}[h]
  \caption{Reconstruction losses.
  $\bmm = [m_1,\dots,m_{L}]$ is the ground-truth presence mask ($m_\ell=1$: active, $m_\ell=0$: inactive).
  $\mcl_{m}$ and $\mcl_{c}$ are defined in eqs.~\eqref{eq:loss_mse}--\eqref{eq:loss_cos}.
  BCE denotes binary cross-entropy.
  Scalar heads operate on z-scored targets.
  Masked heads operate on active paths only.}
  \label{tab:loss}
  \centering
  \small
  \begin{tabular}{clllll}
    \toprule
    $k$ & Task & Symbol & Space & Loss ($\mcl_k$) & Mask \\
    \midrule
    1 & Presence       & $\bmm$                & binary        & $\frac{1}{L}\sum_{\ell}\mathrm{BCE}(m_\ell,\hat{m}_\ell)$ & none \\
    2 & \gls{tof}  & $\tauzero$            & z-score ($\log$ ns) & $(\tauzero - \hat{\tauzero})^2$                           & none \\
    3 & Received power & $\Prx$                & z-score (dB)  & $(\Prx - \hat{P}_{\mathrm{rx}})^2$                        & none \\
    4 & AoA direction  & $\mathbf{D}^{a}$      & unit sphere   & $\mcl_{c}(\mathbf{D}^{a},\widehat{\mathbf{D}}^{a};\bmm)$  & active \\
    5 & AoD direction  & $\mathbf{D}^{d}$      & unit sphere   & $\mcl_{c}(\mathbf{D}^{d},\widehat{\mathbf{D}}^{d};\bmm)$  & active \\
    6 & Excess delay   & $\dtau$               & $[0,1]$       & $\mcl_{m}(\dtau,\widehat{\dtau};\bmm)$                    & active \\
    7 & Baseband gain  & $\Re(\bma),\Im(\bma)$ & $[0,1]$      & $\mcl_m(\Re(\bma),\Re(\widehat{\bma});\bmm)$              & active \\
    \bottomrule
  \end{tabular}
\end{table}

\paragraph{Kendall uncertainty weighting.}
Although all targets are normalized prior to training (Appendix~\ref{app:normalization}), the seven tasks in Tab.~\ref{tab:loss} span fundamentally different output spaces, namely binary classification, unbounded scalar regression, and directional regression on the unit sphere.
Manual tuning of the weights $w_k$ across such heterogeneous outputs is prohibitively expensive.
We instead treat each weight as arising from a per-task noise parameter $\sigma_k$, whose magnitude reflects how difficult the task is to predict~\cite{kendall2018multi}.

For a \textbf{regression} task, assuming a Gaussian observation likelihood $p(\bmx \mid f(\bmz)) = \mathcal{N}(f(\bmz),\, \sigma_k^2)$ and maximizing the log-likelihood yields the weighted task contribution
\begin{equation}\label{eq:kendall_reg}
    \frac{1}{2\sigma_k^2}\,\mathcal{L}_k + \log \sigma_k,
\end{equation}
where the first term down-weights the task loss by $1/\sigma_k^2$ and the second term penalizes excessive uncertainty.

For the \textbf{classification} task (path presence), assuming a Bernoulli likelihood scaled by $1/\sigma_k^2$ gives the analogous contribution~\cite{kendall2018multi}
\begin{equation}\label{eq:kendall_cls}
    \frac{1}{\sigma_k^2}\,\mathcal{L}_k + 2\log \sigma_k.
\end{equation}

In both cases a larger $\sigma_k$ reduces the effective task weight, allowing the model to automatically balance tasks of differing difficulty.
Each $\sigma_k$ is parameterized as $\sigma_k = \exp(s_k)$ where $s_k$ is a learned scalar, initialized to $s_k = 0$ (i.e.\ $\sigma_k = 1$) and jointly optimized with the model weights.

\paragraph{Limitations of \gls{uw}.}
The Kendall formulation in eqs.~\eqref{eq:kendall_reg}--\eqref{eq:kendall_cls} can become degenerate when $\sigma_k \to \infty$ for any task $k$, since the data-fit term $\mathcal{L}_k / \sigma_k^2$ vanishes while the regularizer $\log \sigma_k$ remains unbounded above when the joint loss is allowed to take negative values.
This failure mode is a known limitation of \gls{uw} documented in the multi-task learning literature as task abandonment~\cite{kirchdorfer2024analytical, liebel2018auxiliary}.
Standard mitigations include the Liebel-K\"orner positive regularizer $\log(1 + \sigma_k^2)$~\cite{liebel2018auxiliary}, GradNorm-style direct gradient balancing~\cite{chen_gradnorm_2018}, and analytical \gls{uw} variants with softmax-normalized weights~\cite{kirchdorfer2024analytical}.
We leave the integration of these alternatives to future work.
 
\paragraph{Filtering criterion for quantitative results.}
We exclude runs whose learned \gls{uw} task uncertainty $\sigma_{\mathrm{presence}}$ exceeds $10^{-3}$ at the final training step. 
This threshold separates converged runs (which reach $\approx 10^{-4}$) from runs exhibiting the task abandonment failure mode of \gls{uw}~\cite{kendall2018multi,liebel2018auxiliary,kirchdorfer2024analytical}. 
The criterion uses only training time \gls{uw} parameters and introduces no test set leakage, leaving $n=3,4,4,4,4$ runs for Austin, Dallas, Denver, Fort~Worth, and New~York.

\section{Hardware and training times}
\label{app:hardware}
All models are trained on a high-performance computing cluster.
Each training job is allocated one node with two NVIDIA H100 80GB HBM3 GPUs and 28 CPU cores, using CUDA 12.6 and BF16 mixed precision.
Per-city wall-clock training times for \textsc{CityMPC} are reported in Table~\ref{tab:training_times}.
\begin{table}[h]
  \centering
  \caption{Per-city wall-clock training times for \textsc{CityMPC} on 2$\times$ NVIDIA H100 80GB GPUs.}
  \label{tab:training_times}
  \begin{tabular}{lc}
    \toprule
    City & Training Time (hours) \\
    \midrule
    Austin      & 3.53 \\
    Dallas      & 4.06 \\
    Denver      & 4.55 \\
    Fort Worth  & 3.52 \\
    New York    & 2.78 \\
    \bottomrule
  \end{tabular}
\end{table}

\subsection{Inference timing methodology}
\label{app:timing_methodology}
The per-link inference latency reported in Table~\ref{tab:model_times} is measured on a single NVIDIA H100 80GB HBM3 GPU.
Both Sionna \gls{rt} and \textsc{CityMPC} inference run natively on the \gls{gpu}.
Sionna \gls{rt} uses a \gls{gpu}-optimized ray tracing backend, so the comparison is between two \gls{gpu}-resident pipelines on identical hardware.
For each of the five evaluation cities, three paired runs are executed using independent \gls{rt} sampling seeds and three \textsc{CityMPC} checkpoints trained from different initialization seeds.
Reported values are the mean and standard deviation across the resulting fifteen runs.

We report two \textsc{CityMPC} latency numbers to separate pipeline overhead from model compute.
\textit{End-to-end} (e2e) measures the wall-clock time from reading a test sample off disk to obtaining the predicted \gls{mpc} parameters.
This includes \gls{hdf} data loading, batching, host-to-device transfer, and the model forward pass, and reflects the latency a practitioner observes when running the released inference pipeline as-is.
\textit{Model only} measures the wall-clock time of the model forward pass alone, excluding data loading and host-to-device transfer.
This isolates the cost of the learned surrogate from the surrounding data pipeline.
The Sionna \gls{rt} measurement comprises per-transmitter-group scene preparation and the ray tracing computation, with one-time scene loading excluded.
\textsc{CityMPC} excludes one-time model loading for the same reason.
The end-to-end measurement is dominated by data loading rather than model compute, so the model-only number is the more direct measure of the surrogate's intrinsic speed.

\section{Evaluation metrics}
\label{app:metrics}

All metrics are computed on the held-out test set of each city.
Let $N$ denote the number of test links.
For link $n$, let $\bmm^{(n)} = [m_1, \dots, m_L]$ be the ground-truth presence mask with $m_\ell \in \{0,1\}$, let $\alpha_\ell^{(n)} \in \mathbb{C}$ be the complex gain of path $\ell$, and let $\tau_\ell^{(n)}$ be its absolute delay.

\paragraph{Path presence F1.}
Path presence F1 is the harmonic mean of precision and recall computed over the binary presence mask $\bmm$ across all $L = 25$ slots and all $N$ test links.

\paragraph{\gls{tof} \gls{mae}.}
\Gls{tof} \gls{mae} is the mean absolute error in the predicted $\tauzero = \min\{\tau_\ell : m_\ell = 1\}$,
\begin{equation}
    \mathcal{M}_{\tau_0} = \frac{1}{N} \sum_{n=1}^{N}
    \left| \hat{\tau}_0^{(n)} - \tau_0^{(n)} \right|,
\end{equation}
reported in nanoseconds.

\paragraph{Average delay \gls{mae}.}
The power-weighted mean absolute delay for a link is
\begin{equation}
    \bar{\tau}^{(n)} = \frac{\displaystyle\sum_{\ell=1}^{L} m_\ell^{(n)}\, |\alpha_\ell^{(n)}|^2\, \tau_\ell^{(n)}}
                           {\displaystyle\sum_{\ell=1}^{L} m_\ell^{(n)}\, |\alpha_\ell^{(n)}|^2},
\end{equation}
and the average delay \gls{mae} is
\begin{equation}
    \mathcal{M}_{\bar{\tau}} = \frac{1}{N} \sum_{n=1}^{N}
    \left| \hat{\bar{\tau}}^{(n)} - \bar{\tau}^{(n)} \right|,
\end{equation}
reported in nanoseconds.
This definition is consistent with \citet{orekondy2023winert} and \citet{bian2025genert}.

\paragraph{Received power \gls{mae}.}
Received power \gls{mae} is the mean absolute error in total received power $\Prx = 10\log_{10}(\Ptot)$~dB,
\begin{equation}
    \mathcal{M}_{\Prx} = \frac{1}{N} \sum_{n=1}^{N}
    \left| \hat{P}_{\mathrm{rx}}^{(n)} - P_{\mathrm{rx}}^{(n)} \right|,
\end{equation}
reported in decibels.

\paragraph{Average angular \glspl{mae}.}
Each angle dimension $\phi \in \{\theta^{\mathrm{az}}_{\mathrm{AoD}},\, \theta^{\mathrm{el}}_{\mathrm{AoD}},\, \phi^{\mathrm{az}}_{\mathrm{AoA}},\, \phi^{\mathrm{el}}_{\mathrm{AoA}}\}$ is recovered from the predicted unit vector $\widehat{\mathbf{D}}$ via $\theta_{\mathrm{az}} = \mathrm{atan2}(\hat{d}_2, \hat{d}_1)$ and $\theta_{\mathrm{el}} = \arccos(\hat{d}_3)$, consistent with eq.~\eqref{eq:angle_def}.
For each link, the power-weighted mean angle is computed over active paths as
\begin{equation}
    \bar{\phi}^{(n)} = \frac{\displaystyle\sum_{\ell=1}^{L} m_\ell^{(n)}\, |\alpha_\ell^{(n)}|^2\, \phi_\ell^{(n)}}
                           {\displaystyle\sum_{\ell=1}^{L} m_\ell^{(n)}\, |\alpha_\ell^{(n)}|^2},
\end{equation}
where azimuth averages use the circular mean via $\mathrm{atan2}$, mapped to $[0^\circ, 360^\circ)$.
The average angular \gls{mae} for each dimension is
\begin{equation}
    \bar{\mathcal{M}}_{\phi} = \frac{1}{N} \sum_{n=1}^{N}
    \left| \hat{\bar{\phi}}^{(n)} - \bar{\phi}^{(n)} \right|,
\end{equation}
reported in degrees.
We report four average angular \glspl{mae}: \gls{aod} azimuth, \gls{aod} elevation, \gls{aoa} azimuth, and \gls{aoa} elevation.

\section{Channel realization examples}
\label{app:realizations}
\begin{figure*}[h]
\centering
\scriptsize
\setlength{\tabcolsep}{2pt}
\begin{tabular}{ccccc}
  \gls{cir} & \gls{aod} Azimuth & \gls{aod} Elevation & \gls{aoa} Azimuth & \gls{aoa} Elevation \\[2pt]
  \includegraphics[width=0.24\textwidth]{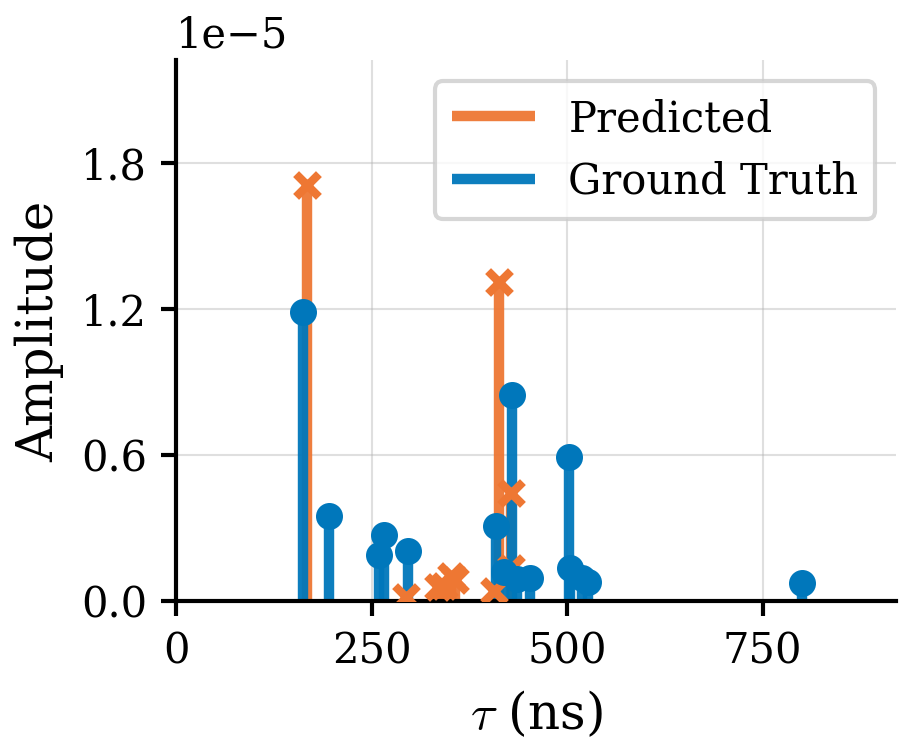} &
  \includegraphics[width=0.19\textwidth]{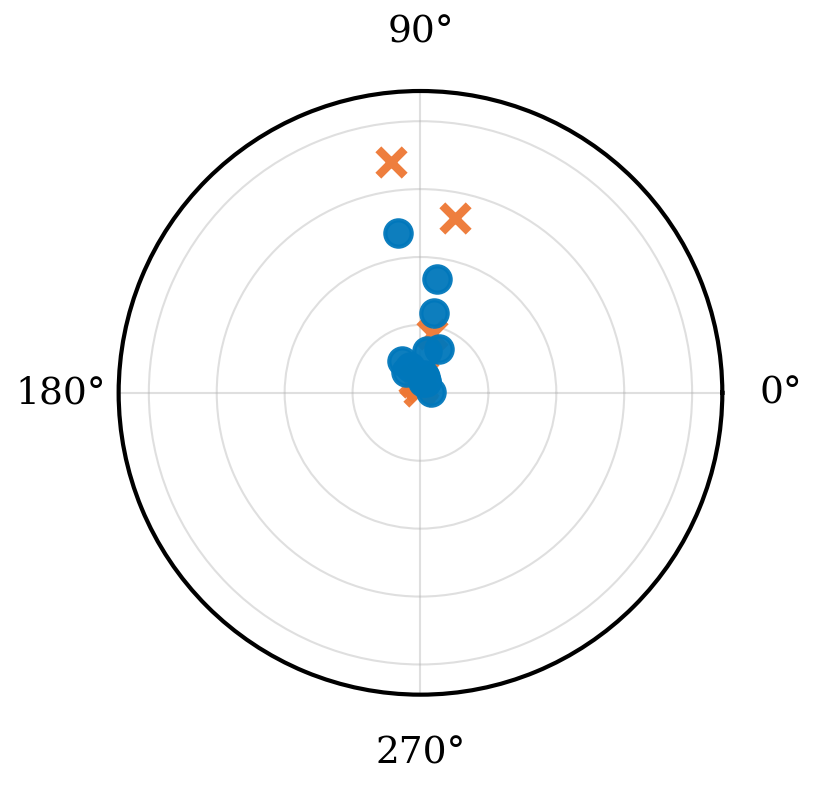} &
  \includegraphics[width=0.14\textwidth]{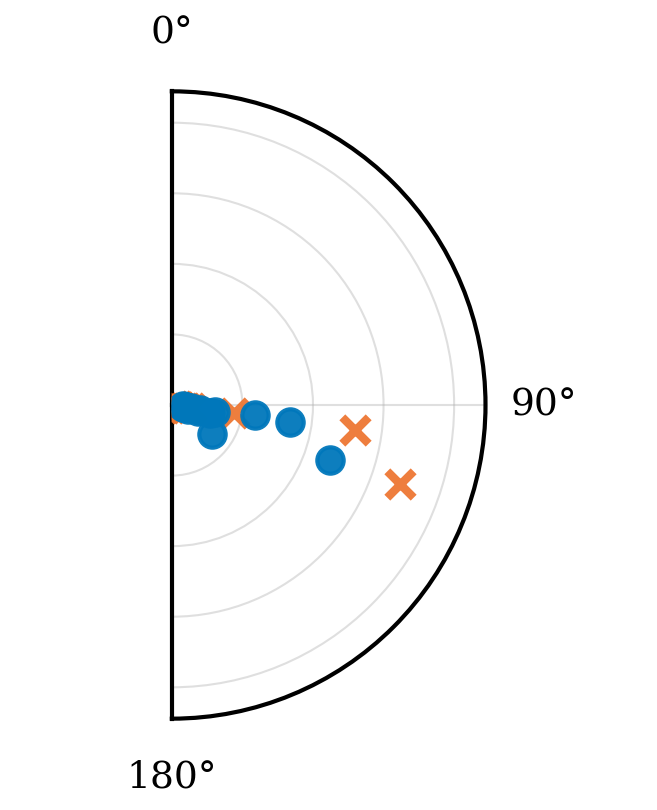} &
  \includegraphics[width=0.19\textwidth]{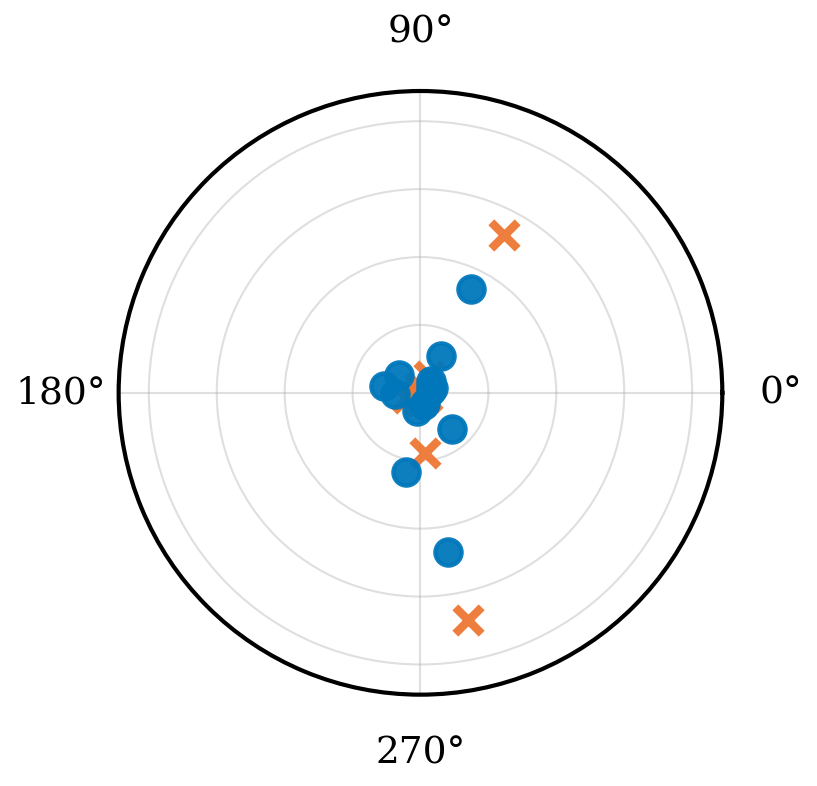} &
  \includegraphics[width=0.114\textwidth]{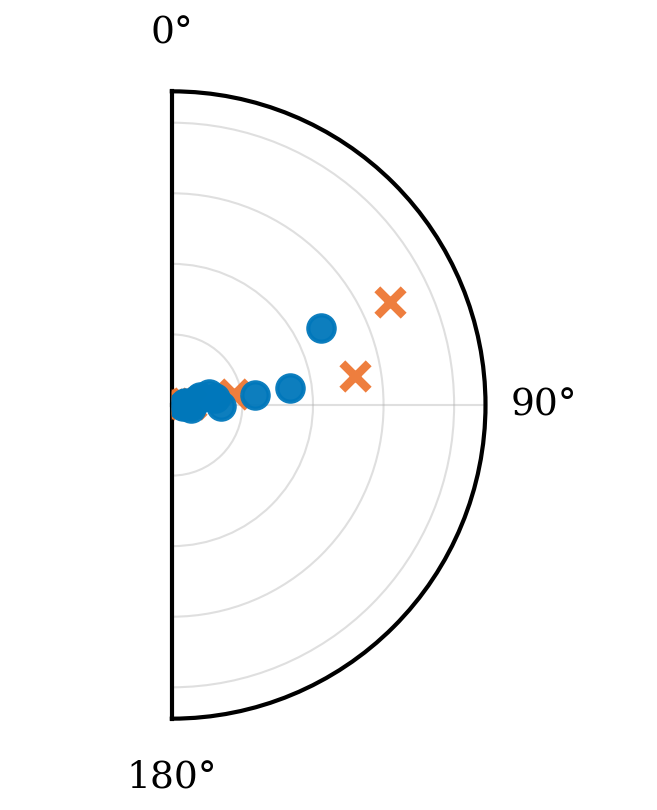} \\[4pt]
  \includegraphics[width=0.24\textwidth]{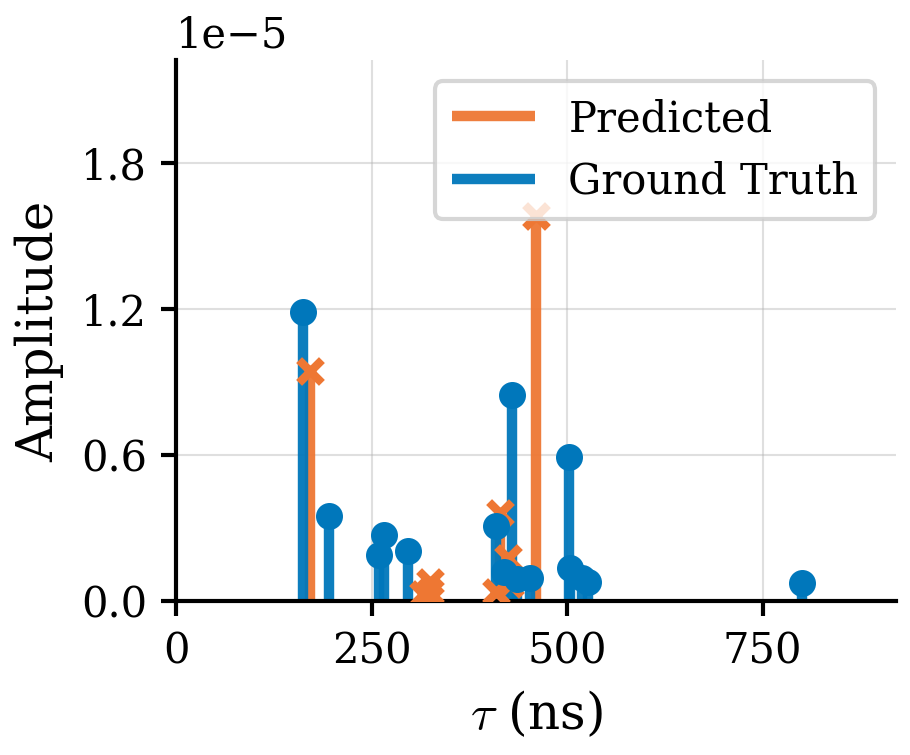} &
  \includegraphics[width=0.19\textwidth]{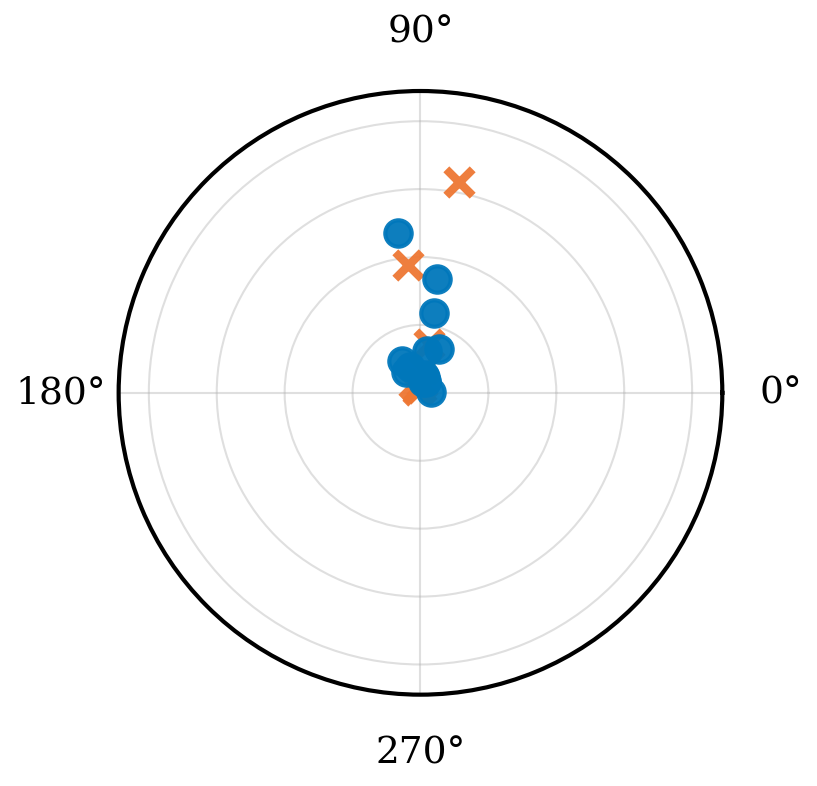} &
  \includegraphics[width=0.14\textwidth]{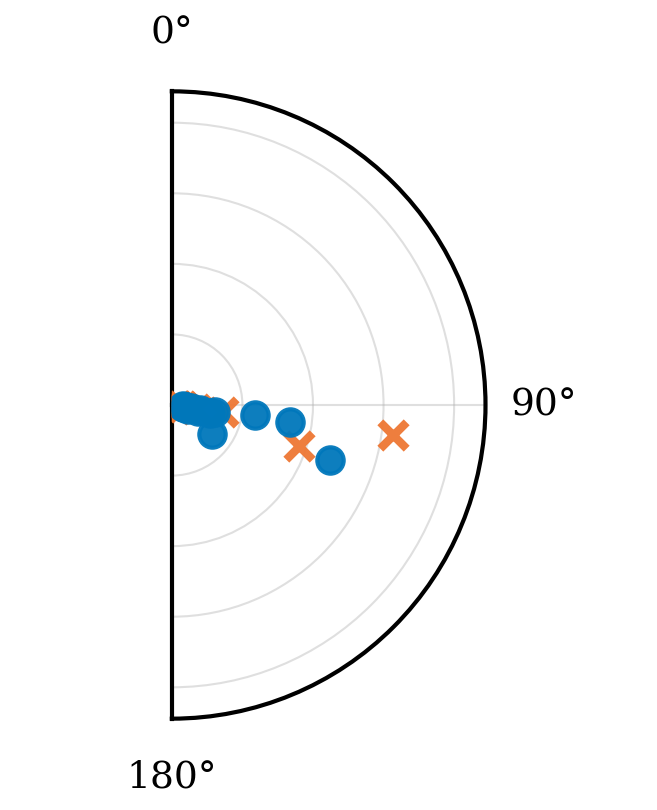} &
  \includegraphics[width=0.19\textwidth]{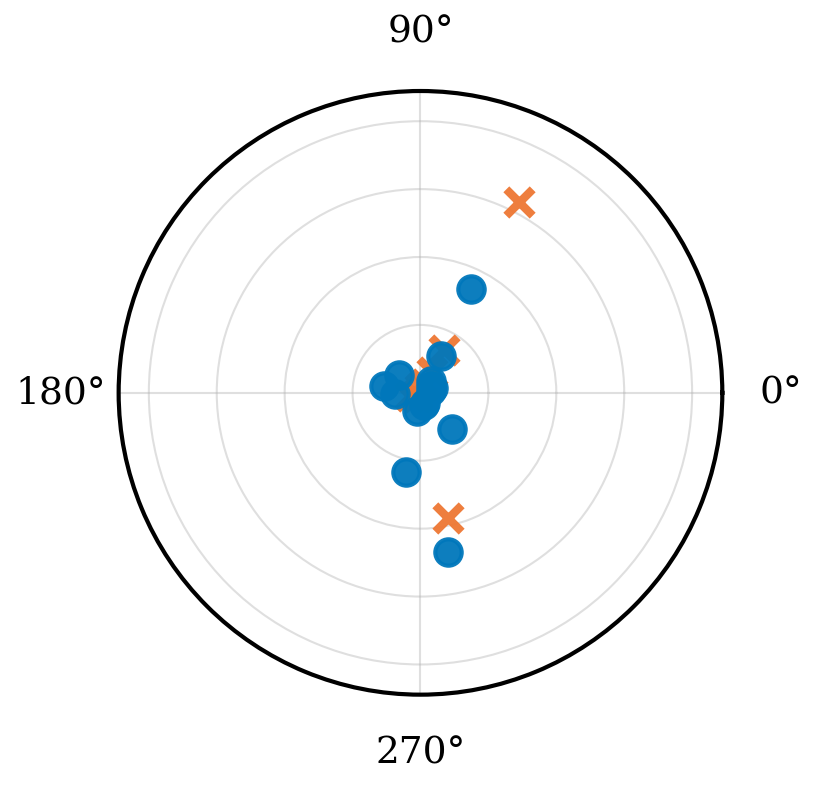} &
  \includegraphics[width=0.14\textwidth]{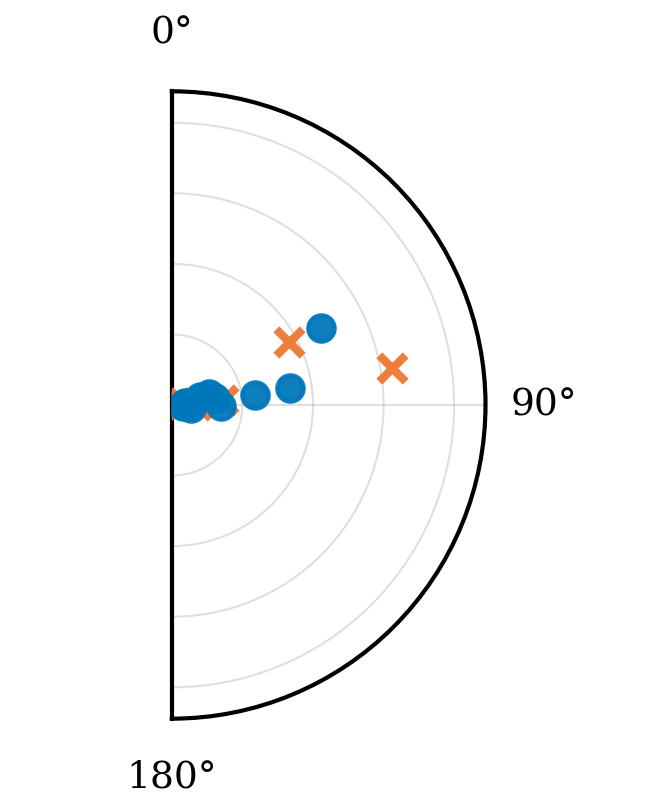} \\
\end{tabular}
\caption{Two additional independent channel realizations generated by \textsc{CityMPC} for the same Austin \gls{tx}-\gls{rx} link shown in Fig.~\ref{fig:cir}.
Each row corresponds to one realization drawn independently from the learned prior $p_\psi(\bmz \mid \bmc)$.
All three realizations, including Fig.~\ref{fig:cir}, differ in their specific path configurations but share the same aggregate channel statistics, including received power, average delay, and mean angles, consistent with the ground truth.}
\label{fig:realizations}
\end{figure*}

Figure~\ref{fig:realizations} shows two independent channel realizations generated by \textsc{CityMPC} for a single \gls{tx}-\gls{rx} link in Austin.
Each row shows one realization: the \gls{cir} $|h(\tau)|$ versus absolute delay $\tau$ (ns), the predicted \gls{aod} azimuth and elevation on a polar plot, and the predicted \gls{aoa} azimuth and elevation on a polar plot.
Ground-truth paths from Sionna \gls{rt} are shown in blue and predicted paths in orange in all panels.
The three realizations are drawn independently from the learned prior $p_\psi(\bmz \mid \bmc)$, producing different path configurations that each remain physically plausible and close to the ground truth.
This stochastic diversity is a key property of the generative model.
\textsc{CityMPC} does not produce a single deterministic channel estimate but rather a distribution over physically consistent realizations conditioned on the scene imagery.

\section{Per-city channel generation results}
\label{app:distributions}

\begin{figure*}[h]
\centering
\scriptsize
\setlength{\tabcolsep}{2pt}
\begin{tabular}{cccc}
\includegraphics[width=0.24\textwidth]{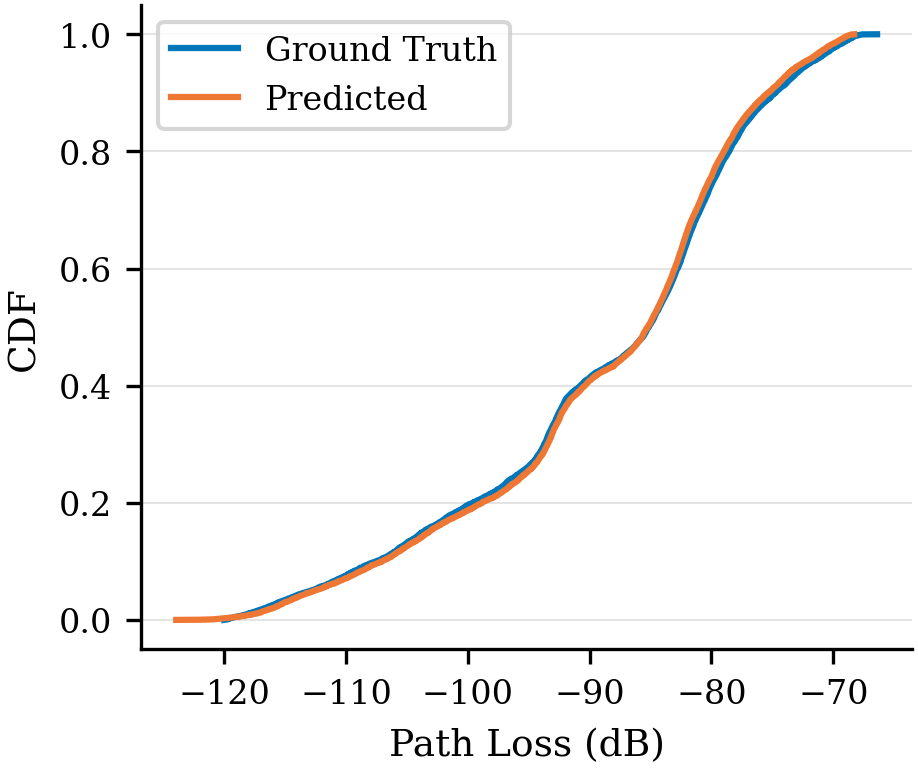} &
\includegraphics[width=0.24\textwidth]{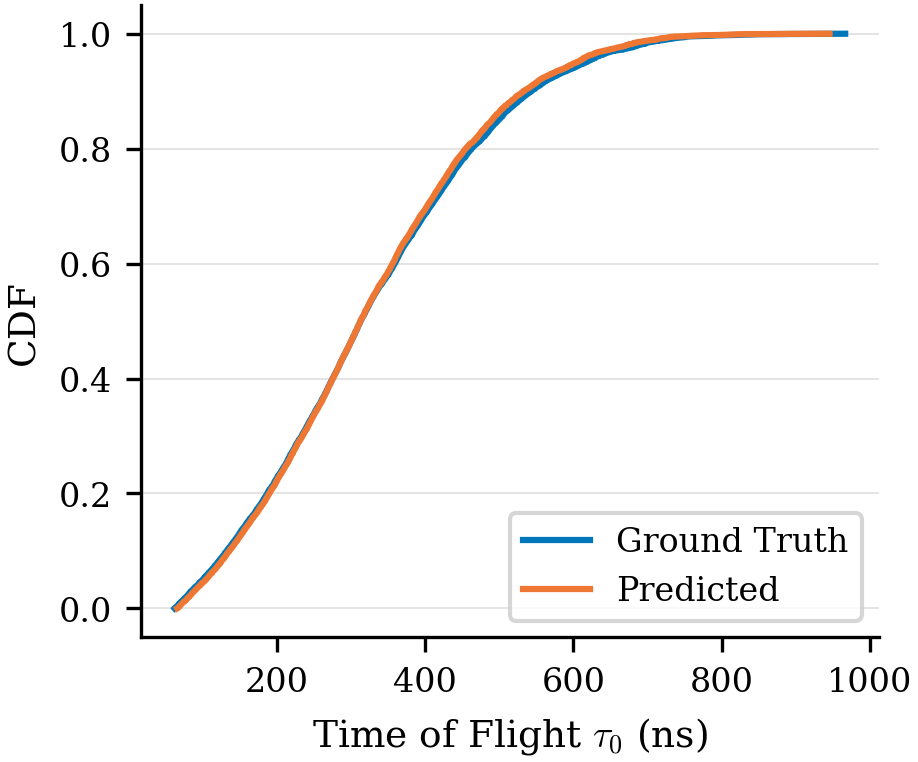} &
\includegraphics[width=0.24\textwidth]{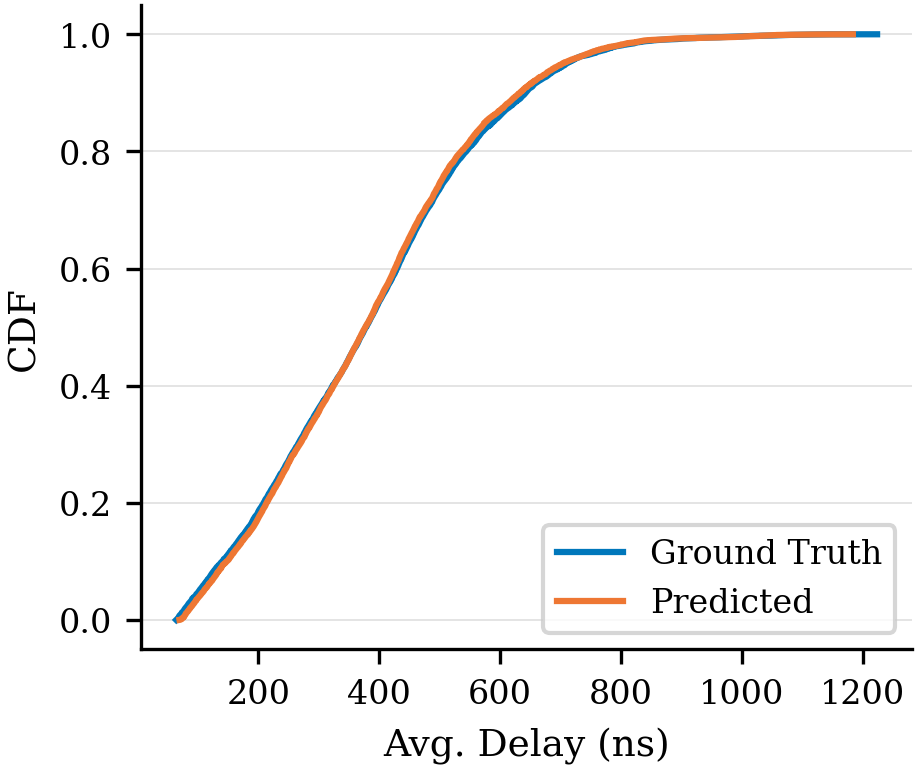} &
\includegraphics[width=0.24\textwidth]{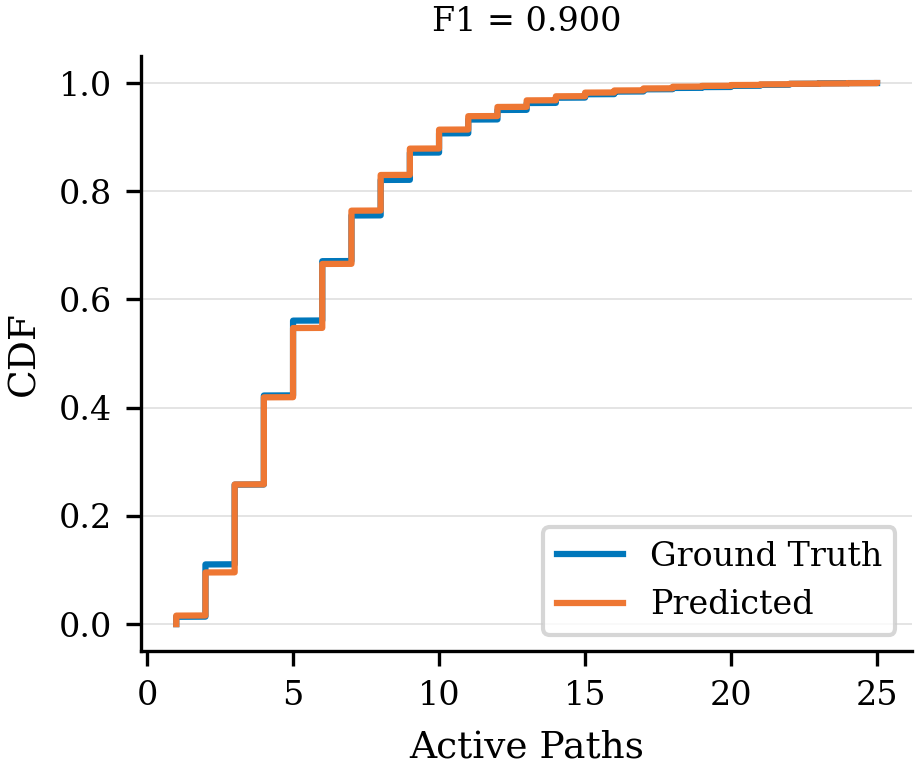} \\
Received Power (dB) & Time of Flight $\tau_0$ (ns) & Avg.\ Delay (ns) & Active Paths \\[4pt]
\includegraphics[width=0.24\textwidth]{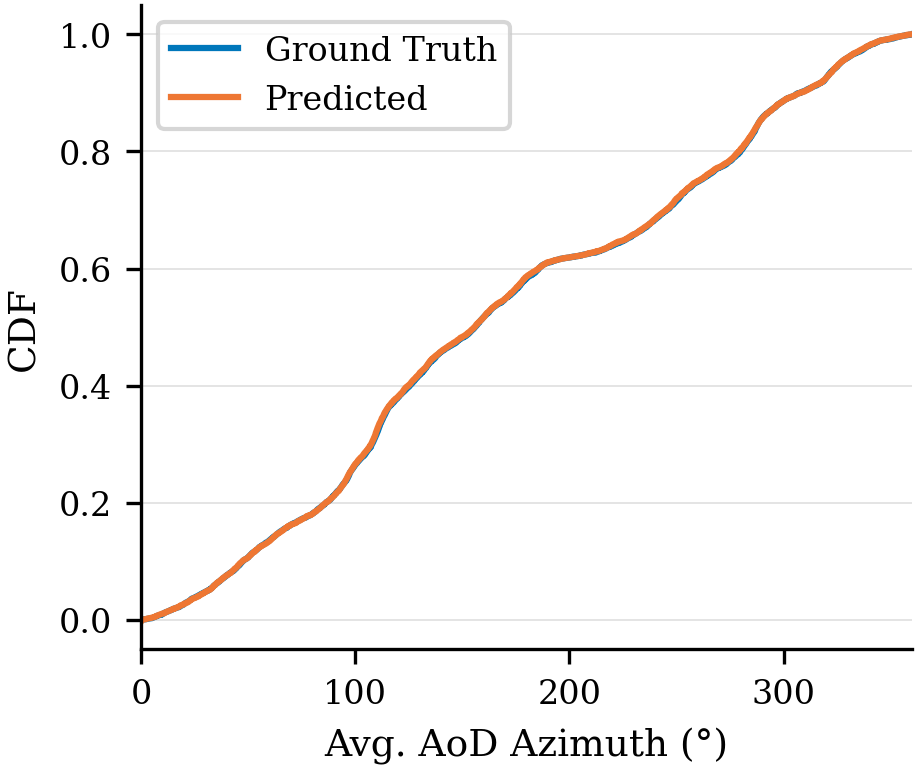} &
\includegraphics[width=0.24\textwidth]{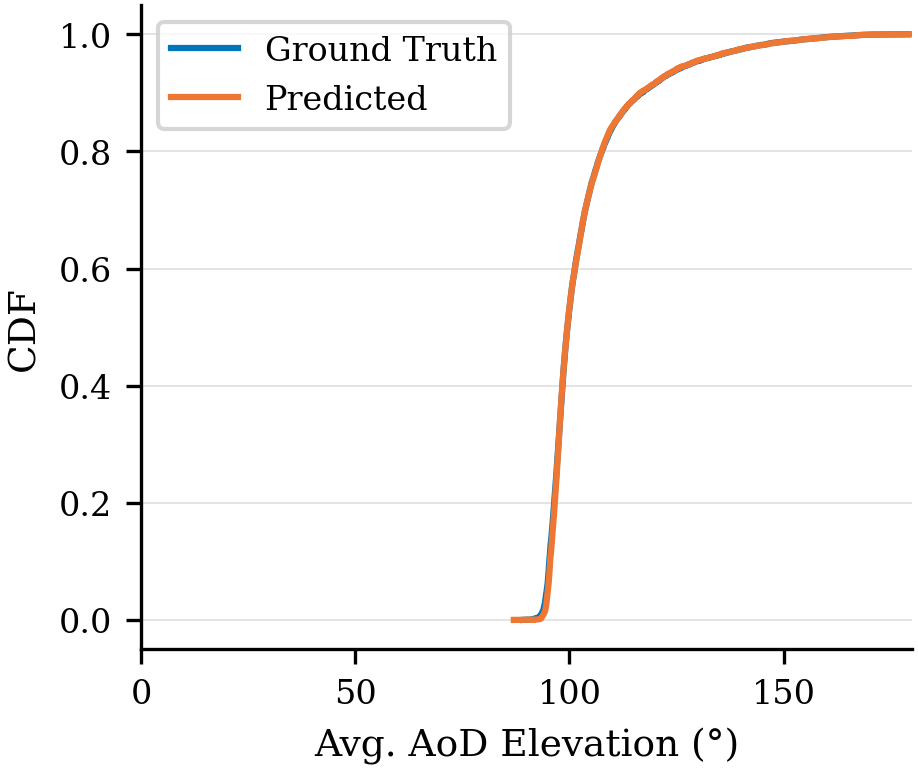} &
\includegraphics[width=0.24\textwidth]{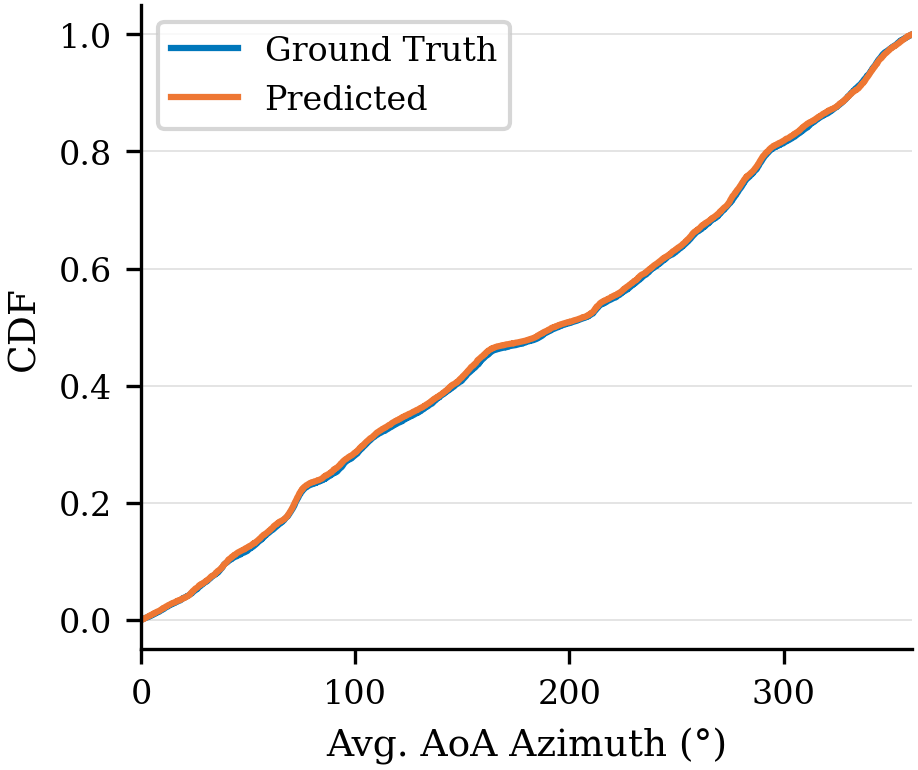} &
\includegraphics[width=0.24\textwidth]{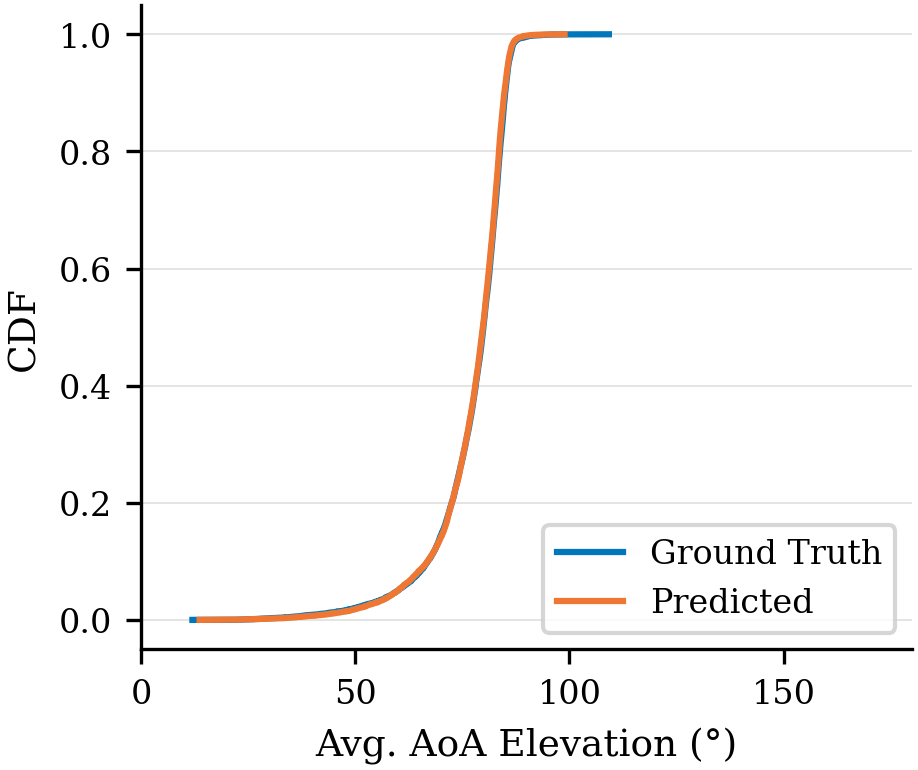} \\
Avg.\ \gls{aod} Azimuth ($^\circ$) & Avg.\ \gls{aod} Elevation ($^\circ$) & Avg.\ \gls{aoa} Azimuth ($^\circ$) & Avg.\ \gls{aoa} Elevation ($^\circ$) \\
\end{tabular}
\caption{Empirical \glspl{cdf} of ground-truth (blue) and generated (orange) channel parameters for Austin.}
\label{fig:cdf_austin}
\end{figure*}

\begin{figure*}[h]
\centering
\scriptsize
\setlength{\tabcolsep}{2pt}
\begin{tabular}{cccc}
\includegraphics[width=0.24\textwidth]{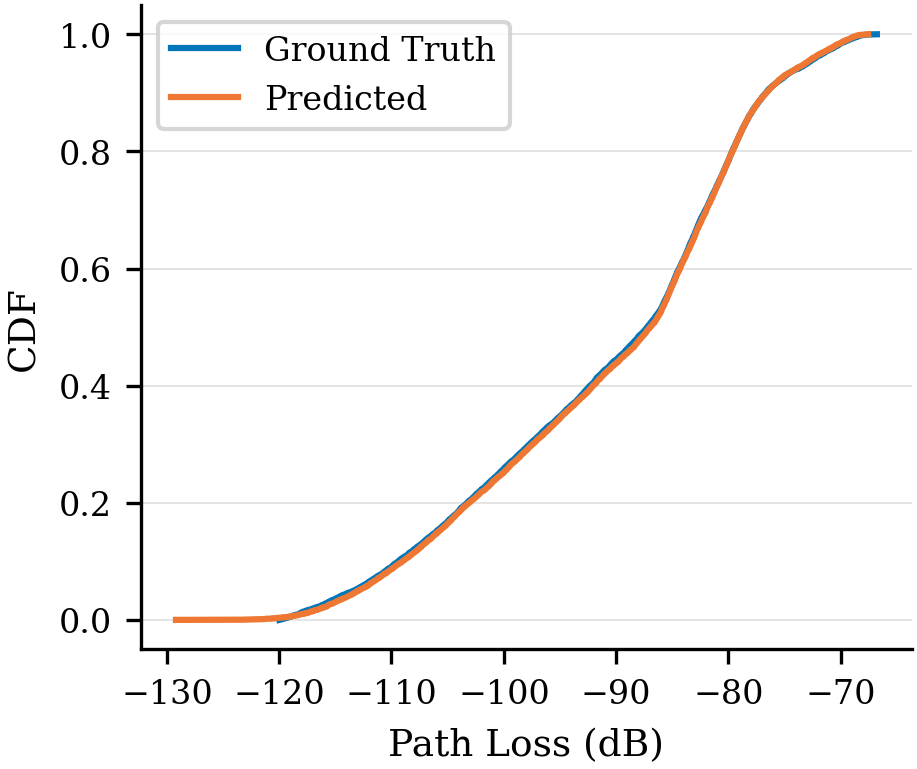} &
\includegraphics[width=0.24\textwidth]{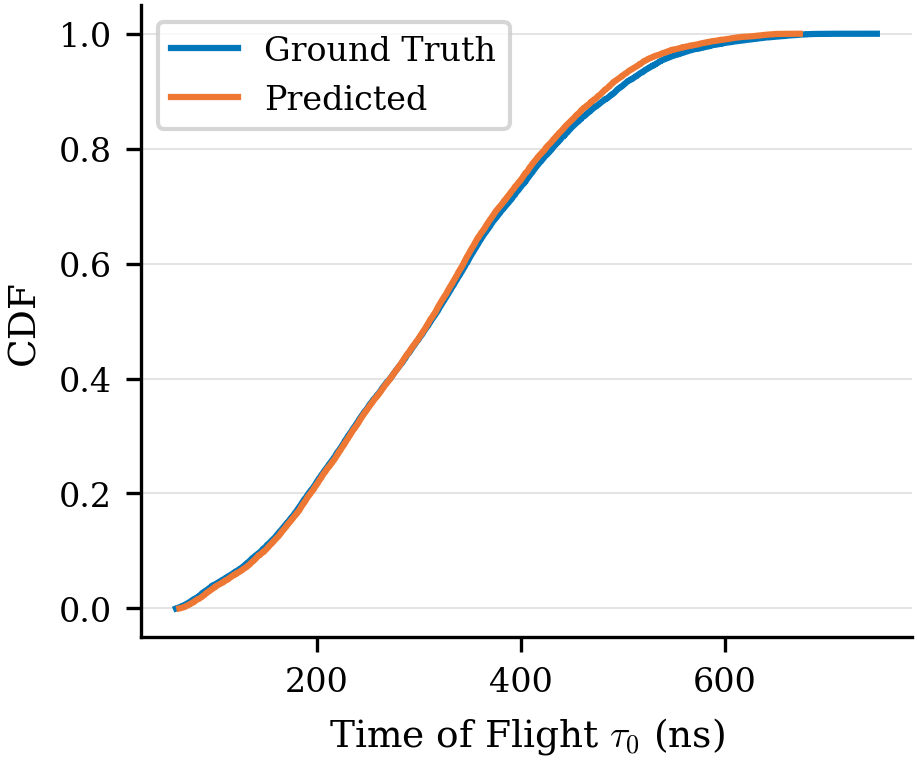} &
\includegraphics[width=0.24\textwidth]{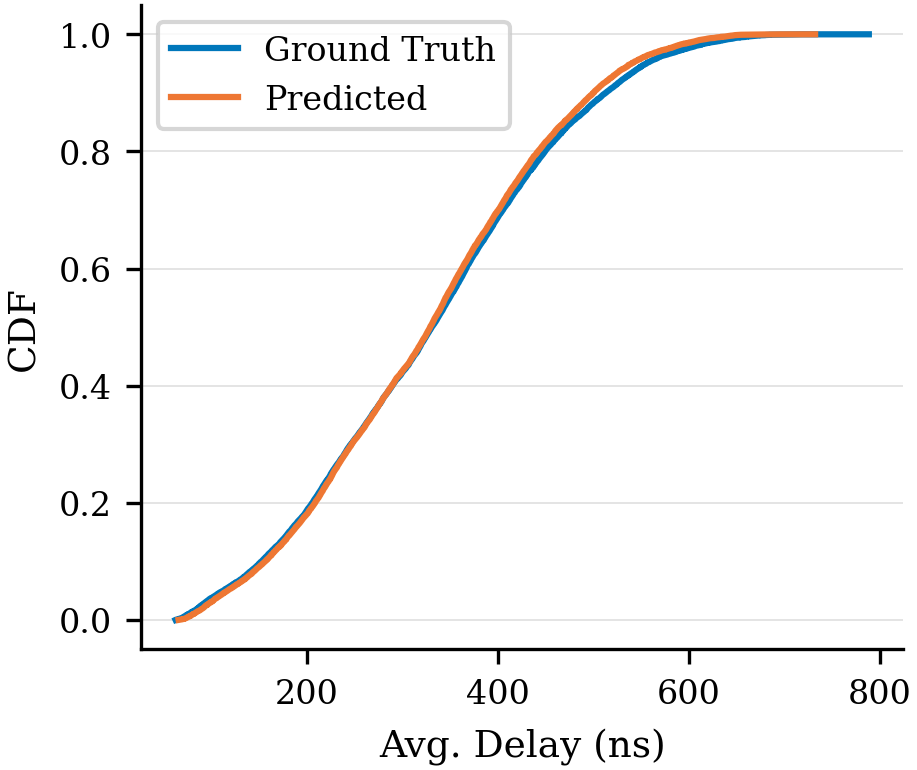} &
\includegraphics[width=0.24\textwidth]{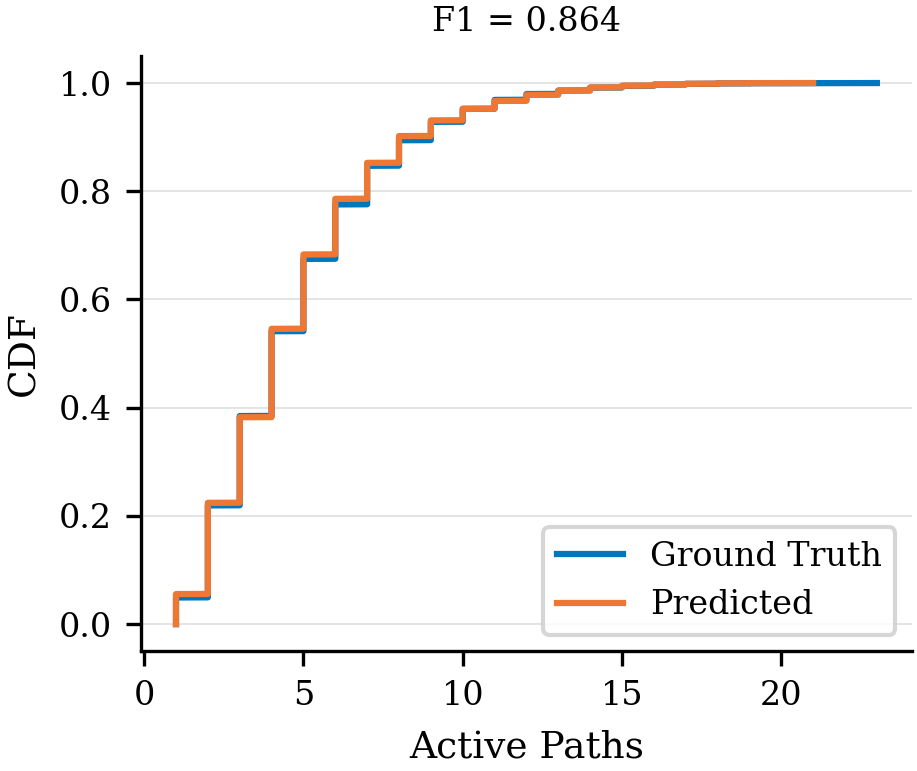} \\
Received Power (dB) & Time of Flight $\tau_0$ (ns) & Avg.\ Delay (ns) & Active Paths \\[4pt]
\includegraphics[width=0.24\textwidth]{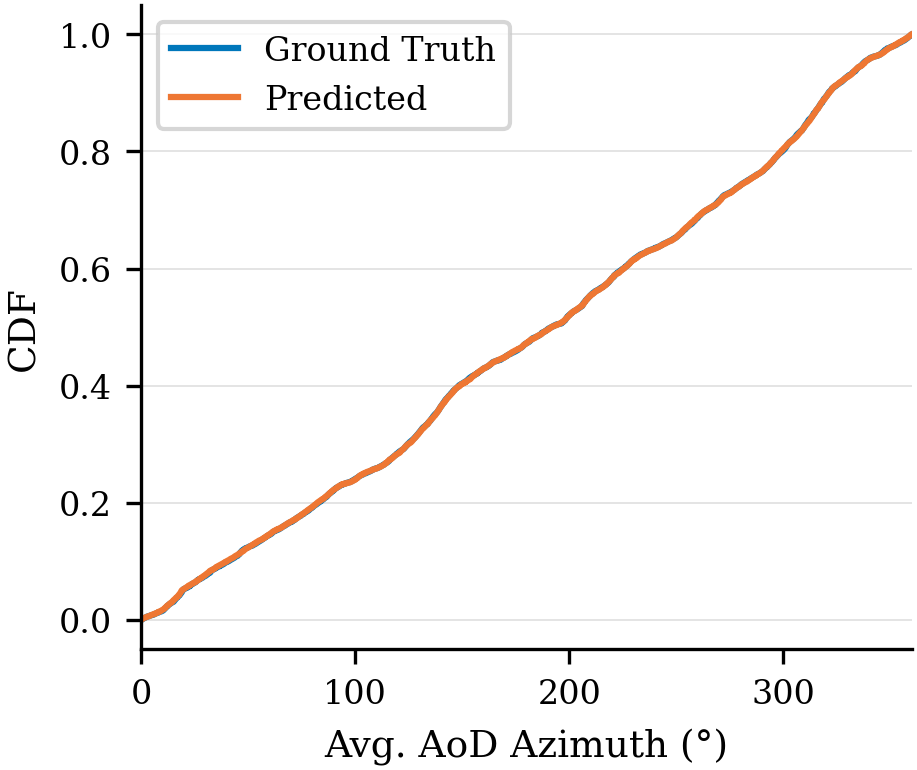} &
\includegraphics[width=0.24\textwidth]{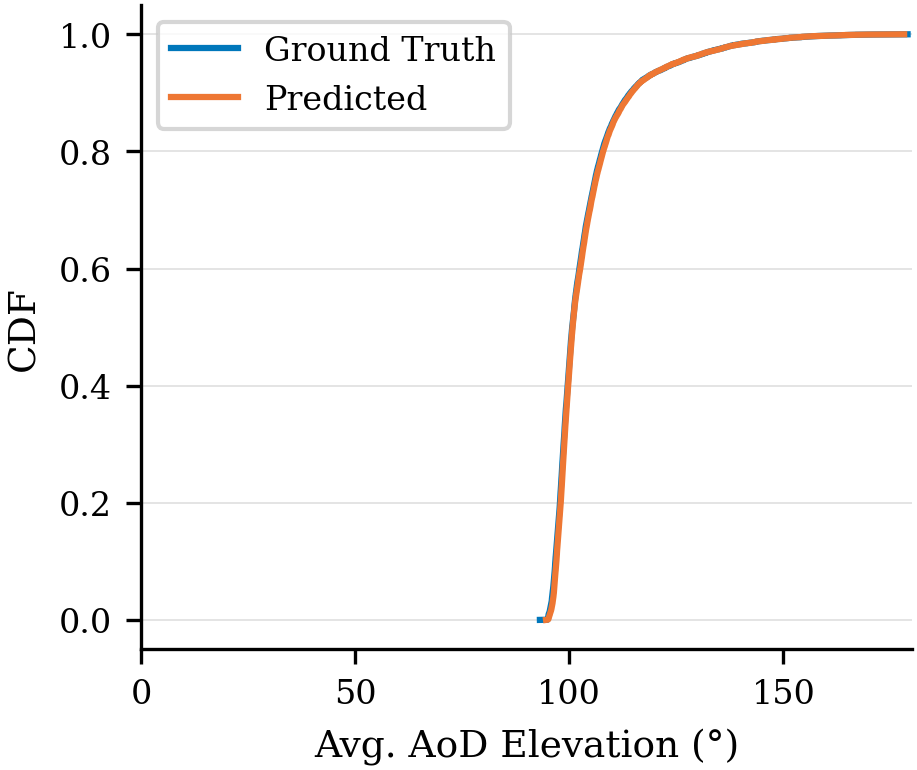} &
\includegraphics[width=0.24\textwidth]{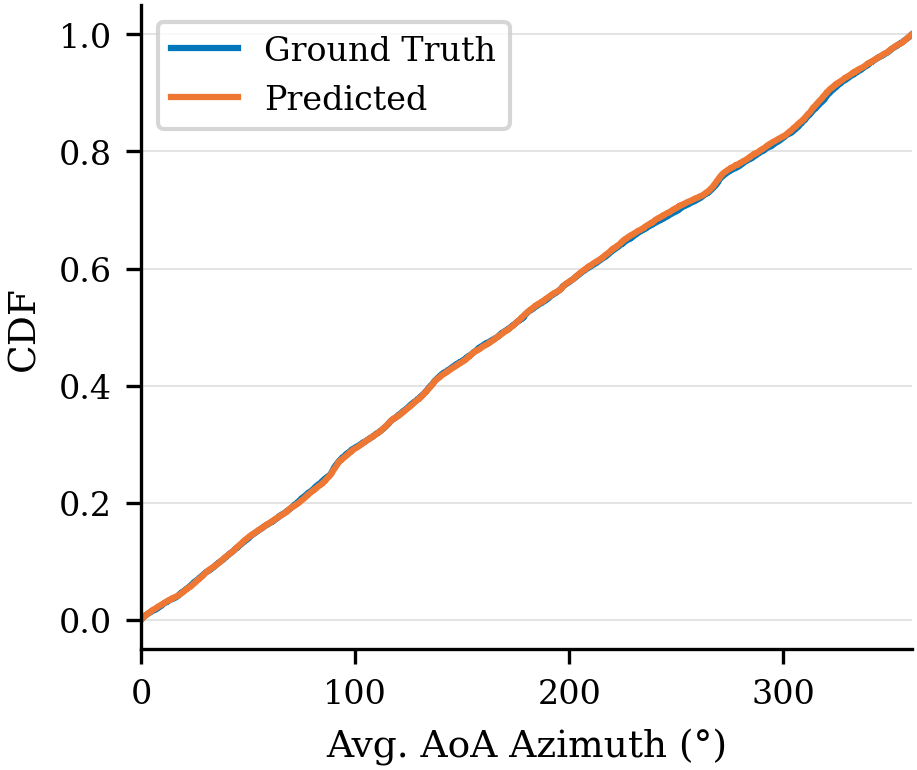} &
\includegraphics[width=0.24\textwidth]{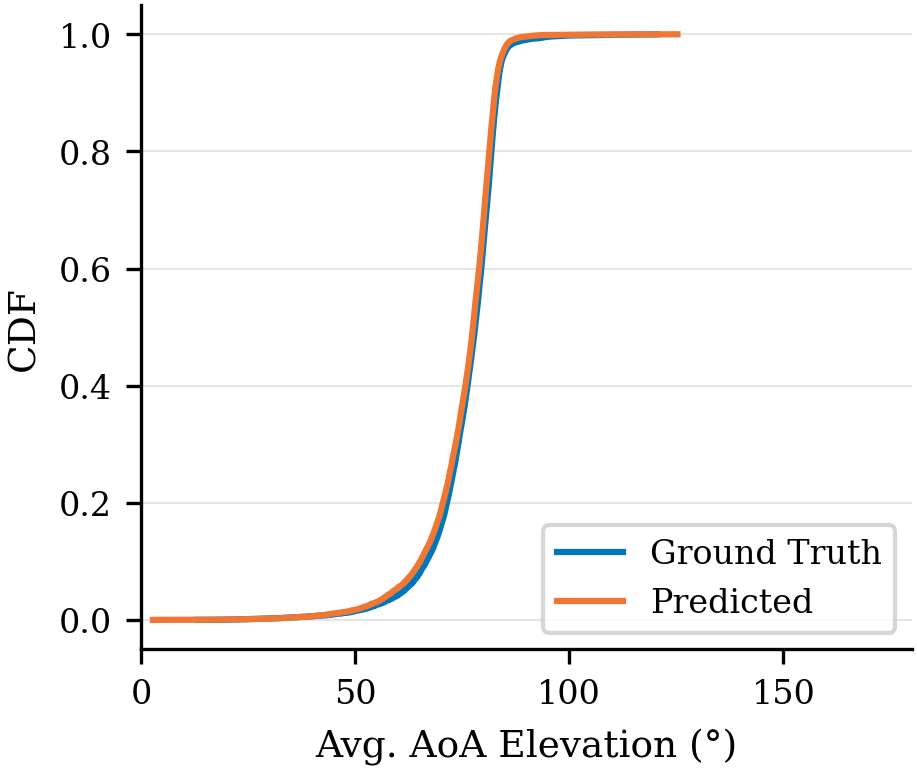} \\
Avg.\ \gls{aod} Azimuth ($^\circ$) & Avg.\ \gls{aod} Elevation ($^\circ$) & Avg.\ \gls{aoa} Azimuth ($^\circ$) & Avg.\ \gls{aoa} Elevation ($^\circ$) \\
\end{tabular}
\caption{Empirical \glspl{cdf} of ground-truth (blue) and generated (orange) channel parameters for Denver.}
\label{fig:cdf_denver}
\end{figure*}

\begin{figure*}[h]
\centering
\scriptsize
\setlength{\tabcolsep}{2pt}
\begin{tabular}{cccc}
\includegraphics[width=0.24\textwidth]{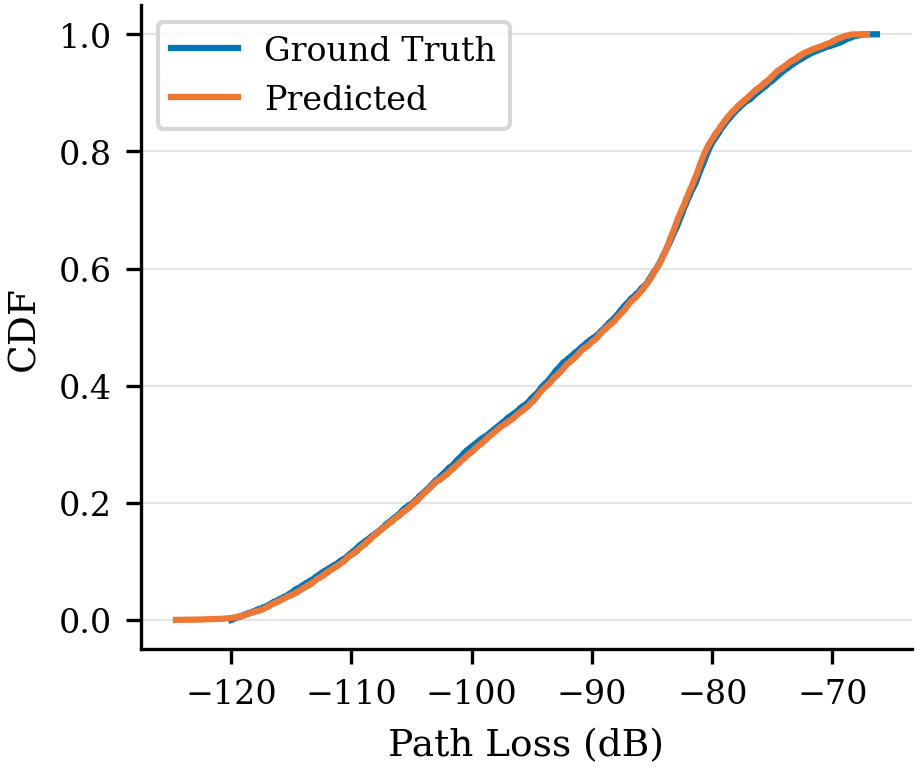} &
\includegraphics[width=0.24\textwidth]{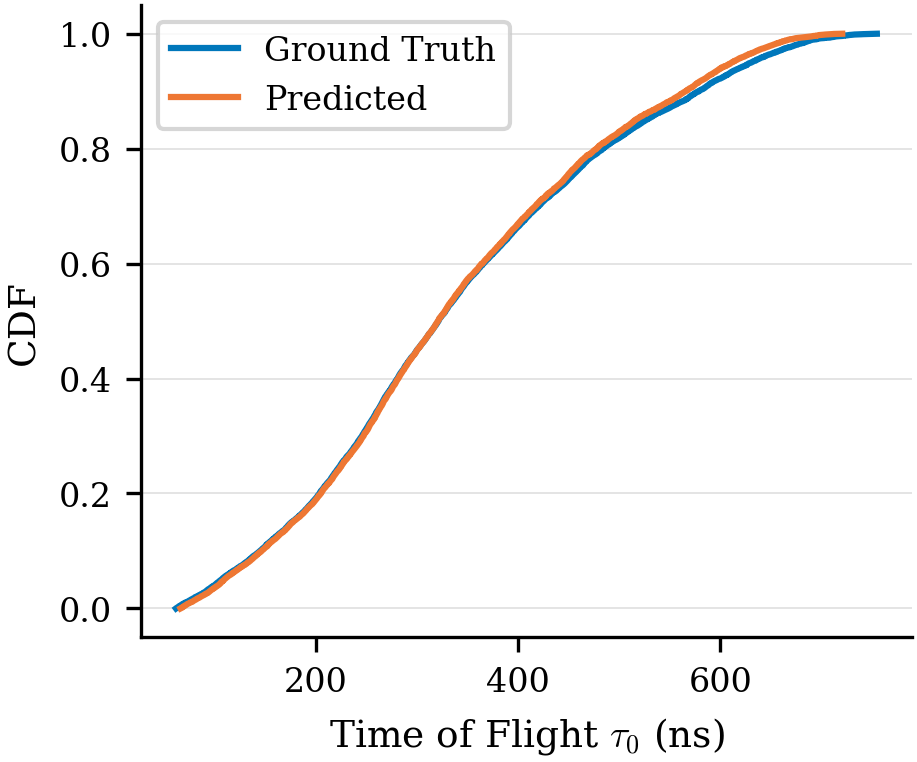} &
\includegraphics[width=0.24\textwidth]{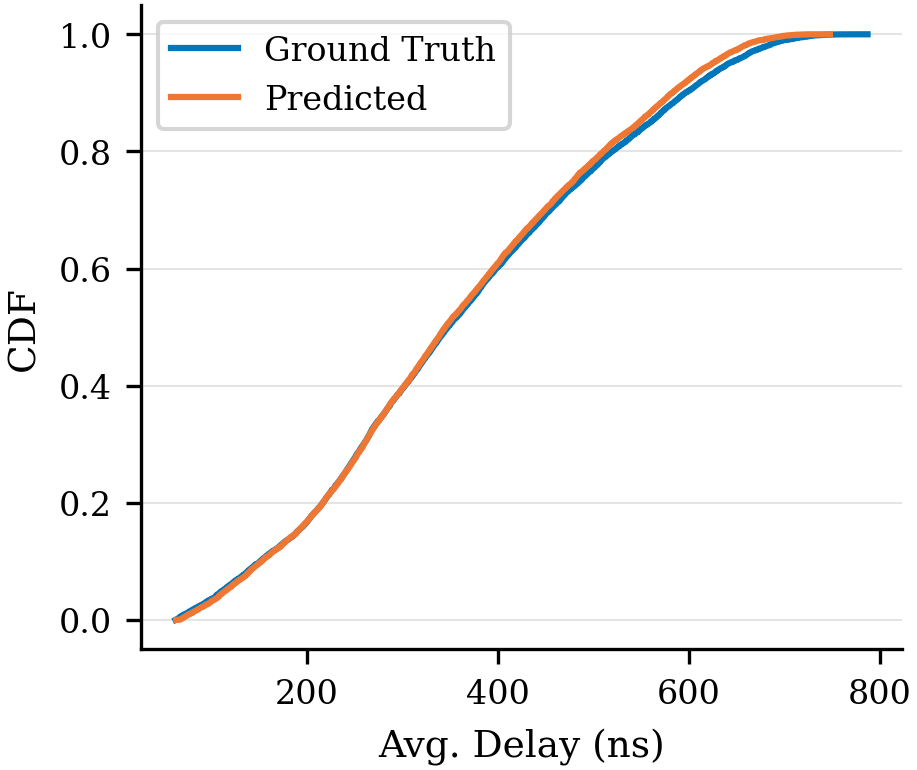} &
\includegraphics[width=0.24\textwidth]{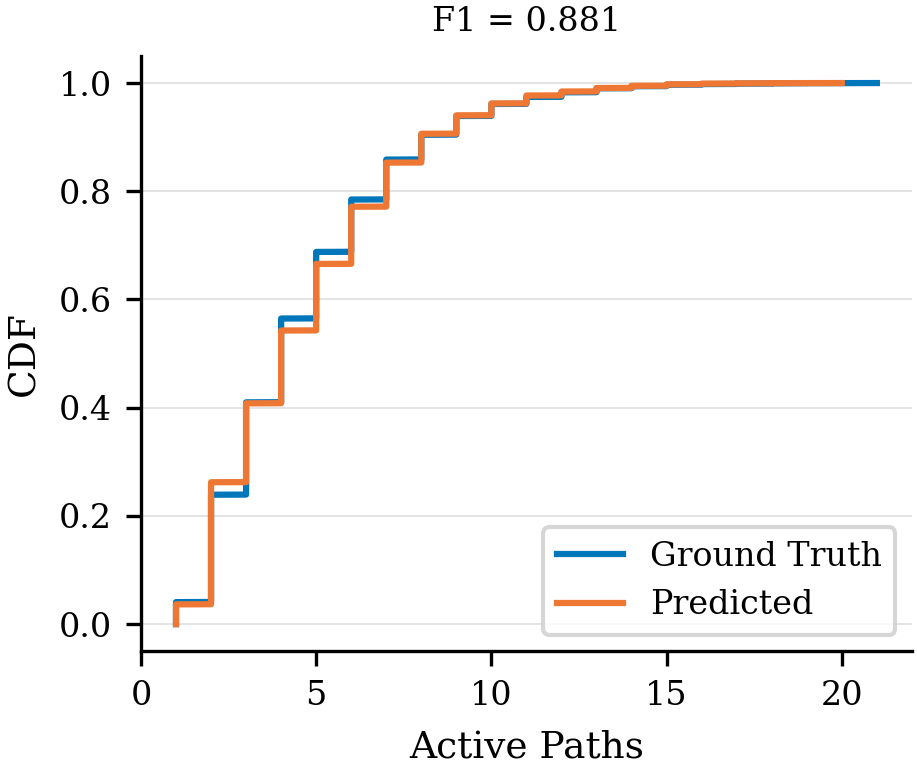} \\
Received Power (dB) & Time of Flight $\tau_0$ (ns) & Avg.\ Delay (ns) & Active Paths \\[4pt]
\includegraphics[width=0.24\textwidth]{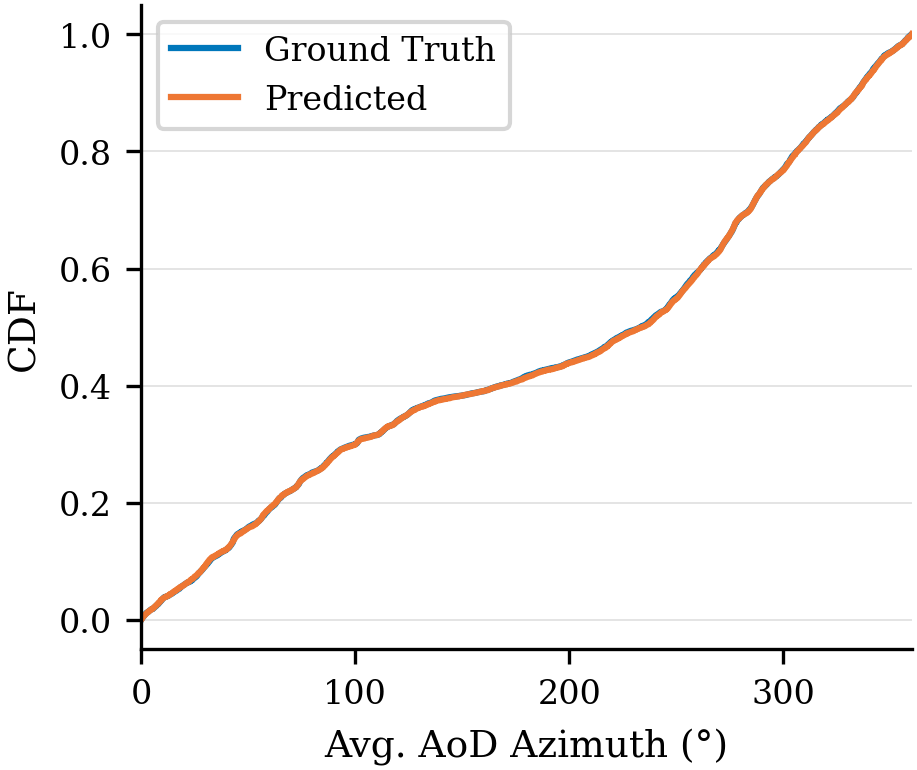} &
\includegraphics[width=0.24\textwidth]{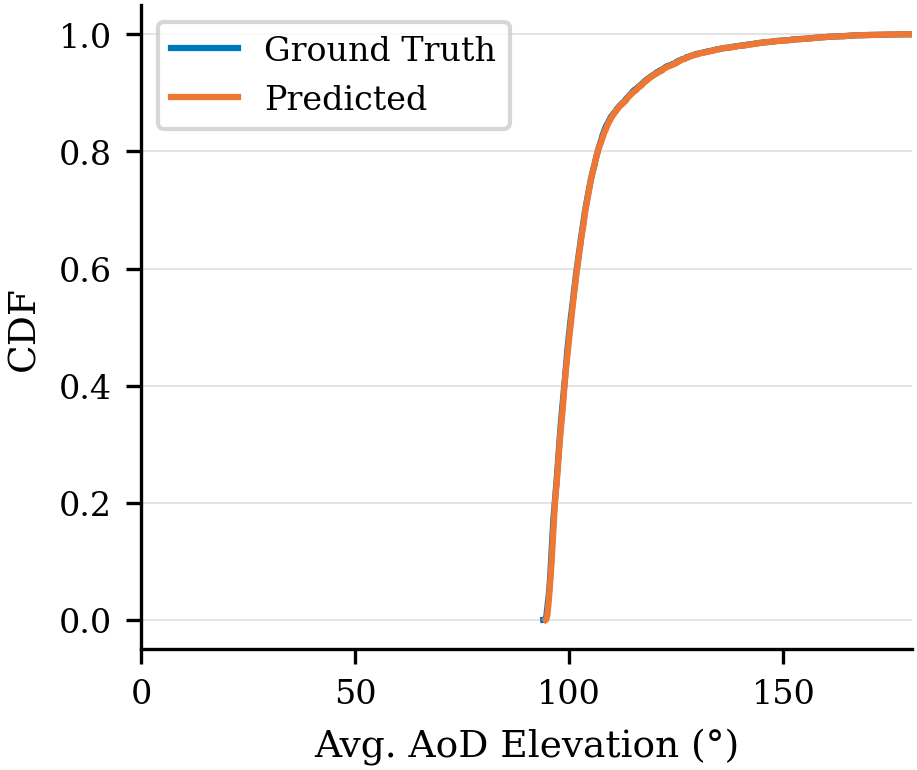} &
\includegraphics[width=0.24\textwidth]{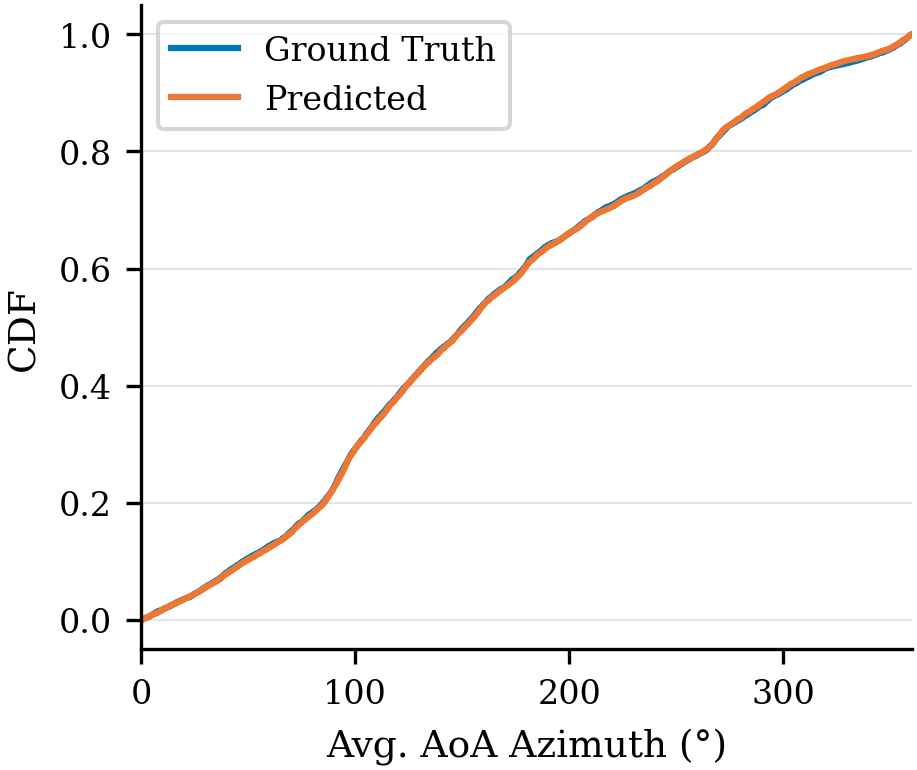} &
\includegraphics[width=0.24\textwidth]{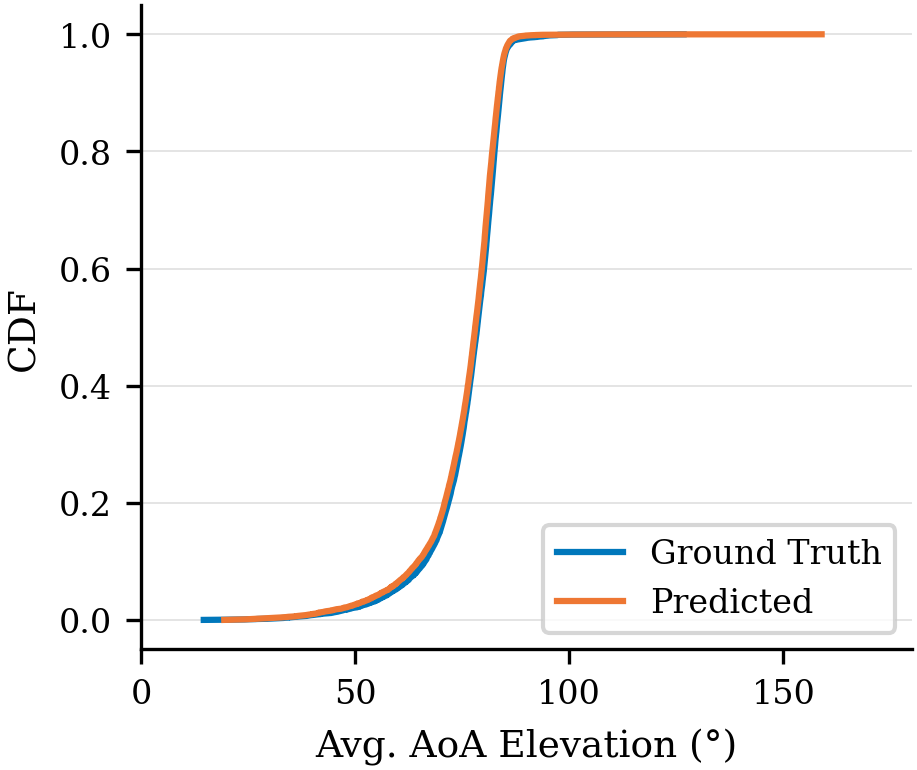} \\
Avg.\ \gls{aod} Azimuth ($^\circ$) & Avg.\ \gls{aod} Elevation ($^\circ$) & Avg.\ \gls{aoa} Azimuth ($^\circ$) & Avg.\ \gls{aoa} Elevation ($^\circ$) \\
\end{tabular}
\caption{Empirical \glspl{cdf} of ground-truth (blue) and generated (orange) channel parameters for Fort Worth.}
\label{fig:cdf_fortworth}
\end{figure*}

\begin{figure*}[h]
\centering
\scriptsize
\setlength{\tabcolsep}{2pt}
\begin{tabular}{cccc}
\includegraphics[width=0.24\textwidth]{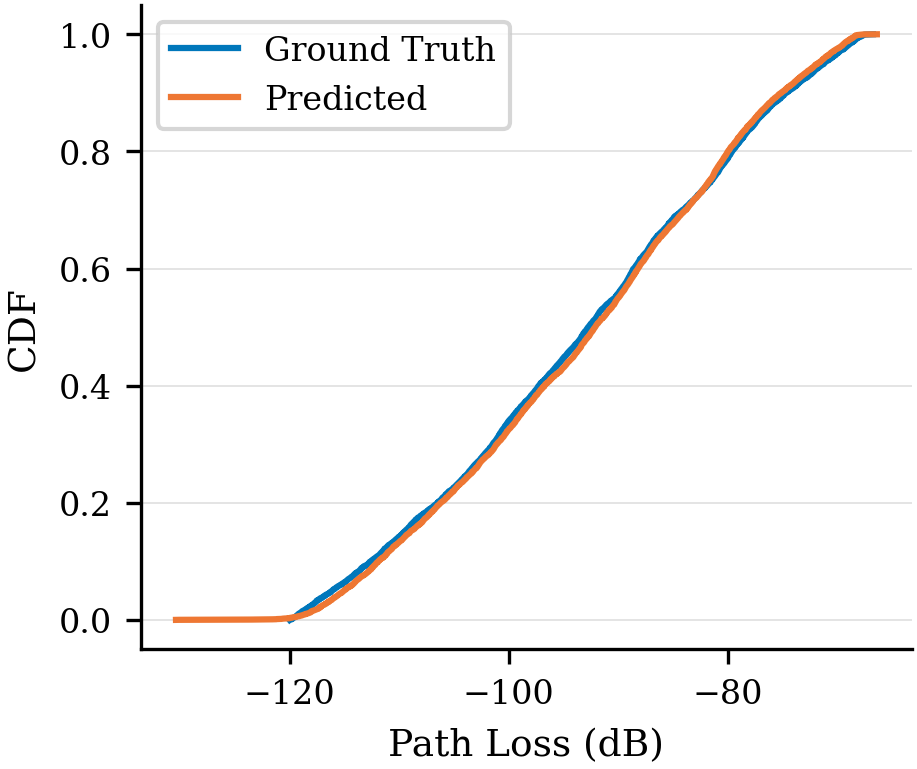} &
\includegraphics[width=0.24\textwidth]{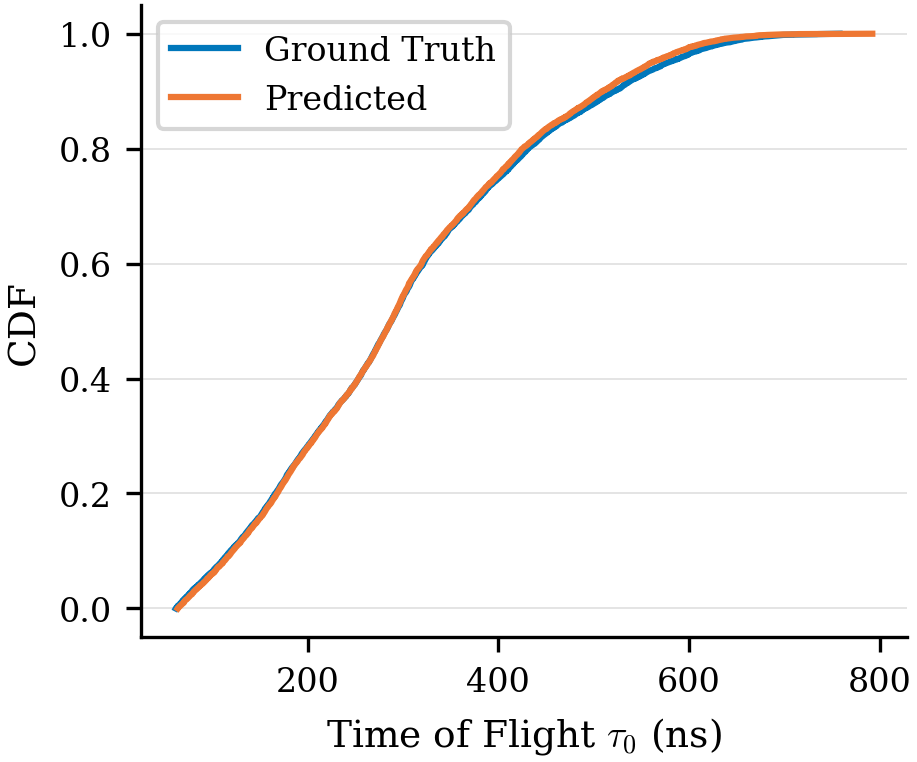} &
\includegraphics[width=0.24\textwidth]{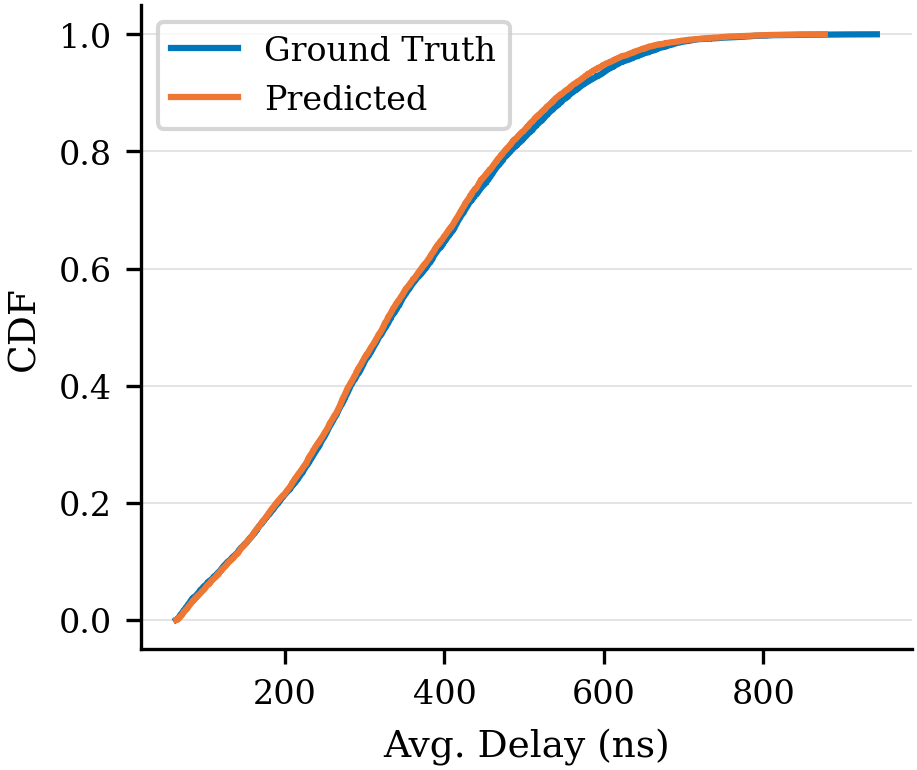} &
\includegraphics[width=0.24\textwidth]{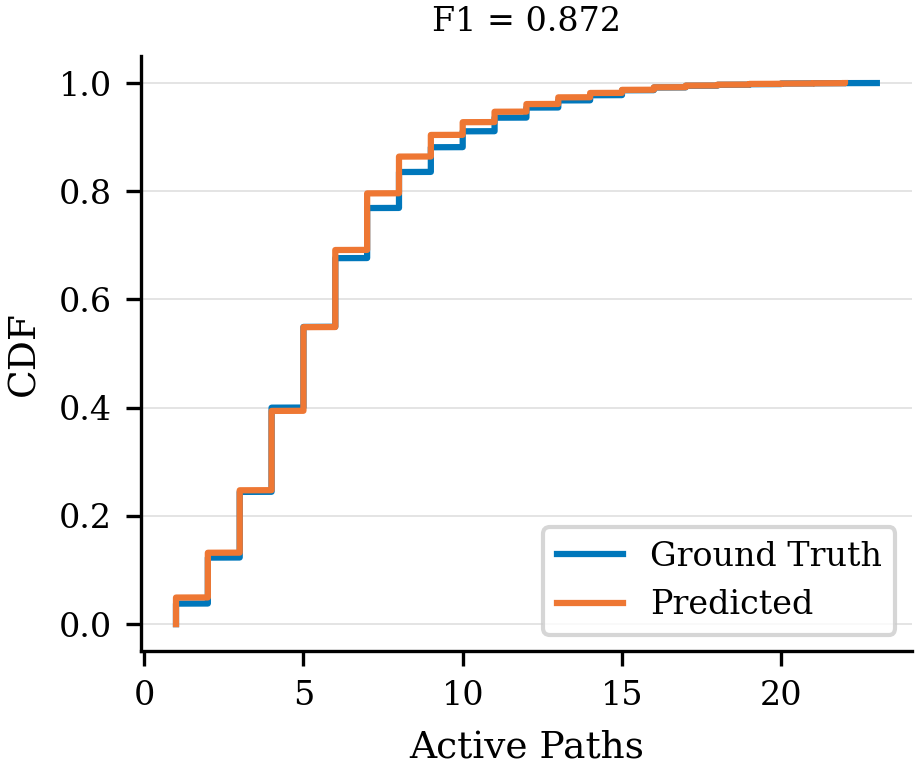} \\
Received Power (dB) & Time of Flight $\tau_0$ (ns) & Avg.\ Delay (ns) & Active Paths \\[4pt]
\includegraphics[width=0.24\textwidth]{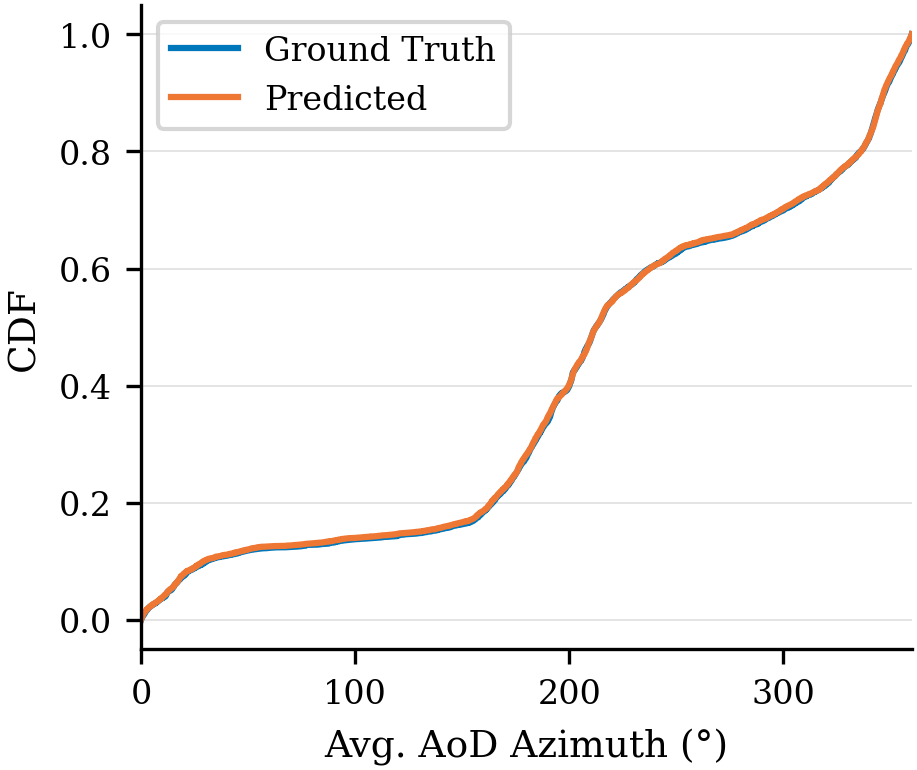} &
\includegraphics[width=0.24\textwidth]{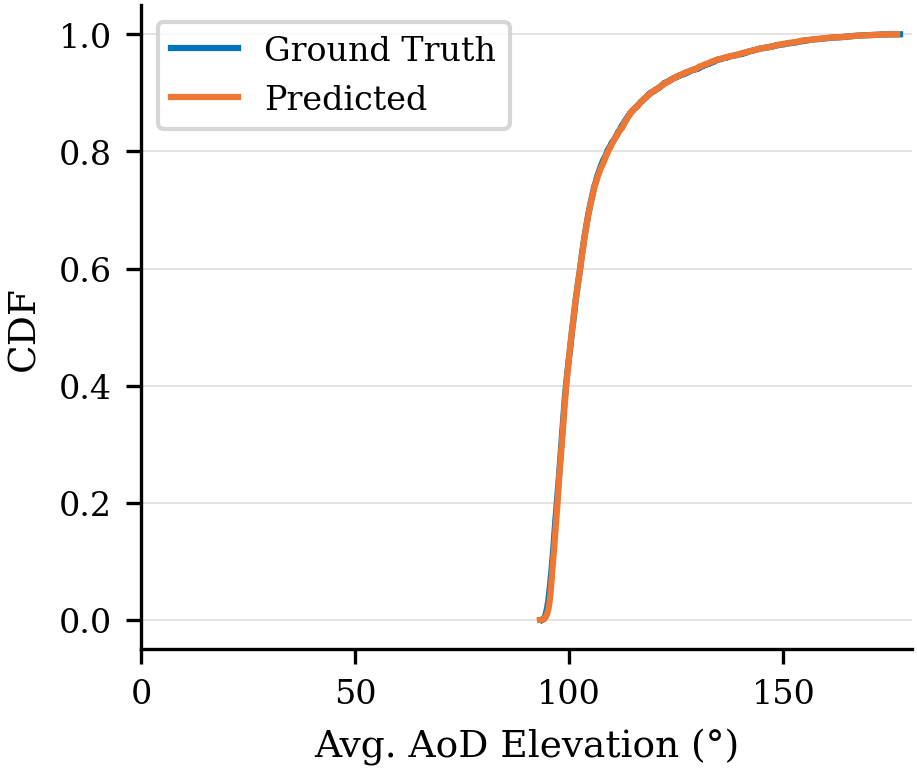} &
\includegraphics[width=0.24\textwidth]{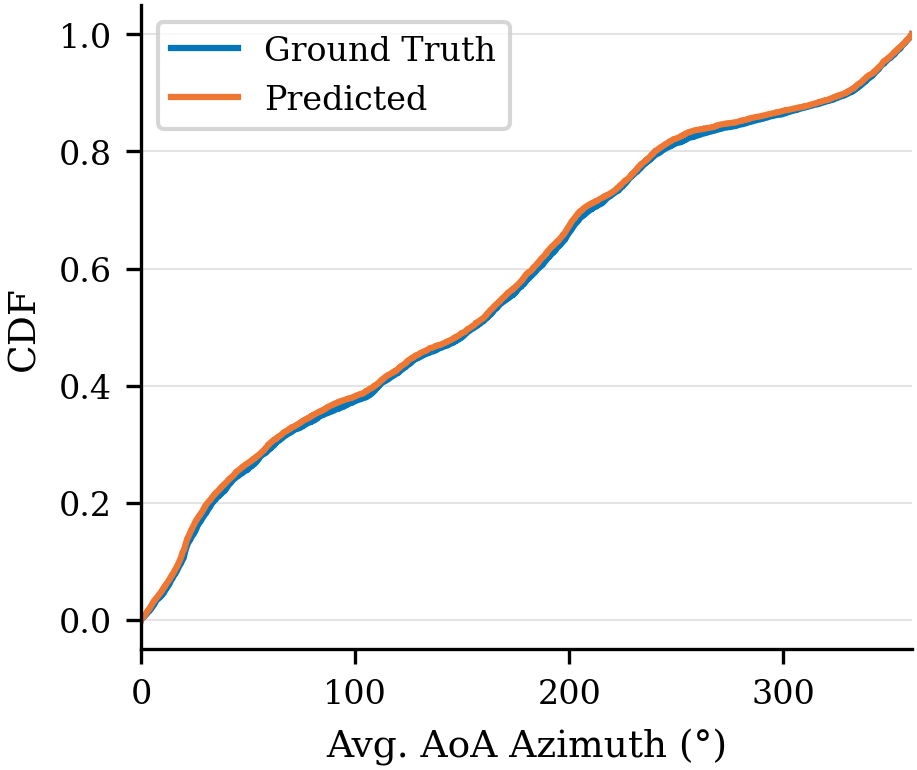} &
\includegraphics[width=0.24\textwidth]{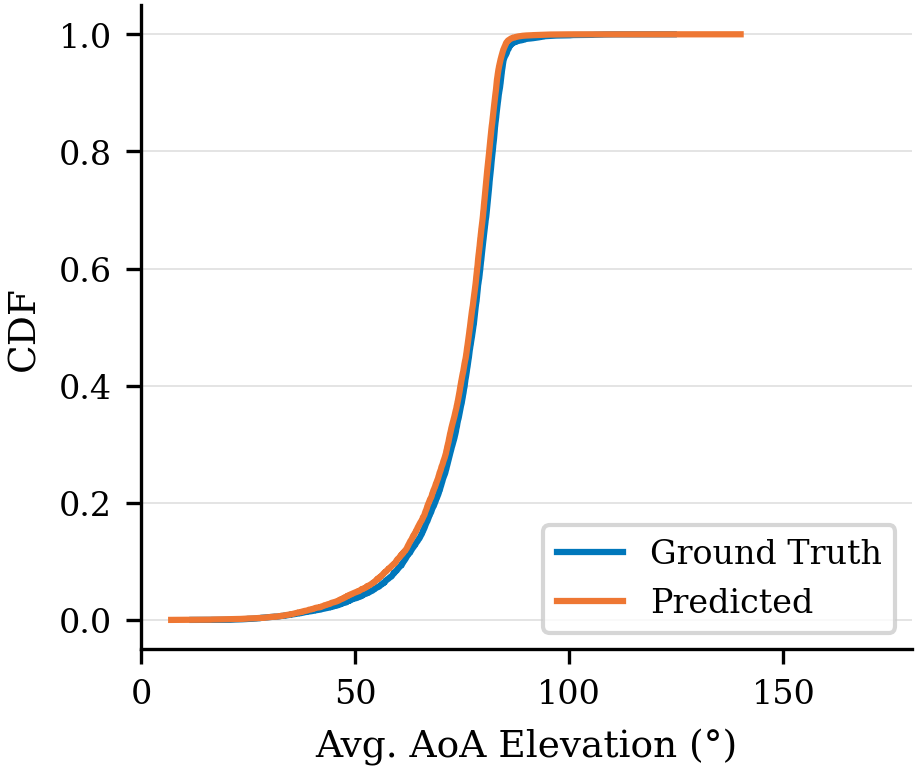} \\
Avg.\ \gls{aod} Azimuth ($^\circ$) & Avg.\ \gls{aod} Elevation ($^\circ$) & Avg.\ \gls{aoa} Azimuth ($^\circ$) & Avg.\ \gls{aoa} Elevation ($^\circ$) \\
\end{tabular}
\caption{Empirical \glspl{cdf} of ground-truth (blue) and generated (orange) channel parameters for New York.}
\label{fig:cdf_newyork}
\end{figure*}

Figures~\ref{fig:cdf_austin}--\ref{fig:cdf_newyork} show the empirical \glspl{cdf} of ground-truth and generated channel parameters for all five cities evaluated in this work.
In each figure, the top row shows scalar and delay parameters and the bottom row shows power-weighted mean angular parameters.
The near-perfect overlap between ground-truth and predicted distributions across all cities demonstrates that \textsc{CityMPC} consistently captures the full marginal channel statistics across diverse urban environments.

\section{Cross-city transfer matrices}
\label{app:crosscity_extra}
\begin{figure*}[h]
\centering
\scriptsize
\setlength{\tabcolsep}{2pt}
\begin{tabular}{ccc}
\includegraphics[width=0.32\textwidth]{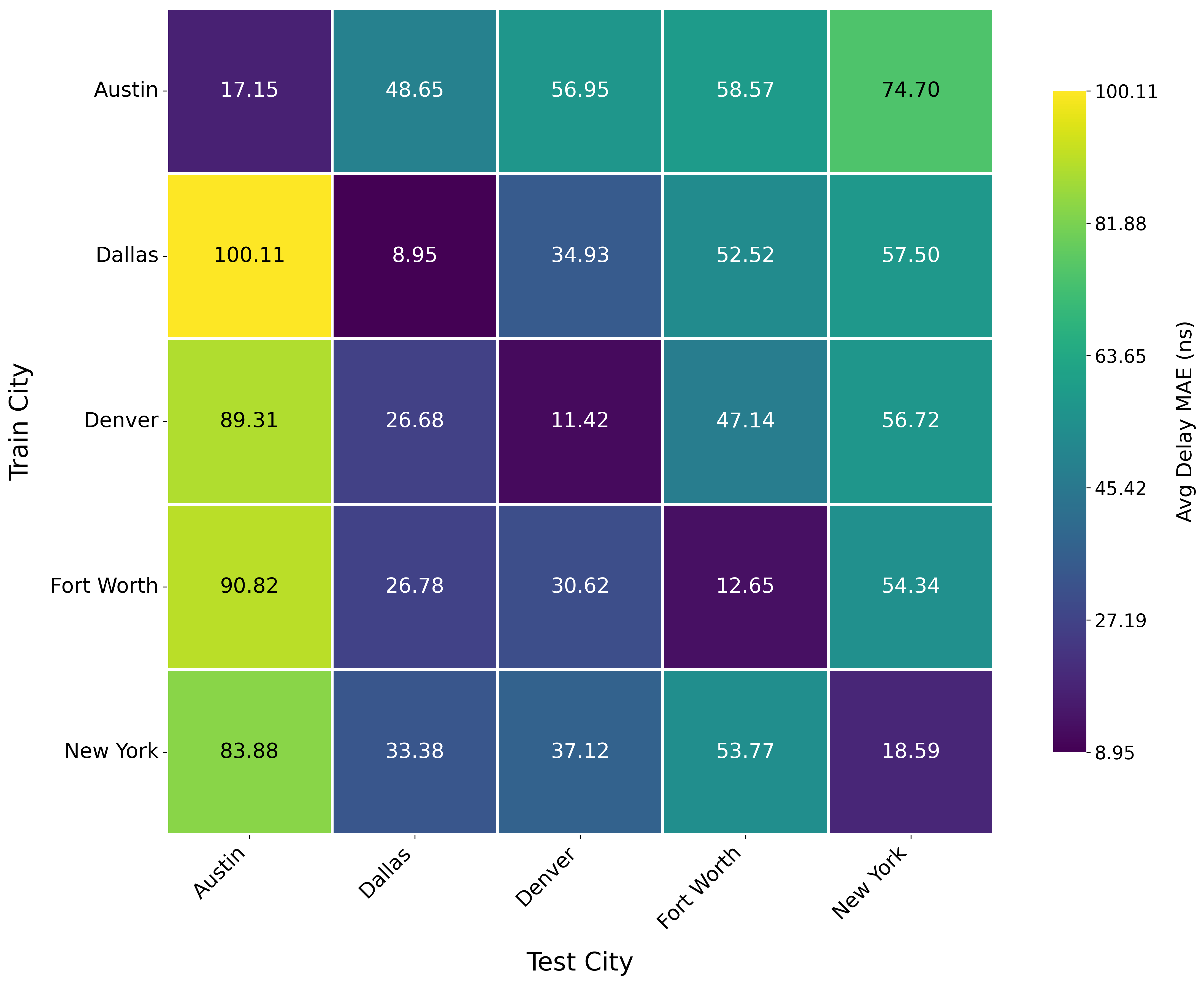} &
\includegraphics[width=0.32\textwidth]{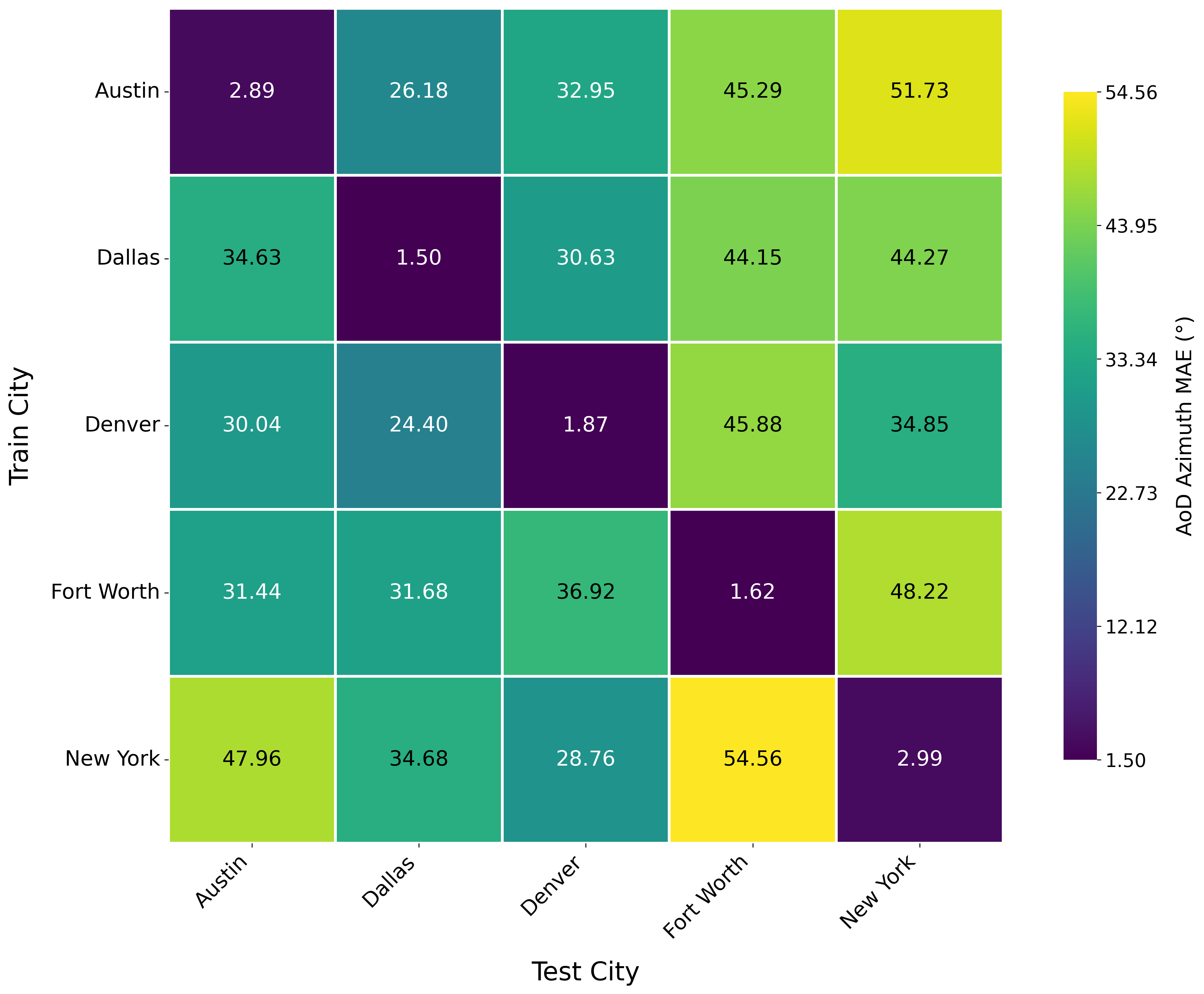} &
\includegraphics[width=0.32\textwidth]{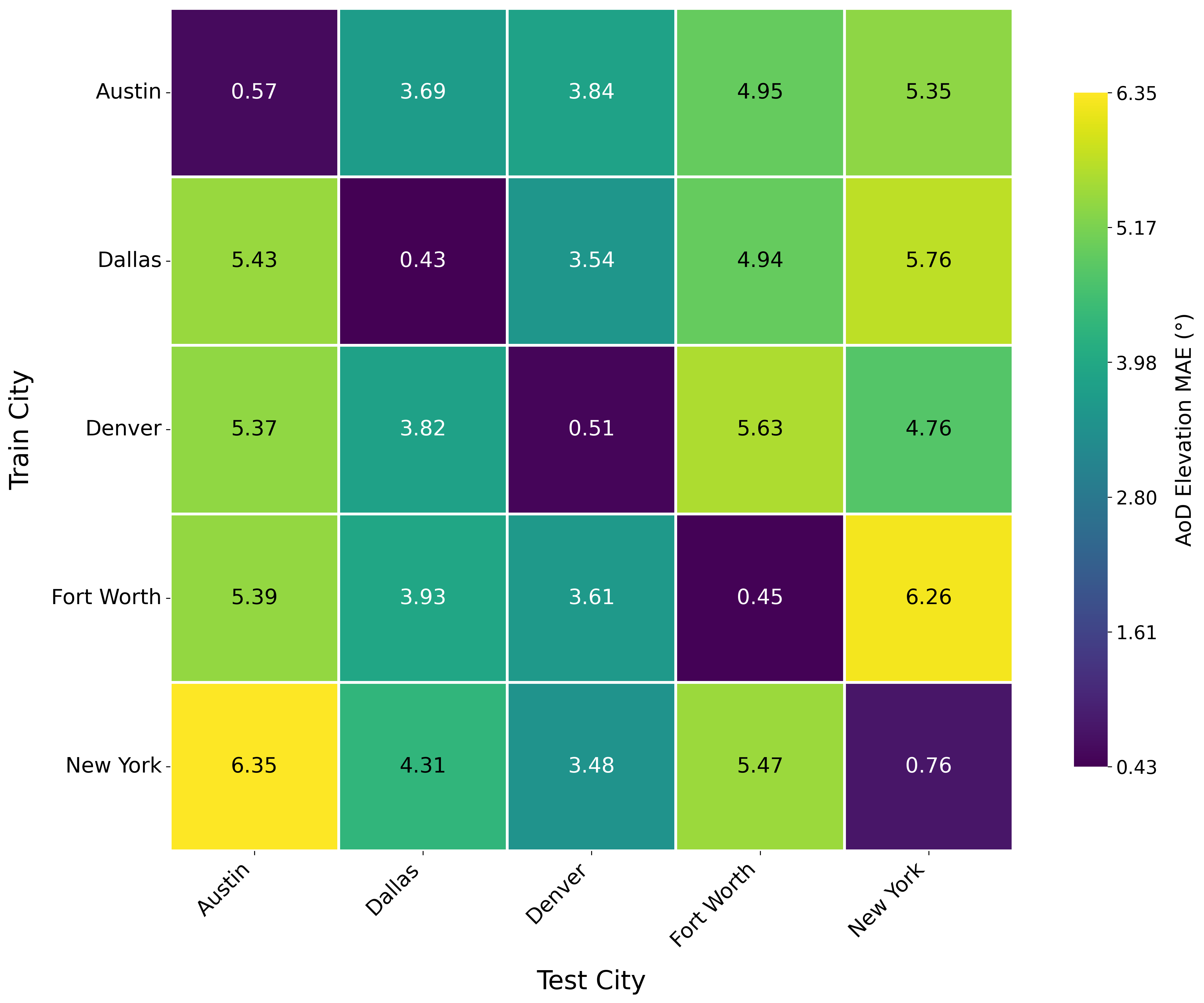} \\
Avg.\ Delay \gls{mae} (ns) & Avg.\ \gls{aod} Azimuth \gls{mae} ($^\circ$) & Avg.\ \gls{aod} Elevation \gls{mae} ($^\circ$) \\[4pt]
\includegraphics[width=0.32\textwidth]{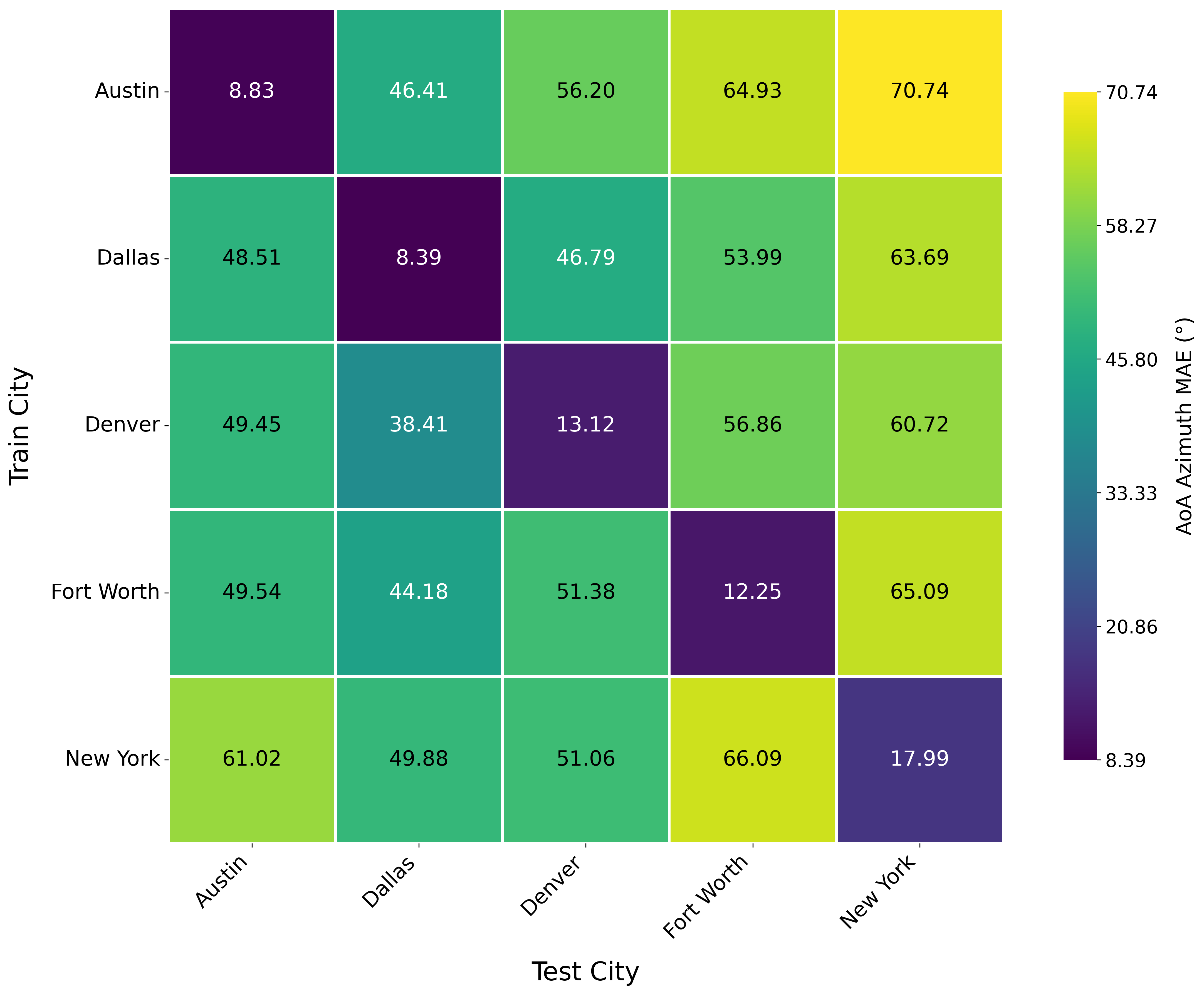} &
\includegraphics[width=0.32\textwidth]{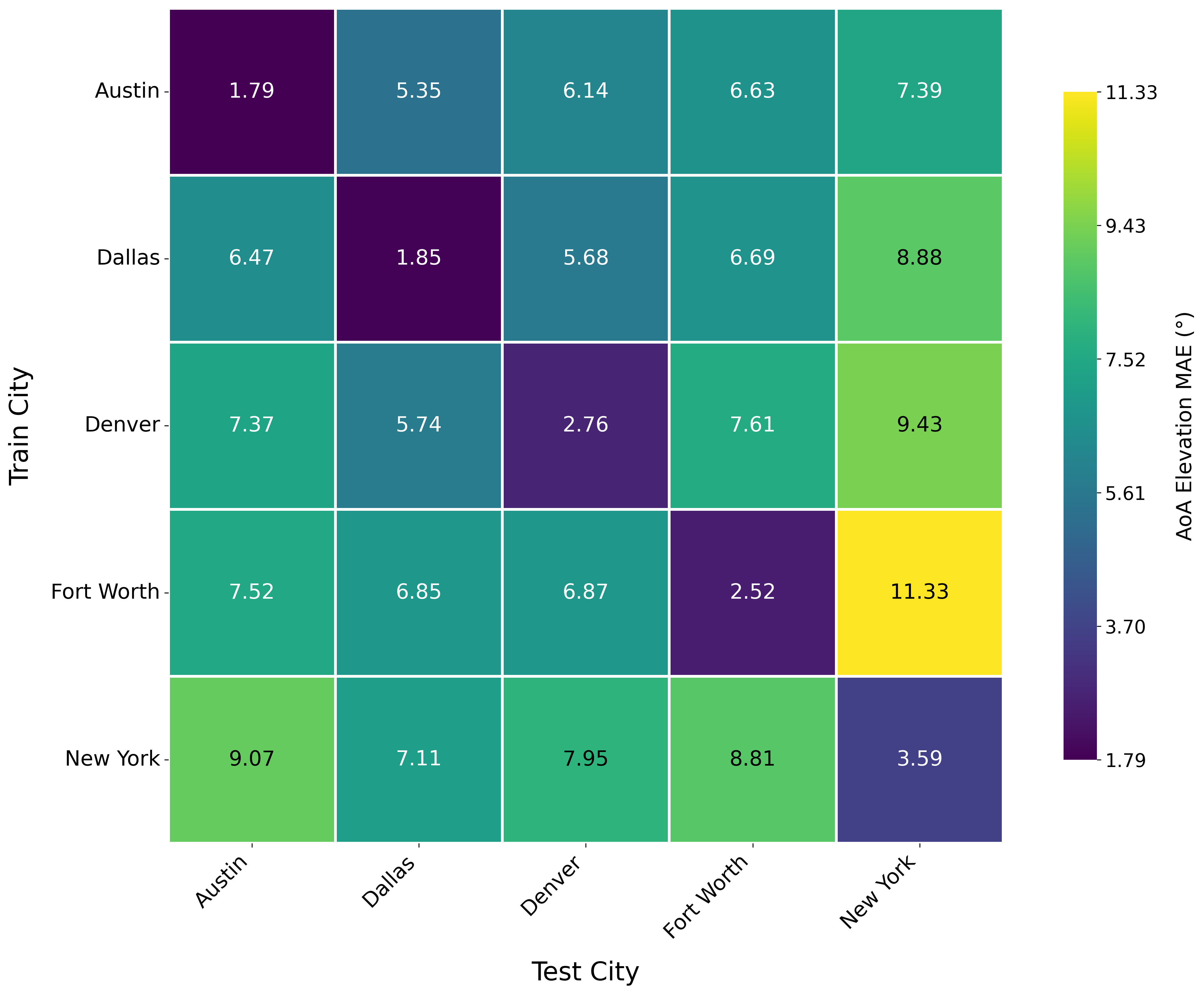} & \\
Avg.\ \gls{aoa} Azimuth \gls{mae} ($^\circ$) & Avg.\ \gls{aoa} Elevation \gls{mae} ($^\circ$) & \\
\end{tabular}
\caption{Cross-city transfer matrices for average delay \gls{mae}, average \gls{aod} azimuth \gls{mae}, average \gls{aod} elevation \gls{mae}, average \gls{aoa} azimuth \gls{mae}, and average \gls{aoa} elevation \gls{mae}.}
\label{fig:heatmap_extra}
\end{figure*}

Figure~\ref{fig:heatmap_extra} shows the cross-city transfer matrices for the remaining five metrics: average delay \gls{mae}, average \gls{aod} azimuth \gls{mae}, average \gls{aod} elevation \gls{mae}, average \gls{aoa} azimuth \gls{mae}, and average \gls{aoa} elevation \gls{mae}.
The pattern of off-diagonal degradation is consistent across all metrics, confirming that the transfer gap is a general phenomenon driven by per-city channel distribution shift rather than metric-specific artifacts.

\end{document}